\documentstyle[12pt]{article}
\begin{document}
\pagestyle{plain}

%  user-defined commands:
\newcommand{\be}{\begin{equation}}
\newcommand{\ee}{\end{equation}}
\newcommand{\bea}{\begin{eqnarray}}
\newcommand{\eea}{\end{eqnarray}}
\newcommand{\vp}{\varphi}
\newcommand{\pr}{\prime}
\newcommand{\sech} {{\rm sech}}
\newcommand{\cosech} {{\rm cosech}}
\newcommand{\psib} {\bar{\psi}}
\newcommand{\cosec} {{\rm cosec}}
\title{
\rightline{LA-UR-94-569}  \vskip .1in
Supersymmetry and Quantum Mechanics}
\author{
Fred Cooper \\
{\small \sl Theoretical Division, Los Alamos National Laboratory,}\\
{\small \sl Los Alamos, NM 87545}\\
{\small \sl   }\\
\and Avinash Khare \\
{\small \sl Institute of Physics,} \\
{\small \sl Bhubaneswar 751005, INDIA}\\
{\small \sl   }\\
\and Uday Sukhatme \\
{\small \sl Department of Physics, University of Illinois at Chicago,} \\
{\small \sl Chicago, IL 60607}}
\maketitle

\begin{abstract}

In the past ten years, the ideas of supersymmetry have been profitably applied
to many nonrelativistic quantum mechanical problems. In particular, there is
now a much deeper understanding of why certain potentials are analytically
solvable and an array of powerful new approximation methods for handling
potentials which are not exactly solvable. In this report, we review the
theoretical formulation of supersymmetric quantum  mechanics and discuss many
applications. Exactly solvable potentials can be  understood in terms of a few
basic ideas which include supersymmetric partner potentials, shape invariance
and operator transformations. Familiar solvable potentials all have the
property of shape invariance. We describe new exactly solvable shape invariant
potentials which include the recently discovered self-similar
potentials as a special case. The connection between inverse scattering,
isospectral potentials and supersymmetric quantum mechanics is discussed
and multi-soliton solutions of the KdV equation are constructed. Approximation
methods are also discussed within the framework of supersymmetric quantum
mechanics and in particular it is shown that a supersymmetry inspired WKB
approximation is exact for a class of shape invariant potentials.
Supersymmetry ideas give particularly nice results for the tunneling rate in a
double well potential and for improving large $N$ expansions.  We also discuss
the problem of a charged Dirac particle in an  external magnetic field and
other
potentials in terms of supersymmetric quantum mechanics.  Finally, we discuss
structures more general than supersymmetric quantum mechanics such as
parasupersymmetric quantum mechanics in which there is a symmetry between a
boson and a para-fermion of order $p$.
\end{abstract}
\pagebreak
\tableofcontents
\pagebreak

\section{Introduction}
\label{sec1}

Physicists have long strived to obtain a
unified description of all basic interactions of nature, i.e. strong,
electrowea
 k,
and gravitational interactions.  Several ambitious attempts have been made in
the last two decades, and it is now widely felt that supersymmetry (SUSY)
is a necessary
ingredient in any unifying  approach. SUSY relates bosonic and
fermionic
degrees of freedom and  has the virtue of taming ultraviolet divergences.
It was discovered in 1971 by Gel'fand and Likhtman \cite{Gelfand71},
Ramond \cite{Ramond71} and Neveu and Schwartz \cite{Neveu71} and
later was rediscovered by several groups \cite{Volkov73,Wess74,Sohnius85}.
The algebra involved in  SUSY is a graded
Lie algebra which closes under a combination of commutation and
anti-commutation relations. It  was first introduced in the context of the
string models to unify the bosonic and the fermionic sectors.  It was later
shown by Wess and Zumino \cite{Wess74} how to construct a $3+1$
dimensional field theory which
was invariant under this symmetry and which had very interesting properties
such
as a softening of ultraviolet divergences as well as having paired fermionic
and bosonic degrees of freedom. For particle theorists, SUSY offered a
possible way of unifying space-time and internal symmetries of the S-matrix
which avoided the no-go theorem of Coleman and Mandula \cite{Coleman67} which
was based on the assumption of a Lie algebraic realization of symmetries
(graded Lie algebras being unfamiliar to particle theorists at the time of the
proof of the no-go theorem). Gravity was generalized by incorporating  SUSY
to a theory called supergravity \cite{Ferrara76,Deser76}. In such theories,
Einstein's general theory of relativity  turns out to be a necessary
consequence of a local gauged SUSY. Thus, local SUSY theories provide a
natural framework for the unification of gravity with the other
fundamental interactions of nature.

   Despite the beauty of all these unified theories, there has so far been
no experimental evidence of SUSY being realized in nature.  One of the
important predictions of unbroken SUSY theories is the existence of SUSY
partners of quarks, leptons and gauge bosons which have the same masses as
their
SUSY counterparts. The fact that no such particles have been seen implies that
SUSY must be spontaneously broken. One hopes that the scale of this breaking
is of the order of the electroweak scale of 100 GeV in order that it can
explain the hierarchy problem of mass differences.  This leads to a conceptual
problem since the natural scale of symmetry breaking is the gravitational or
Planck scale which is of the order of $10^{19}$ GeV.  Various schemes have
been invented to try to resolve the hierarchy problem, including the idea
of non-perturbative breaking of SUSY. It was in the context of this
question that SUSY was first studied in the simplest case of SUSY quantum
mechanics (SUSY QM) by Witten \cite{Witten81} and Cooper and Freedman
\cite{Cooper83}.
In a subsequent paper, a topological index was introduced (the Witten index)
by Witten \cite{Witten82} for studying SUSY breaking and several people
studied the possibility that
instantons provide the non-perturbative mechanism for SUSY breaking. In the
work of Bender et al. \cite{Bender83} a new critical index was introduced to
study
non-perturbatively the breakdown of SUSY in a lattice regulated theory. Thus,
in
the early days, SUSY was studied in quantum mechanics as a testing ground for
the non-perturbative methods of seeing SUSY breaking in field theory.

   Once people started studying various aspects of SUSY QM, it was soon clear
that this field was interesting in its own
right, not just as a model for testing field theory methods. It was
realized that SUSY gives insight into the factorization
method of Infeld and Hull \cite{Infeld51} which was the first method to
categorize the analytically solvable potential problems.  Gradually a whole
technology was evolved based on SUSY to understand the solvable potential
problems. One purpose of this article is to review some of the
major developments in this area.

      Before we present a brief historical development of supersymmetric
quantum mechanics, let us
note another remarkable aspect. Over the last 10 years, the ideas
of SUSY  have stimulated new approaches to other branches
of physics \cite{Kostelecky85}.  For example, evidence has been found for a
dynamical SUSY relating even-even and even-odd nuclei. The Langevin
equation and the method of stochastic quantization has a path integral
formulation which embodies SUSY.  There have also been applications of SUSY
in atomic, condensed matter and statistical physics \cite{Kostelecky85}.

     In SUSY QM one is considering a simple realization
of a SUSY algebra involving the fermionic and the bosonic operators.  Because
of the existence of the fermionic operators which commute
with the Hamiltonian, one obtains
specific relationships between the energy eigenvalues, the eigenfunctions and
the S-matrices of the component parts of the full SUSY Hamiltonian. These
relationships will be exploited in this article to categorize analytically
solvable
potential problems.  Once the algebraic structure is understood, the results
follow and one never needs to return to the origin of the Fermi-Bose symmetry.
In any case,
the interpretation of SUSY QM as a degenerate Wess-Zumino field
theory in one dimension has not led to any further insights into the workings
of
SUSY QM.

   The introduction by Witten \cite{Witten82} of a topological invariant to
study dynamical SUSY breaking led to a flurry of interest in the topological
aspects of SUSY QM.  Various properties of the Witten index
were studied in SUSY QM and it was shown that in theories with
discrete as well as continuous spectra, the index could display
anomalous behavior
\cite{Akhoury84,Niemi84,Khare84a,Cecotti83,Boyanovsky84,%
Freedman85,Girardello83,Kihlberg85}.
Using the
Wigner-Kirkwood $\hbar$ expansion it was shown that
for systems in one and two dimensions, the first term in the $\hbar$ expansion
 gives the exact Witten index
 \cite{Bhaduri89}. Further, using the methods of SUSY QM, a derivation of the
Atiyah-Singer index theorem was also given
\cite{Alvarez83,Alvarez84,Alvarez85,Friedan85}.
In another development, the relationship between SUSY and the stochastic
differential equations such as the Langevin equation was elucidated and
exploited by Parisi and Sourlas \cite{Parisi82} and Cooper and Freedman
\cite{Cooper83}.  This connection, which implicitly existed  between the
Fokker-Planck equation, the path integrals and
the Langevin equation was then used to prove algorithms about the stochastic
quantization as well as to solve non-perturbatively for the correlation
functions of
SUSY QM  potentials using the Langevin equation.

  A path integral formulation of SUSY QM was first given by Salomonson
and van Holten \cite{Salomonson82}.  Soon afterwards it was shown by using
SUSY methods, that the tunneling rate through double well barriers could be
accurately determined \cite{Bernstein84,Kumar86,Marchesoni88,Keung88}. At the
same time, several workers extended ideas of SUSY QM  to higher dimensionsal
systems as well as to systems with large numbers of particles with
a motivation to understand the potential problems of widespread interest
in nuclear, atomic,
statistical and condensed matter physics
\cite{Crumbrugghe83,Urrutia83,Ui84,Andrianov85,%
Khare84,Balantekin85,Kostelecky84,Rau86,Blockley85}.

    In 1983, the concept of a shape invariant potential (SIP)
within the structure of SUSY QM was introduced by Gendenshtein
\cite{Gendenshtein83}. This Russian paper remained largely unnoticed for
several years. A potential is said to be shape invariant if its
SUSY partner potential has the same spatial dependence
as the original potential  with possibly altered
parameters. It is readily shown that for any SIP, the energy eigenvalue
spectra could be obtained algebraically \cite{Gendenshtein83}.  Much later, a
list of SIPs was given and it was shown that the energy
eigenfunctions as well as the scattering matrix could also be obtained
algebraically for these potentials \cite{Dutt86,Dabrowska88,Cooper88a,Khare88}.
It was soon realized that the formalism of SUSY QM plus shape invariance
(connected with translations of parameters) was intimately connected to
the factorization method of Infeld and Hull \cite{Infeld51}.

It is perhaps appropriate at this point to digress a
bit and talk about the history of the factorization method.
The factorization  method was first introduced by
Schr\"{o}dinger \cite{Schrodinger40} to solve the hydrogen atom problem
algebraically. Subsequently, Infeld and Hull \cite{Infeld51} generalized
this method and obtained
a wide class of solvable potentials by considering six different forms of
factorization. It turns out that the factorization method as well as the
methods of SUSY QM including the
concept of  shape invariance (with translation of parameters),
are both reformulations \cite{Reid72} of  Riccati's idea of
using the equivalence between the solutions
of the Riccati equation and a related second order linear differential
equation.
This method was supposedly used for the first time by Bernoulli
and the history is discussed in detail by Stahlhofen
\cite{Stahlhofen89}.

The general problem of the classification of SIPs
has not yet been solved. A partial classification of
the SIPs involving a translation of parameters was done by Cooper et al.
\cite{Cooper87,Chuan91}. It turns out that in this
case one only gets all the standard
analytically solvable potentials contained in the list given by
 Dutt et al. \cite{Dutt88} except for one which was later pointed out
by  Levai \cite{Levai89}. The connection between SUSY, shape
invariance and solvable potentials \cite{Natanzon71,Ginocchio84} is also
discussed in the paper of Cooper et al. \cite{Cooper87} where these authors
show that shape invariance even though sufficient, is not necessary for
exact solvability. Recently, a new class of SIPs has been discovered which
involves a scaling of parameters \cite{Khare93a}. These new potentials as
well as multi-step
SIPs \cite{Barclay93}  have been studied, and
their connection with self-similar potentials as well as with q-deformations
has been explored \cite{Shabat92,Spiridonov92a,Spiridonov92b,Degasperis92}.

        In yet another development, several people showed that SUSY QM
 offers a simple way of obtaining isospectral potentials by using either the
Darboux \cite{Darboux82} or Abraham-Moses \cite{Abraham80} or
Pursey \cite{Pursey86} techniques, thereby offering glimpses of the
deep connection
between the methods of the inverse quantum scattering \cite{Chadan77},
and SUSY QM \cite{Nieto84,Sukumar85a,Khare89a,Khare89b,Keung89}. The intimate
connection between the soliton solutions of the KdV hierarchy and SUSY QM
was also  brought out at this time \cite{Kwong86,Sukumar86,Wang90,Kwong89}.

        Approximate methods based on SUSY QM have also been developed.
Three of the
notable ones are  the $1/N$ expansion within SUSY QM \cite{Imbo85b}, $\delta$
expansion for the superpotential \cite{Cooper90} and a SUSY inspired WKB
approximation (SWKB) in quantum mechanics for the  case of unbroken SUSY
\cite{Comtet85,Khare85,Dutt91a}. It turns out that the SWKB approximation
preserves the exact SUSY relations between the energy  eigenvalues as well as
the scattering amplitudes of the partner potentials \cite{Dutt91b}. Further,
it is not only exact for large $n$ (as any WKB approximation is) but by
construction it is also exact for the ground state of $V_1(x)$. Besides it
has been proved \cite{Dutt86} that the
lowest order SWKB approximation necessarily gives the exact spectra for all SIP
 (with translation).  Subsequently a systematic higher order SWKB expansion
has been developed and it has been explicitly
shown that to $O(\hbar^{6})$  all the higher order corrections are zero for
these SIP \cite{Adhikari88}.  This has subsequently been  generalised to
all orders
in $\hbar$ \cite{Raghunathan87,Barclay91}. Energy eigenvalue spectrum has
also been obtained for several non-SIP \cite{Dutt87,Roy88,DeLaney90,Varshni92}
and it turns out that in many of the cases the SWKB does better than
the usual WKB approximation. Based on a study of these and other examples, it
has been suggested
that shape invariance is not only sufficient
but perhaps necessary for the lowest order SWKB to give the exact bound
state spectra \cite{Khare89}. Some attempts have also been made to obtain the
bound state eigenfunctions
within the SWKB formalism \cite{Fricke88,Murayama89,Sukhatme90}.

        Recently, Inomata and Junker \cite{Inomata92} have derived the
lowest order SWKB quantization condition (BSWKB) in case SUSY is broken. It
has recently been
shown that for the cases of shape invariant three dimensional oscillator
as well as for P\"{o}schl-Teller I and
II potentials with broken SUSY, this lowest order BSWKB calculation gives
the exact spectrum \cite{Dutt93a,Inomata93}. Recently, Dutt et al. \cite
{Dutt93b} have also developed a systematic higher order BSWKB expansion and
using it have shown that in all the three (shape invariant) cases, the higher
order corrections to $O(\hbar^6)$  are zero. Further, the energy eigenvalue
spectrum has also been obtained in the case of several non-SIP and it turns
out that in many cases BSWKB does as well as (if not better than) the
usual WKB approximation \cite{Dutt93b}.

        Recursion relations between the propagators pertaining to the SUSY
partner potentials have been obtained and  explicit expressions for
propagators of several SIP have been obtained \cite{Das90,Jayannavar93}.

        Several aspects of the Dirac equation have also been studied within
SUSY QM formalism \cite{Thaller92,Cooper88,Nogami93}.  In particular, it has
been shown using the
results of SUSY QM and shape invariance that whenever there is an analytically
solvable Schr\"{o}dinger problem in 1-dimensional QM then there always exists a
corresponding Dirac problem with scalar interaction in 1+1 dimensions which is
also analytically solvable. Further, it has been shown that there is always
SUSY
for massless Dirac
equation in two as well as in four Euclidean dimensions. The celebrated problem
of the Dirac particle in a Coulomb field has also been solved algebraically by
using the concepts of SUSY and shape invariance \cite{Sukumar85b}. The SUSY of
the Dirac electron in the field of a magnetic monopole has also been
studied \cite{dHoker84,Yamagishi84}. Also, the classic calculation of Schwinger
on pair production from strong fields can be dramatically simplified by
exploiting SUSY.

   The formalism of SUSY QM has also been recently
extended and models for parasupersymmetric QM
\cite{Rubakov88,Khare92,Khare93b} as
well as orthosupersymmetric QM \cite{Khare93c} of arbitrary order have been
written down. The question of singular superpotentials has also been
discussed in some detail within SUSY QM formalism
\cite{Jevicki84,Shifman88,Casahorran91,Roy85,Panigrahi93}.  Very recently
it has been shown that SUSY
QM offers a systematic method \cite{Pappademos93} for constructing bound
states in the continuum \cite{von Neumann29,Stillinger75,Ballentine90}.

        As is clear from this (subjective) review of the field, several
aspects of SUSY QM have been explored in great detail in the last ten years
and it is almost impossible to cover all these topics and do proper justice to
them.  We have therefore, decided not to pretend to be objective but cover only
those topics which we believe to be important and which we believe have not so
far been discussed in great detail in other review  articles.  We have,
however,
included in Sec. 14  a list of the important topics missed in
this review and given some references so that the intersested reader can trace
back and study these topics further.  We have been fortunate in the sense that
review articles already exist in this field where several of these missing
topics have been discussed
\cite{Gendenshtein85,Haymaker86,Lancaster84,Stedman85,%
Dutt88,Lahiri90,deLange91,
Khare91,Thaller92}.
We must also apologize to several authors whose work may not have been
adequately quoted in this review article in spite of our best attempts.

The plan of the article is the following:  In
Sec. 2, we discuss the Hamiltonian formalism of SUSY QM. We have
deliberately kept this section at a pedagogical level so that a graduate
student
should be able to understand and work out all the essential details.  The SUSY
algebra is given and the connection between the energy eigenvalues,
the eigenfunctions and the S-matrics of the two SUSY partner
Hamiltonians are derived.
The question of unbroken vs. broken SUSY is also introduced at a pedagogical
level using polynomial potentials of different parity and the
essential ideas of partner potentials are illustrated using  the example
of a one
dimensional infinite square well.  The ideas of SUSY are made more explicit
through the example of one dimensional SUSY harmonic oscillator.

 In Sec. 3, we discuss the connection between SUSY QM and factorization and
show how one can always construct a hierarchy of $p \geq 1$ Hamiltonians
 with known energy eigenvalues, eigenfunctions and $S$-matrices by starting
from
any given Hamiltonian with $p$ bound states whose energy eigenvalues,
eigenfunctions and S-matrices (reflection coefficient $R$ and transmission
coefficient $T$) are known.

Sec. 4 is in a sense the heart of the article.  We first show that if
the SUSY partner potentials satisfy an integrability condition called shape
invariance then the energy eigenvalues, the eigenfunctions and the
S-matrices for these potentials can be obtained algebraically.  We then
discuss satisfying the condition of shape
invariance with translations and show that in this case the classification of
SIP can be done and the resultant list of solvable
potentials include essentially all the popular ones that are
included in the standard QM textbooks. Further, we discuss the newly
discovered one
and multi-step shape invariant potentials when the partner potentials are
related by a change of parameter of a  scaling rather than translation type.
It
turns out that in most of these cases the resultant potentials are
reflectionless and contain an infinte number of bound states.  Explicit
expressions for the energy  eigenvalues, the eigenfunctions and the
transmission
coefficients are obtained in various cases. It is further shown that the
recently discovered self-similar potentials which also statisfy $q$-SUSY,
constitute a special case of the SIP. Finally, we show that a wide class
of noncentral but separable potential problems are also algebraically
solvable by using the results obtained for the SIP \cite{Khare93f}. As a
by product, exact solutions of a number of three body problems in one
dimension are obtained analytically \cite{Khare93g}.

 In Sec. 5, we
discuss the solvable but non-shape invariant Natanzon and Ginocchio potentials
and show that using the ideas of SUSY QM, shape invariance and operator
transformations, their spectrum can be obtained algebraically. We also
show that the Natanzon potentials are not the most general solvable potentials
in nonrelativistic QM.

Sec. 6 is devoted to a discussion of the SUSY inspired WKB approximation
(SWKB) in quantum mechanics both when SUSY is unbroken and when it is
spontaneously
broken. In the unbroken case, we first develop a systematic higher order
$\hbar$ -expansion for the energy eigenvalues and then show that for the SIP
with translation, the lowest order term in the
$\hbar$-expansion gives the exact bound state spectrum. We also show here
that even for many of the non-SIP, the SWKB does as well as if
not better than the WKB approximation. We then discuss the
broken SUSY case and in that case too we develop a systematic $\hbar$-expansion
(BSWKB) for the energy eigenvalues. We show that even in the broken case the
lowest order BSWKB gives the exact bound state spectrum for
the SIP with translation.

Sec. 7 contains a description of how SUSY QM can be used to construct
multiparameter
families of isospectral and strictly isospectral potentials.  As an
illustration
we give plots of the one continuous parameter family of isospectral potentials
corresponding to  the one-dimensional harmonic oscillator.  From here
we are immediately able to construct the two
as well as multisoliton solutions of the KdV equation.

ln Sec. 8, we discuss
more formal aspects of SUSY QM. In particular, we discuss the path integral
formulation of SUSY QM as well as various subtleties associated with the Witten
index  \cite{Witten82} $\Delta = (-1)^F$. We also discuss in some detail
the connection of SUSY QM with
classical stochastic processes and discuss  how one can develop a
systematic strong coupling and $\delta$ -expansion for the Langevin equation.

Sec. 9 contains a description of several approximation schemes like the
variational method, the $\delta$-expansion, large-$N$ expansion, energy
splitting in double well potentials within the SUSY QM framework.

In Sec. 10, we discuss the question of SUSY QM in
higher dimensions.  In particular, we discuss the important problem of a
charged particle in a magnetic field (Pauli equation) in two dimensions
and show that there is
always a SUSY in the problem so long as the gyromagnetic ratio is 2.

In Sec. 11,  we show that there is always a SUSY in the case of massless
Dirac equation in two or four Euclidean dimensions in the background of
external
electromagnetic fields.
Using the results of SUSY QM we then list a number of problems with nonuniform
magnetic field which can be solved analytically. We also show here
that whenever a Schr\"{o}dinger problem in 1-dimensional QM is analytically
solvable, then one can always obtain an exact solution of a corresponding Dirac
problem with the scalar coupling. We also show how the calculation of the
fermion propagator in an external field can be simplified by exploiting SUSY.

Sec. 12 contains a comprehensive discussion of
the general problem of singular superpotentials, explicit breaking
of SUSY, negative energy states and unpaired positive energy eigenstates.  We
also show here  how to construct bound states in the continuum within the
formalism of SUSY QM.

Quantum mechanical models relating bosons and  parafermions of order $p$ are
described in Sec. 13. It is shown that such models encompass $p$
supersymmetries.  Various consequences of such models are discussed including
the  connection with the hierarchy of Hamiltonians as well as with strictly
isospectral potentials.  We also discuss a quantum mechanical model where
instead there is a symmetry between a boson and an orthofermion of order $p$.

Finally, in Sec. 14,  we give a list of
topics related to SUSY QM which we have not discussed and provide some
reference
 s
for each of these topics.

\section{Hamiltonian Formulation of Supersymmetric Quantum Mechanics}
\label{sec2}

One of the key ingredients in solving exactly for the spectrum
of  one dimensional potential problems is the connection between the
bound state wave functions and the
potential.  It is not usually appreciated that once one knows the
ground state wave function (or any other bound state wave function)
then one knows exactly the potential (up to a constant). Let us choose
the ground state energy for the moment to be zero. Then one has from
the Schr\"{o}dinger equation that the ground state wave function $\psi_0 (x)$
obeys
\be
H_1 \psi_0 (x) = -{ \hbar^2 \over 2m} {d^2 \psi_0 \over dx^2 }
+ V_1(x) \psi_0(x)  = 0 ~,
\ee
so that
\be
V_1(x) = { \hbar^2 \over 2m} {\psi_0^{\prime \prime}(x) \over  \psi_0(x)}~.
\ee
This allows a global
reconstruction of the potential $V_1(x)$ from a knowledge of its ground state
wave function which has no nodes (we will discuss the case of using the excited
wave functions  later in Sec. 12). Once we
realize this, it is now  very simple to factorize the Hamiltonian using the
following ansatz:
\be \label{eq2.3}
H_1= A^{\dag} A
\ee
where
\bea \label{aadag}
A = {\hbar \over \sqrt{2m}}{d \over dx} + W(x)~,~
A^{\dag} = { -\hbar \over \sqrt{2m}}{d \over dx} + W(x)~.
\eea
This allows us to identify
\be
V_1(x) = W^2(x) - {\hbar \over \sqrt{2m}}W^{\prime}(x)~.
\ee
This equation is the well-known Riccati equation. The quantity $W(x)$ is
generally referred to as the ``superpotential" in SUSY QM
literature.
The  solution for $W(x)$ in terms of the ground state wave function is
\be \label{defW}
W(x) = - {\hbar \over \sqrt{2m}}{ \psi_0^{\prime}(x)  \over \psi_0 (x)}~~.
\ee
This solution is obtained by recognizing that once we satisfy
$A \psi_0 = 0$, we automatically have a solution to $H_1
\psi_0=A^{\dag}A \psi_0=0$.

The next step in constructing the SUSY theory related to
the original Hamiltonian $H_1$ is to define the operator $H_2=AA^{\dag}$
obtained by reversing the order of $A$ and $A^{\dag}$.  A little simplification
shows that the operator $H_2$ is in fact a Hamiltonian corresponding to a new
potential $V_2(x)$.
\be
H_2 = -{ \hbar^2 \over 2m} {d^2\over dx^2}+V_2(x)~, ~ V_2(x)=
 W^2(x) + {\hbar\over \sqrt{2m}} W^\prime (x)~.
\ee
The potentials $V_1(x)$ and $V_2(x)$ are known as supersymmetric partner
potentials.

 As we shall see, the
energy eigenvalues, the wave functions and the S-matrices  of $H_1$ and $H_2$
are related.
To that end notice that the energy eigenvalues of both $H_1$ and  $H_2$
are positive semi-definite  ($E_n^{(1,2)} \geq 0$) .
For $n>0$, the Schr\"{o}dinger equation for $H_1$
\be
H_1 \psi_n^{(1)} = A^{\dag} A \psi_n^{(1)} =  E_n^{(1)}\psi_n^{(1)}
\label{deg1}
\ee
implies
\be
H_2 (A \psi_n^{(1)})  = A A^{\dag} A \psi_n^{(1)} =  E_n^{(1)}(A
\psi_n^{(1)})~.
\label{deg2} \ee
Similarly, the Schr\"{o}dinger equation for $H_2$
\be
H_2 \psi_n^{(2)} = A A^{\dag}  \psi_n^{(2)} =  E_n^{(2)}\psi_n^{(2)}
\label{deg3} \ee
implies
\be
H_1 (A^{\dag}  \psi_n^{(2)})  =  A^{\dag} A A^{\dag} \psi_n^{(2)} =  E_n^{(2)}
(A^{\dag}
\psi_n^{(2)})~. \label{deg4}
\ee
{}From eqs. (\ref{deg1})-(\ref{deg4}) and the fact that
$E_0^{(1)} =0 $, it is clear that the eigenvalues and eigenfunctions
of the two Hamiltonians $H_1$ and $ H_2$ are related by
$(n=0,1,2,...)$

\be
E_n^{(2)} = E_{n+1}^{(1)}, \hspace{.2in} E_0^{(1)} = 0,
\ee
\be \label{psin2}
\psi_n^{(2)} = [E_{n+1}^{(1)}]^{-1/2} A \psi_{n+1}^{(1)},
\ee
\be \label{psin1}
\psi_{n+1}^{(1)} = [E_{n}^{(2)}]^{-1/2} A^{\dag}  \psi_{n}^{(2)}.
\ee
Notice that if $\psi_{n+1}^{(1)}$ ( $\psi_n^{(2)}$) of $H_1$ ($H_2$) is
normalized then the wave function  $\psi_n^{(2)}$ ($\psi_{n+1}^{(1)}$)
in eqs. (\ref{psin2}) and (\ref{psin1}) is also normalized. Further, the
operato
 r
$A$ ($A^{\dag}$) not only converts an eigenfunction of $H_1$ $(H_2)$ into an
eigenfunction of $H_2 (H_1)$ with the same energy, but it also destroys
(creates) an extra node in the eigenfunction.
Since the ground state wave function of $H_1$ is annihilated by the operator
$A$, this state has no SUSY partner.  Thus the picture we get is that
knowing all the eigenfunctions of $H_1$ we can determine the eigenfunctions
of $H_2$ using the operator $A$, and vice versa using $A^{\dag}$ we can
reconstruct all the eigenfunctions of $H_1$ from those of $H_2$ except
for the ground state. This is illustrated in Fig. 2.1.

  The underlying reason for the degeneracy of the spectra of $H_1$ and $H_2$
can be understood most easily from the properties of the  SUSY
algebra.  That is we can consider a matrix SUSY Hamiltonian of the form
\be
 H =  \left[\matrix{ H_1 &0\cr
0&H_2\cr}\right]
\ee
which contains both $H_1$ and $H_2$.  This matrix Hamiltonian is part of a
closed algebra which contains both bosonic and fermionic
operators with commutation and anti-commutation relations. We consider
the operators
\be \label{q}
Q= \left[\matrix{ 0 & 0\cr
A &0\cr}\right],
\ee
\be \label{qdag}
Q^{\dag}= \left[\matrix{ 0 & A^{\dag}\cr
0&0\cr}\right]
\ee
in conjunction with $H$.  The following commutation and anticommutation
relation
 s
then describe the closed superalgebra $sl(1/1)$:

\bea
[H,Q] &=&[H,Q^{\dag}] = 0 ~,\nonumber \\
\{Q,Q^{\dag}\} &=& H, \hspace{.2in}  \{Q,Q \} = \{Q^{\dag},Q^{\dag}\}=0~.
\label{susyalg}
 \eea

The fact that the supercharges $Q$ and $Q^{\dag}$ commute with $H$ is
responsibl
 e
for the  degeneracy.
The operators $Q$ and $Q^{\dag}$ can be interpreted as operators which
change bosonic degrees of freedom into fermionic ones and vice versa.
This will be elaborated further below  using the example of the SUSY
harmonic oscillator.

Let us look at a well known potential, namely the infinite square well
and determine its SUSY partner potential. Consider a particle of mass $m$ in an
infinite square well potential of width $L$.
\bea
V(x) &=& 0, \hspace{.5in} 0 \leq x \leq L ~, \nonumber \\
   &=& \infty, \hspace{.5in} -\infty < x <0, x > L~.
\eea
The ground state wave function is known to be
\be
\psi_0^{(1)} = (2/L)^{1/2} \sin(\pi x/L), \hspace{.3in} 0 \leq x \leq L~,
\ee
and the ground state energy is $E_0= {\hbar^2 \pi^2 \over 2mL^2}$.

Subtracting off the ground state energy so that we can factorize the
Hamiltonian
we have for $H_1$ = $H- E_0$ that the energy eigenvalues are
\be
E_n^{(1)} = {n(n+2) \over 2mL^2} \hbar^2 \pi^2
\ee
and the eigenfunctions are
\be
\psi_n^{(1)} = (2/L)^{1/2} \sin {(n+1)\pi x \over L}~,
\hspace{.5in} 0 \leq x \leq L ~.
\ee
The superpotential for this problem is readily obtained using eq. (\ref{defW})

\be
W(x) = - {\hbar \over \sqrt{2m}} {\pi \over L} {\rm cot} (\pi x /L)
\ee
and hence the supersymmetric partner potential $V_2$ is
\be
V_2(x) = {\hbar^2 \pi^2 \over 2mL^2} [2 {\rm cosec}^2 (\pi x /L) -1 ]~.
\ee
The wave functions for $H_2$ are obtained by applying the operator $A$
to the wave functions of $H_1$. In particular we find that
\be
\psi_0^{(2)} \propto \sin^2 (\pi x /L) ,\hspace{.3in}
\psi_1^{(2)} \propto \sin (\pi x /L)\sin (2\pi x /L)~.
\ee

Thus we have shown using SUSY that two rather different
potentials corresponding to $H_1$ and $H_2$ have exactly the same spectra
except
for the fact that $H_2$ has one fewer bound state. In Fig. 2.2 we show
the supersymmetric partner potentials $V_1$ and $V_2$ and the first few
eigenfunctions. For  convenience we have chosen $L=\pi$ and $\hbar=2m=1$.

  Supersymmetry also allows one to relate the reflection and transmission
coefficients in situations where the two partner potentials have
continuum spectra. In order for scattering to take place in both
of the partner potentials, it is necessary that the potentials $V_{1,2}$
are finite as $x \rightarrow  - \infty $ or as  $x \rightarrow  + \infty $
or both.  Define:
\be
W(x \rightarrow \pm \infty) = W_{\pm}~.
\ee
Then
\be
V_{1,2} \rightarrow W_{\pm}^2 \hspace{.5in} {\rm as} \hspace{.1in} x
\rightarrow
\pm \infty.
\ee

Let us consider an incident plane wave $ e^{ikx}$ of energy $E$ coming from
the direction $ x \rightarrow - \infty$. As a result of scattering from the
potentials
$V_{1,2}(x)$ one would obtain transmitted waves $T_{1,2}(k) e^{ik'x}$
and reflected waves $R_{1,2}(k) e^{-ikx}$.
Thus we have
\bea
\psi^{(1,2)}(k, x \rightarrow - \infty) && \rightarrow  e^{ikx} + R_{1,2}
e^{-ikx}~, \nonumber \\ \psi^{(1,2)}(k', x \rightarrow + \infty) && \rightarrow
T_{1,2} e^{ik'x}~.
\eea
SUSY connects continuum wave functions of $H_1$ and $H_2$ having the
same energy analogously to what happens in the discrete spectrum.
Thus we have the relationships:
\bea
 e^{ikx} + R_{1} e^{-ikx} &=& N [ (-ik+W_{-}) e^{ikx} +(ik+W_{-}) e^{-ikx}
R_2],
\nonumber \\
T_{1} e^{ik'x} &=&N[ (-ik'+W_{+}) e^{ik'x} T_2],
\eea
where $N$ is an overall normalization constant.  On equating terms with the
same
exponent and eliminating $N$, we find:
\bea \label{rt}
R_1(k) &=& \left( {W_{-} + ik \over W_{-} - ik }\right) R_2(k), \nonumber \\
T_1(k) &=& \left( {W_{+} - ik^{\prime} \over W_{-} - ik }\right) T_2(k),
\eea
where $ k$ and $k^{\prime}$ are given by
\be
k= (E- W_{-}^2)^{1/2}, \hspace{.2in} k^{\prime}= (E- W_{+}^2)^{1/2}.
\ee

A few remarks are now in order at this stage.\\
(1) Clearly $ |R_1|^2 =|R_2|^2$ and  $ |T_1|^2 =|T_2|^2$, that is the
partner potentials have identical reflection and transmission probabilities.\\
(2) $R_1(T_1)$ and $R_2(T_2)$ have the same poles in the complex plane
except that $R_1 (T_1)$ has an extra pole at $k=-iW_{-}$.  This pole is on
the positive imaginary axis only if $W_{-} <0$ in which case it corresponds to
a zero energy bound state.\\
(3) In the special case that $W_{+} =W_{-}$, we have that $T_1(k) = T_2(k)$.\\
(4) When $W_{-} =0$ then $R_1(k)= -R_2(k)$.

It is clear from these remarks that if one of the partner potentials  is a
constant  potential (i.e. a free particle), then the other partner will be
of necessity reflectionless.  In this way we can understand the reflectionless
potentials of the form $ V(x) =  A {\sech}^2\alpha x $ which play a
critical role in understanding the soliton solutions of the KdV hierarchy.
Let us consider the superpotential
\be
W(x) = A \ \rm{tanh}~ \alpha x~.
\ee
The two partner potentials are
\bea
V_1 &=& A^2 - A \ (A+\alpha {\hbar \over \sqrt{2m} }) \rm{sech}^2 \alpha x~,
\nonumber \\   V_2 &=& A^2 - A(A-\alpha {\hbar \over \sqrt{2m} }) \rm{sech}^2~.
\alpha x  \eea
We see that for $A=\alpha {\hbar \over \sqrt{2m} }$,  $ V_2(x)$ corresponds to
a constant potential so that the corresponding $V_1$ is a reflectionless
potential. It is worth noting that $V_1$ is $\hbar$-dependent. One can in fact
rigorously show,though it is not mentioned in most text books,that the
reflectionless potentials are necessarily $\hbar$-dependent.

So far we have discussed SUSY QM on the full line $(-\infty <x<\infty)$. Many
of these results have analogs for the $n$-dimensional potentials with spherical
symmetry.  For example, in three dimensions one can make a partial wave
expansion in terms of the wave functions:
\be
\psi_{nlm}(r,\theta,\phi) = {1 \over r} R_{nl}(r) Y_{lm} (\theta, \phi)~.
\ee
Then it is easily shown \cite{Landau77} that the reduced radial wave function
R satisfies the one-dimensional Schr\"{o}dinger equation $(0 <r<\infty)$
\be
-{ \hbar^2 \over 2m} {d^2 \psi(r) \over dr^2 } + [ V(r)+{l(l+1)\hbar^2  \over
2mr^2}] \psi (r) =E \psi(r)
\ee
We notice that there is an effective one dimensional potential which contains
the original potential plus an angular momentum barrier.  The
asymptotic form of the radial wave function for the $l'$th partial
wave is
\be
\psi(r,l) \rightarrow { 1 \over 2k^{\prime}}[S^{l}(k') e^{ik'r}- (-1)^{l}
e^{-ik'r}]~,
\ee
where $S^l$ is the scattering function for the $l'$th partial wave.
i.e. $S^l(k) = e^{i\delta_l(k)}$ and $\delta$ is the phase shift.

For this case we find the relations:
\be \label{eq2.36}
S_1^l(k') = \left( {W_{+} - ik' \over W_{+} + ik' }\right) S_2^{l}(k')~.
\ee
Here $W_{+} = W(r \rightarrow \infty)$.

We thus have seen that when $H_1$ contained a known
ground state wave function then we could factorize the Hamiltonian and
find a SUSY partner Hamiltonian $H_2$.  Now let us consider
the converse problem. Suppose we are given a superpotential $W(x)$. In this
case there are two possibilities. The candidate ground state wave function is
the ground state for $H_1$ (or $H_2$)  and can be obtained from:
\bea
 A \psi_0^{(1)}(x) & = &0  \hspace{.3in} \Rightarrow   \psi_0^{(1)}(x)  = N
\exp\left(-{\sqrt{2m} \over \hbar} \int^x W(y) \ dy \right)~, \nonumber \\
 A^{\dag} \psi_0^{(2)}(x) & = & 0 \hspace{.3in} \Rightarrow\psi_0^{(2)}(x)= N
\exp\left(+{\sqrt{2m} \over \hbar} \int^x W(y) \ dy \right)~. \label{gst}
\eea

By convention, we shall always choose $W$ in such a way that amongst $H_1,H_2$
only
$H_1$ (if at all) will have a normalizable zero energy ground state
eigenfunction.
This is ensured by choosing $W$ such that $W(x)$ is positive(negative) for
large
positive(negative) $x$. This defines $H_1$ to have fermion number zero in our
later formal treatment of SUSY.

If there are no normalizable solutions of this form, then $H_1$ does not have
a zero eigenvalue and SUSY is broken.  Let us now be more
precise. A symmetry of the Hamiltonian (or Lagrangian) can be spontaneously
broken if the lowest energy solution does not respect that symmetry, as for
example in a ferromagnet, where rotational invariance of the Hamiltonian
is broken by the ground state.  We can define the ground state in our system
by a two dimensional column vector:
\be \label{eq2.38}
|0 > = \psi_0(x) = \left[\matrix{\psi_0^{(1)}(x) \cr
\psi_0^{(2)}(x)\cr} \right]
\ee
For SUSY to be unbroken requires
\be
Q |0 > = Q^{\dag} |0 > = 0 |0 >
\ee

Thus we have immediately from eq. (\ref{susyalg} ) that the ground state energy
must be zero in this case.  For all the cases we discussed previously,
the ground state energy was indeed zero and hence the ground state wave
function for the matrix Hamiltonian can be written:

 \be
\psi_0(x) = \left[\matrix{\psi_0^{(1)}(x) \cr
0 \cr} \right]
\ee
where  $\psi_0^{(1)}(x)$  is given by eq. (\ref{gst}).

If we consider superpotentials of the
form
\be
W(x) = g x^{n}~~,
\ee
then for $n$ odd and $g$ positive one always has a normalizable ground state
wave function.  However for the case $n$ even and $g$ arbitrary, then
there is no normalizable  ground state wave function.
In general when one has a superpotential $W(x)$ so that
neither $Q$ nor $Q^{\dag}$ annihilates the ground state as
given by eq. (\ref{eq2.38}) then SUSY is broken and the potentials $V_1$ and
$V_2$ have degenerate positive
ground state energies.
Stated another way, if the
ground state energy of the matrix Hamiltonian is non zero then SUSY is
broken. For the case of broken SUSY the operators $A$ and $A^{\dag}$ no longer
change the number of nodes and there is a 1-1 pairing of all the eigenstates of
$H_1$ and $H_2$. The precise relations that one now obtains are: \be
E_n^{(2)} = E_{n}^{(1)} >0,~~~ n=0,1,2,...
\ee
\be \label{psin22}
\psi_n^{(2)} = [E_{n}^{(1)}]^{-1/2} A \psi_{n}^{(1)}~~,
\ee
\be \label{psin12}
\psi_{n}^{(1)} = [E_{n}^{(2)}]^{-1/2} A^{\dag}  \psi_{n}^{(2)}.
\ee
while the relationship between the scattering amplitudes is still given by
eqs. (\ref{rt}) or (\ref{eq2.36}).
The breaking of SUSY can be described by a topological quantum number
called the Witten index \cite{Witten82} which we will discuss later.
Let us however remember that
in general if the sign of $W(x)$ is opposite  as we approach infinity from
the positive and the negative sides, then SUSY is unbroken, whereas in the
other cases it is always broken.

\subsection{State Space Structure of the SUSY Harmonic Oscillator}
\label{sec2.1}

For the usual quantum mechanical harmonic oscillator one can introduce
a Fock space of boson occupation numbers where we label the states by the
occupation number $n$.  To that effect one introduces instead of P and q
the creation and annihilation operators $a $ and $a^{\dag}$. The usual harmonic
oscillator Hamiltonian  is
\be
{\cal H}={P^2 \over 2m}  + {1 \over 2} m \omega^2 q^2.
\ee
Let us rescale the Hamiltonian in terms of dimensionless coordinates and
momenta $x$ and $p$ so that we measure energy in units of $\hbar \omega$.
We put
\be
{\cal H}= H \hbar \omega, \hspace{.2in} q=({\hbar \over 2m \omega})^{1/2} x,
\hspace{.2in} P=(2m \hbar \omega)^{1/2} p.
\ee
Then
\be
H= (p^2+{x^2\over 4}), \hspace{.2in} [x,p] = i.
\ee
Now introduce
\be
 a=({x\over 2} +ip) , \hspace{.2in}  a^{\dag}=({x\over 2} -ip).
\ee
Then
\bea
[a,a^{\dag}] & = & 1, \hspace{.2in}
\left[ N,a  \right] = -a , \hspace{.2in} [N,a^{\dag}] = a^{\dag},
\nonumber \\
 N           & = & a^{\dag} a  , \hspace{.4in}  H= N+{1 \over 2}~~.
\eea

The usual operator formalism for solving the harmonic oscillator potential is
to
define the ground state by requiring
\be
     a|0> = 0 ~,
\ee
which leads to a first order differential equation for the ground state
wave function. The $n$ particle state (which is the $n$'th excited wave
function
in the coordinate representation) is then given by:

\be
|n_b>  = {  a^{\dag \ n }  \over  \sqrt{(n!)} } |0>~,.
\ee
where we have used the subscript $b$ to refer to the boson sector as distinct
from the fermions we will introduce below.
For the case of the SUSY harmonic oscillator one can rewrite the
operators $Q$  ($Q^{\dag})$ as a product of the bosonic operator $a$ and
the fermionic operator $\psi$. Namely we write $ Q= a \psi^{\dag} $ and
$ Q^{\dag}=  a^{\dag}\psi$ where the matrix fermionic creation and
annihilation operators are defined via:

\be
\psi =  \sigma_{+}
= \left[\matrix{ 0 & 1\cr
0 &0\cr}\right],
\label{psi}
\ee

\be
\psi^{\dag} =  \sigma_{-}=
 \left[\matrix{ 0 & 0\cr
1 &0\cr}\right].
\label{psidag}
\ee

Thus $\psi$ and $\psi^{\dag}$ obey the usual algebra of the fermionic creation
and annihilation operators, namely,
\be
\{\psi^{\dag},\psi\} =1, \hspace{.2in} \{\psi^{\dag},\psi^{\dag}\}=
\{\psi,\psi\} =0,
\ee
as well as obeying the commutation relation:

\be
[\psi,\psi^{\dag}] = \sigma_3 = \left[\matrix{ 1 & 0\cr
0 &-1\cr}\right].
\ee
The SUSY Hamiltonian can be rewritten in the form
\be
H= QQ^{\dag} + Q^{\dag}Q = (- {d^2 \over dx^2} +{x^2\over 4}) I
-{1\over 2} [\psi,\psi^{\dag}].
\label {susyh1}
\ee
The effect of the last term is to remove the zero point energy.

The state vector can be thought of as a matrix in the
Schr\"{o}dinger picture
or as the state   $|n_b , n_f>$ in this Fock space picture.
Since the fermionic creation and annihilation operators obey
anti-commutation relations hence the fermion number is either zero or one.
As stated before, we will choose the ground state of $H_1$ to have
zero fermion number. Then we can introduce the fermion
number operator
\be
   n_F =  {1-\sigma_3  \over 2} = { 1- [\psi,\psi^{\dag}] \over 2}~.
\ee

Because of the anticommutation relation, $n_f$ can only take on the values 0
and 1. The action of the operators $a,a^{\dag}, \psi , \psi^{\dag}$ in
this Fock space are then:
\bea
a |n_b ,n_f > & = &  |n_b-1, n_f>,  \hspace {.2in}  \psi |n_b, n_f >
= |n_b, n_f-1 > ,\nonumber \\
a^{\dag} |n_b, n_f > & = &  |n_b+1, n_f>,  \hspace {.2in}
\psi^{\dag} |n_b, n _f > = |n_b, n_f+1 >.
\eea
We now see that the operator $Q^{\dag}= -i a \psi^{\dag}$ has the
property of changing a boson into a fermion without changing the energy of the
state. This is the boson-fermion degeneracy characteristic of all SUSY
theories.

For the general case of SUSY QM, the operator $a$ gets replaced by $A$
in the definition of $Q$, $Q^{\dag}$, i.e. one writes $ Q= A \psi^{\dag} $ and
$ Q^{\dag}=  A^{\dag}\psi$.
The effect of $ Q$ and $ Q^{\dag}$ are now to relate the wave functions of
$H_1$ and $ H_2$ which have fermion number zero and one respectively but now
there is no simple Fock space description in the bosonic  sector because the
interactions are non-linear.
Thus in the general case, we can rewrite the SUSY Hamiltonian in the form

\be
H= (- {d^2 \over dx^2} + W^2)   I - [\psi,\psi^{\dag}] W^{\prime}.
\label {susyh2}
\ee
This form will be useful later when we discuss the Lagrangian formulation of
SUSY QM in Sec. 8.

\subsection{Broken Supersymmetry}
\label{sec2.2}

As discussed earlier,
for SUSY to be a good symmetry, the operators $Q$ and $Q^{\dag}$
must annihilate the vacuum. Thus the ground state energy of the
super-Hamiltonian must be zero since
\[ H= \{ Q^{\dag}, Q \}.
\]
Witten \cite{Witten82}  proposed an index to determine whether SUSY is broken
in
supersymmetric field theories. The index is defined by
\be
\triangle = {\rm Tr} (-1)^F ~,
\ee
where the trace is over all the bound states and continuum states of the
super-Hamiltonian. For SUSY QM, the fermion number $n_F \equiv F$
is defined by $ {1 \over 2} [1- \sigma_3]$ and we can represent $(-1)^F$ by the
matrix $\sigma_3$.
 If we write the eigenstates of H as the vector:
 \be
\psi_n(x) = \left[\matrix{\psi_n^{(+)}(x) \cr
\psi_n^{(-)}(x) \cr} \right]
\ee
then the $\pm$ corresponds to the eigenvalues of $(-1)^F$ being $\pm 1$. For
our conventions the eigenvalue +1 corresponds to $H_1$ and the bosonic
sector and the
eigenvalue $-1$  corresponds to $H_2$ and the fermionic sector.  Since the
bound
states of $H_1$ and $H_2$ are paired, except for the case of unbroken
SUSY where there is an extra state in the bosonic sector with $E=0$ we
expect for the quantum mechanics situation that $\triangle =0$ for
broken SUSY and $\triangle=1$ for unbroken SUSY.  In the general field theory
case, Witten gives arguments that in general the index measures $N_+(E=0)  -
N_- (E=0)$  which is the difference $\triangle$ between the number of Bose
states and Fermi states of zero energy. In field theories the Witten index
needs
to be regulated to be well defined so that one considers instead

\be
\triangle(\beta) = {\rm{Tr}} (-1)^F e^{-\beta H}~,
\ee
which for SUSY quantum mechanics becomes
\be
\triangle(\beta) ={\rm{Tr}}[ e^{-\beta H_1} -e^{-\beta H_2}].
\ee

In field theory it is quite hard to determine if SUSY is broken
non-perturbatively, and thus SUSY quantum mechanics became a testing
ground for finding different methods to understand non-perturbative
SUSY breaking.  In the quantum mechanics case, the breakdown of SUSY is related
to the question of whether there is a normalizable wave function solution to
the
equation $Q|0 > = 0|0>$ which implies
\be
 \psi_0(x) = N e^{-\int W(x) dx}.
\ee

As we said before, if  this candidate ground state wave function  does not
fall off fast enough at $\pm \infty$ then $Q$ does not annihilate the vacuum
and SUSY is spontaneously broken.  Let us show using a trivial calculation
that for two simple polynomial potentials the Witten index does indeed provide
the correct answer to the question of SUSY breaking.  Let us consider
\be
\triangle(\beta) = Tr \sigma_3 \int [{dp dx \over 2 \pi}] e^{-\beta[p^2/2 +
W^2/2 - \sigma_3 W^{\prime} (x)/2}.
\ee
Expanding the term proportional to $\sigma_3$ in the exponent and taking the
trace we obtain
\be
\triangle(\beta) =  \int [{dp dx \over \pi}] e^{-\beta[p^2/2 +
W^2/2} {\rm{sinh}}(\beta  W^{\prime} (x) /2).
\ee
We are interested in the regulated index as $\beta$ tends to $0$, so that
practically we need to evaluate
\be
\triangle(\beta) =  \int [{dp dx \over 2 \pi}] e^{-\beta[p^2/2 +
W^2/2]} (\beta  W^{\prime} (x) /2).
\ee
If we directly evaluate this integral for any potential of the form
$W(x) = g x^{2n+1}$, which leads to a normalizable ground state wave
function, then all the integrals are gamma functions and we explictly
obtain $\triangle = 1$.
If instead $W(x) = g x^{2n}$ so that  the candidate ground state wave function
is not normalizable then the integrand becomes an odd function of $x$ and
therefore vanishes. Thus we see for these simple cases that the Witten
index immediately coincides with the direct method available in the quantum
mechanics case.

Next let us discuss a favorite type of regularization scheme for field
theory-- namely the heat kernel method.
(Later we will discuss a path integral formulation for the regulated Witten
Index).

Following Akhoury and Comtet \cite{Akhoury84} one defines the heat kernels
$K_{\pm}(x,y;\beta)$  which satisfy
\be
({d \over d \beta} - {d^2 \over dx^2} + W^2 \mp W^{\prime} ) K_{\pm} =0.
\ee
These have the following eigenfunction representation:
\[
K_{\pm}(x,y;\beta) = <y | e^{- \beta H_{\pm}} |x> \]
\be
 = \sum_n e^{-\beta E_n} \psi_n^{\pm} (x)  \psi_n^{\pm} (x) +
\int dE e^{-\beta E_n} \psi_E^{\pm}(x) \psi_E^{\pm} (x).
\ee

In terms of the heat kernels one has
\[ \triangle (\beta) = \int dx [K_+(x,x;\beta) - K_-(x,x;\beta)] \],
or
\be
\triangle(\beta) = N_+(E=0) - N_-(E=0) + \int_{E_D} ^{\infty} dE  e^{-\beta E}
(\rho_+(E) - \rho_-(E)),
\ee
where $\rho_{\pm}$ corresponds to the density of states.
What Akhoury and Comtet were able to show, was that in cases when $W(x)$ went
to different constants at plus and minus infinity, then the density of
states factors for the continuum did not cancel and that $\triangle(\beta)$
could depend on $\beta$ and be fractional at $\beta =0$. We refer the
interested
reader to the original paper for further details.

Another non-perturbative method for studying SUSY breaking in field theory
is to explicitly break SUSY by placing the theory on a lattice and either
evaluating the path integral numerically or via some lattice non weak-
perturbative method such as the strong coupling (or high temperature)
expansion. This method was studied in detail in \cite{Bender83,Freedman85} and
we will summarize the results here.  The basic idea is to introduce a new
parameter, namely the lattice spacing $a$. This parameter explicitly breaks
SUSY
so that the ground state energy of the system will no longer be zero, even in
the unbroken case.  One hopes that as the lattice spacing is taken to zero then
the ground state energy will go to zero as a power of the
lattice spacing if SUSY is unbroken, that is we expect
\be
E_0(a)  = c a^{\gamma},
\ee
where $\gamma$ is a critical index which if greater than zero should be easy
to measure in a Monte Carlo calculation.  Measuring the ground state energy
at two different lattice spacings one studies:
\be
\gamma = {\rm{ln}} { E_0(a') \over E_0(a)} / {\rm{ln}}{a' \over a}
\ee
in the limit $a'$ and $a \rightarrow 0$.  The case of broken SUSY is more
difficult because then we expect $\gamma =0$ which is a hard measurement
to make numerically.  In this latter case it is easier to directly measure
the ground state energy and show that it remains non-zero as one takes
the lattice spacing to zero. To see how this works in quantum mechanics one
can do a lattice strong coupling expansion of the Langevin equation which
allows one to determine the ground state wave function of $H_1$ as we shall
show later.

For the superpotential $W(x) = gx^3$, we expect to find a positive critical
index since here the candidate ground state wave function is proportional to
$e^{-gx^4/4}$ and is normalizable.
The ground state expectation values of $x^n$ for the Hamiltonian  $H_1$ can be
determined by first solving the Langevin equation
\be
{dx \over dt} + W(x) = \eta(t)
\ee
and then averaging $ x(\eta(t))$ over Gaussian noise whose width is
related to $\hbar$.
Since by the virial theorem
\[  E_0 = 3g^2 <x^6> - 3 g <x^2> ~, \]
knowing the corelation functions will allow us to calculate the ground
state energy. We first put the Langevin equation on a time lattice ($t_n=na$):
\be
\epsilon (x_n - x_{n-1}) + gx^3 = \eta_n,
\ee
where
$\epsilon = 1/a $,
which allows a solution by  strong coupling expansion for large $g$. The result
is
\be
x_n = ({\eta_n \over g})^{1/3} + { \epsilon \over 3 g^{2/3}} ( \eta_{n-1}^{1/3}
\eta_n^{-2/3} - \eta_n^{-1/3}) + 0(\epsilon^2).
\ee
As we will demonstrate in Sec. 8,
the quantum mechanical expectation values of $< x^n(t)>$ are the
same as the noise averaged expection values of $x_n(\eta)$
\be
< x^n(t)>= \int[D\eta] P[\eta] x^n (\eta(t)).
\ee
On the lattice the path integral becomes a product of ordinary integrals which
can be performed with:
\be
P[\eta] = \Pi_i e^{-a \eta^2(i) /2} \sqrt {a \over 2 \pi}.
\ee
The ground state energy on the lattice regulated theory then has the form
\be
E_0 = \sqrt{g} z \sum_{n=0} ^m C_n z^n,
\ee
where
\[  z^3 = {2 \over a \sqrt{g}} \]
is a dimensionless length.  The critical index can be determined from the
logarithmic derivative of $E_0$ with respect to $z$.
Using Pad\'e approximants to extrapolate the lattice series to small lattice
spacing we found that \cite{Freedman85}
\be
E_0 = c a^{1.16}
\ee
verifying that SUSY is unbroken in the continuum limit.  Using
the same methods for the case $W(x) = gx^2/2$ we were able to verify
that the ground state energy was not zero as we took the continuum limit.
After verifying the applicability of this method in SUSY QM,
Bender et al. then successfully used this method to study non-perturbative
SUSY breaking in Wess-Zumino models of field theory \cite{Bender83}.

 \section{Factorization and the Hierarchy of Hamiltonians}
\label{sec3}

In the previous section we found that once we know the ground state wave
function corresponding to a Hamiltonian $H_1$ we can find the superpotential
$W_1(x)$ from eq. (\ref{defW}). The resulting operators $A_1$ and $A_1^{\dag}$
obtained from eq. (\ref{aadag}) can be used to factorize Hamiltonian $H_1$.  We
also know that the ground state wave function of the partner Hamiltonian $H_2$
i
 s
determined from the first excited state of $H_1$ via the application of the
operator $A_1$.  This allows a refactorization  of the second Hamiltonian in
terms of $W_2$. The partner of this refactorization is now another Hamiltonian
$H_3$.  Each of the new Hamiltonians has one fewer bound state, so that this
process can be  continued until the number of bound states is exhausted.  Thus
if one has an exactly solvable potential problem for $H_1$, one can solve for
th
 e
energy eigenvalues and wave functions for the entire hierarchy of Hamiltonians
created by repeated refactorizations.   Conversely if we know the ground state
wave functions for all the Hamiltonians in this hierarchy, we can reconstruct
th
 e
solutions of the original problem.  Let us now be more specific.

  From the last section we have seen that if the ground state energy of a
Hamiltonian $H_1$ is zero then it can always be written in a factorizable form
as a product of a pair of linear differential operators. It is then clear that
if  the ground  state energy of a Hamiltonian $H_1$ is $E_0^{(1)}$ with
eigenfunction $\psi_0^{(1)}$ then in view of eq. (\ref{eq2.3}), it can always
be
written in the form (unless stated otherwise, from now on
we set $\hbar=2m=1$ for simplicity):
\be
H_1 = A_1^{\dag} A_1 + E_0^{(1)} = - {d^2 \over dx^2} + V_1(x),
\ee
where
\be
A_1 ={d  \over dx} + W_1(x) ~,~A_1^{\dag}  = -{d  \over dx}  +
W_1(x)~,~
W_1(x) = -{d~ {\rm{ln}} \psi_0^{(1)} \over dx}.
\ee
The SUSY partner Hamiltonian
is then given by
\be
H_2 = A_1  A_1^{\dag}+ E_0^{(1)} = -{d^2 \over dx^2} + V_2(x),
\ee
where
\be
V_2(x) = W_1^2+ W_1^{\prime} + E_0^{(1)} = V_1(x) + 2 W_1^{\prime}
= V_1(x) - 2{d^2\over dx^2} {\rm{ln}} \psi_0^{(1)}.
\ee

We will introduce the notation that in $E_n^{(m)}$, $n$ denotes the energy
level and $(m)$ refers to the $m$'th Hamiltonian $H_m$.
{}From Sec. 2, the energy eigenvalues and eigenfunctions of
the two Hamiltonians $H_1$ and $H_2$ are related by
\be
E_{n+1} ^{(1)} = E_n^{(2)}, \hspace{.3in} \psi_n^{(2)} =
(E_{n+1} ^{(1)} - E_0^{(1)})^{-1/2} A_1 \psi_{n+1}^{(1)}.
\ee

Now starting from $H_2$ whose ground state energy is $E_0^{(2)}=E_1^{(1)}$ one
can similarly  generate a third Hamiltonian $H_3$ as a SUSY partner of $H_2$
since we can write $H_2$ in the form:
\be
H_2 = A_1  A_1^{\dag}+ E_0^{(1)} =  A_2^{\dag}A_2 + E_1^{(1)},
\ee
where
\be
A_2 = {d  \over dx} + W_2(x)~,~A_2^{\dag}=-{d  \over dx} + W_2(x)~,~
 W_2(x) = -{d~ {\rm{ln}} \psi_0^{(2)} \over dx}.
\ee
Continuing in this manner we obtain
\be
H_3 = A_2  A_2^{\dag}+ E_1^{(1)} = -{d^2 \over dx^2} + V_3(x),
\ee
where
\bea
V_3(x) &=& W_2^2+ W_2^{\prime} + E_1^{(1)} = V_2(x) -2 {d^2 \over dx^2}
{\rm{ln}}\psi_0^{(2)}  \nonumber \\
        &=& V_1(x) -2 {d^2 \over dx^2}{\rm{ln}}(\psi_0^{(1)} \psi_0^{(2)}).
\eea
Furthermore

\bea
E_{n} ^{(3)} &=&E_{n+1} ^{(2)} =  E_{n+2}^{(1)},  \nonumber \\
 \psi_n^{(3)} &=&
(E_{n+1} ^{(2)} - E_0^{(2)})^{-1/2} A_2 \psi_{n+1}^{(2)} \nonumber \\
 &=&
(E_{n+2} ^{(1)} - E_1^{(1)})^{-1/2} (E_{n+2} ^{(1)} - E_0^{(1)})^{-1/2}A_2
A_1 \psi_{n+2}^{(1)} .
\eea

In this way, it is clear that if the original Hamiltonian $H_1$ has $p (\geq
1)$
bound
states with eigenvalues $E_n^{(1)}$, and eigenfunctions $\psi_n^{(1)}$ with
$0 \leq n \leq (p-1)$, then we can always generate a hierarchy
of $(p-1)$ Hamiltonians $H_2, ...H_p$ such
that the $m$'th member of the hierarchy of Hamiltonians ($H_m$) has the same
eigenvalue spectrum as $H_1$ except that the first $(m-1)$ eigenvalues of $H_1$
are missing in $H_m$. In particular, we can always write ($m=2,3,...p$):

\be
H_m = A_m^{\dag} A_m+ E_{m-1}^{(1)} = -{d^2 \over dx^2} + V_m(x),
\ee
where
\be
A_m = {d  \over dx} + W_m(x)~,~
W_m(x) = -{d~ {\rm{ln}} \psi_0^{(m)} \over dx}.
\ee
One also has
\bea
E_{n} ^{(m)} &=&E_{n+1} ^{(m-1)}=... =  E_{n+m-1}^{(1)}~,  \nonumber \\
 \psi_n^{(m)} &=&
(E_{n+m-1} ^{(1)} - E_{m-2}^{(1)})^{-1/2}... (E_{n+m-1} ^{(1)} -
E_0^{(1)})^{-1/2}A_{m-1}... A_1 \psi_{n+m-1}^{(1)} \nonumber \\
V_m(x)&=& V_1(x) -2 {d^2 \over dx^2}{\rm{ln}}(\psi_0^{(1)}... \psi_0^{(m-1)})~.
\eea

In this way, knowing all the eigenvalues and eigenfunctions of $H_1$ we
immediately know all the energy eigenvalues and eigenfunctions of the hierarchy
of $p-1$ Hamiltonians.  Further the
reflection and transmission coefficients (or phase shifts) for the hierarchy of
Hamiltonians can be obtained in terms of $R_1,T_1$ of the first Hamiltonian
$H_1$ by a repeated use of eq. (\ref{rt}). In particular we find

\bea
R_m(k) &=& \left( {W_{-}^{(1)} - ik \over W_{-}^{(1)} + ik }\right)...
\left( {W_{-}^{(m-1)} - ik \over W_{-}^{(m-1)} + ik }\right) R_1(k),\nonumber
\\
T_m(k) &=& \left( {W_{-}^{(1)} - ik \over W_{+}^{(1)} - ik' }\right)...
\left( {W_{-}^{(m-1)} - ik \over W_{+}^{(m-1)} - ik' }\right) T_1(k),
\eea
where $ k$ and $k^{\prime}$ are given by
\be
k= [E- (W_{-}^{(1)}) ^2]^{1/2}, \hspace{.3in} k'= [E- (W_{+}^{(1)}) ^2]^{1/2}.
\ee

\section{Shape Invariance and Solvable Potentials}
\label{sec4}

Most text books on quantum mechanics describe how the one dimensional
harmonic oscillator problem can be elegantly solved using the raising and
lowering operator method. Using the ideas of SUSY QM developed in Sec. 2
and an integrability condition called the shape invariance condition
 \cite{Gendenshtein83}, we now show that the operator method for
the harmonic oscillator can be generalized to the whole class of shape
invariant potentials (SIP) which include all the popular, analytically
solvable potentials. Indeed, we shall see that for such potentials,
the generalized operator method quickly yields all the bound state
energy eigenvalues, eigenfunctions as well as the scattering matrix. It
turns out that this approach is essentially equivalent to
Schr\"{o}dinger's method of factorization \cite{Schrodinger40,Infeld51}
although the language of SUSY is more appealing.

Let us now explain precisely what one means by shape invariance.
If the pair of SUSY partner potentials $V_{1,2}(x)$
defined in Sec. 2 are similar in shape and differ only in the
parameters that appear in them, then they are said to be shape invariant.
More precisely, if the partner potentials $V_{1,2}(x;a_1)$ satisfy the
condition
\be \label{eq4.1}
V_2(x;a_1) = V_1(x;a_2) + R(a_1),
\ee
where $a_1$ is a set of parameters, $a_2$ is a function of $a_1$ (say
$a_2=f(a_1)$) and the remainder $R(a_1)$ is independent of $x$, then
$V_{1}(x;a_1)$ and $V_{2}(x;a_1)$ are said to be shape invariant. The shape
invariance condition (\ref{eq4.1}) is an integrability condition. Using
this condition and the hierarchy of Hamiltonians discussed in Sec. 3, one
can easily obtain the energy eigenvalues and eigenfunctions of any SIP
when SUSY is unbroken.

\subsection{General Formulas for Bound State Spectrum, Wave
Functions and S-Matrix}
\label{sec4.1}

Let us start from the SUSY partner Hamiltonians $H_1$ and $H_2$ whose
eigenvalues and eigenfunctions are related by SUSY.
Further, since SUSY is unbroken we know that
\be \label{eq4.2}
E^{(1)}_0(a_1)=0,\quad \psi^{(1)}_0(x;a_1)
= N \exp\left[-\int^xW_1(y;a_1)dy\right] .
\ee
We now show that the entire spectrum of $H_1$ can be very easily
obtained algebraically by using the shape invariance condition (\ref{eq4.1}).
To that purpose, let us construct a series of Hamiltonians $H_s$,
$s=1,2,3$... In particular, following the discussion of the last
section it is clear that if $H_1$ has $p$ bound states then one can
construct $p$ such Hamiltonians $H_1, H_2.....H_p$ and the $n$'th
Hamiltonian $H_n$ will have the same spectrum as $H_1$ except that
the first $n-1$ levels of $H_1$ will be absent in $H_n$. On repeatedly
using the shape invariance condition (\ref{eq4.1}), it is then clear that
\be \label{eq4.3}
H_s = - {d^2\over dx^2} + V_1(x;a_s) + \sum^{s-1}_{k=1}R(a_k),
\ee
where $a_s=f^{s-1}(a_1)$ i.e. the function $f$ applied $s-1$ times. Let us
compare the spectrum of $H_s$ and $H_{s+1}$. In view of eqs. (\ref{eq4.1}) and
 (\ref{eq4.3}) we have
\bea \label{eq4.4}
H_{s+1} = -{d^2\over dx^2}+V_1 (x;a_{s+1})+\sum^s_{k=1} R(a_k) \nonumber\\
        =-{d^2\over dx^2}+V_2 (x;a_s)+\sum^{s-1}_{k=1} R(a_k).
\eea
Thus $H_s$ and $H_{s+1}$ are SUSY partner Hamiltonians and hence have
identical bound state spectra except for the ground state of
$H_s$ whose energy is
\be \label{eq4.5}
E^{(s)}_0 = \sum^{s-1}_{k=1} R(a_k).
\ee
This follows from eq. (\ref{eq4.3}) and the fact that $E^{(1)}_0=0$. On going
back from $H_s$ to $H_{s-1}$ etc, we would eventually reach $H_2$ and
$H_1$ whose ground state energy is zero and whose $n$'th level is
coincident with the ground state of the Hamiltonian
$H_n$. Hence the complete eigenvalue spectrum of $H_1$ is
given by
\be \label{eq4.6}
E^-_n(a_1) = \sum^n_{k=1} R(a_k); \hspace{.2in}  E^-_0(a_1)=0.
\ee

We now show that, similar to the case of the one dimensional harmonic
oscillator, the bound state wave functions $\psi^{(1)}_n(x;a_1)$ for
any shape invariant potential can also be easily obtained from its
ground state wave function $\psi^{(1)}_0(x;a_1)$ which in turn is
known in terms of the superpotential. This is possible
because the operators $A$ and $A^{\dag}$ link up the eigenfunctions of the
same energy for the SUSY partner Hamiltonians $H_{1,2}$. Let us
start from the Hamiltonian $H_s$ as given by eq. (\ref{eq4.3}) whose ground
state eigenfunction is then given by
 $\psi^{(1)}_0(x;a_s)$. On going from
$H_s$ to $H_{s-1}$ to $H_2$ to $H_1$ and using eq. (\ref{psin1}) we then find
that the $n$'th state unnormalized, energy eigenfunction
$\psi^{(1)}_n(x;a_1)$ for the original Hamiltonian $H_1(x;a_1)$ is
given by
\be \label{eq4.7}
\psi^{(1)}_n(x;a_1) \propto A^{\dag}(x;a_1)A^{\dag}(x;a_2)...A^{\dag}(x;a_n)
\psi^{(1)}_0(x;a_{n+1}),
\ee
which is clearly a generalization of the operator method of
constructing the energy eigenfunctions for the one dimensional
harmonic oscillator.

It is often convenient to have explicit expressions for the wave
functions. In that case, instead of using the above equation, it is
far simpler to use the identify \cite{Dabrowska88}
\be \label{eq4.8}
\psi^{(1)}_n(x;a_1) = A^{\dag}(x;a_1) \psi^{(1)}_{n-1}(x;a_2).
\ee

Finally, it is worth noting that in view of the shape invariance
condition (\ref{eq4.1}), the relation (\ref{rt}) between scattering amplitudes
takes a particularly simple form
\be \label{eq4.9}
R_1(k;a_1)=\left({W_-(a_1)+ik \over W_-(a_1)-ik} \right) R_1(k;a_2),
\ee
\be \label{eq4.10}
T_1(k;a_1)=\left({W_+(a_1)-ik'\over W_-(a_1)-ik} \right) T_1(k;a_2),
\ee
thereby relating the reflection and transmission coefficients of the
same Hamiltonian $H_1$ at $a_1$ and $a_2(=f(a_1))$.

\subsection{Shape Invariance in More Than One Step}
\label{sec4.2}

We can expand the list of solvable potentials by extending the shape
invariance idea to the more general concept of shape invariance in
two and even multi-steps. We shall see later that in this way we will
be able to go much beyond the factorization method and obtain a huge
class of new solvable potentials \cite{Barclay93}.

Consider the unbroken SUSY case of two superpotentials $W(x;a_1)$ and
$\tilde W(x;a_1)$ such that $V_2(x;a_1)$ and $\tilde V_1(x;a_1)$ are
same up to an additive constant i.e.
\be \label{eq4.11}
V_2(x;a_1) = \tilde V_1(x;a_1) + R(a_1)
\ee
or equivalently
\be \label{eq4.12}
W^2(x;a_1)+W'(x;a_1)=\tilde W^2(x;a_1)-\tilde W'(x;a_1)+R(a_1).
\ee
Shape invariance in two steps means that
\be \label{eq4.13}
\tilde V_2(x;a_1) = V_1 (x;a_2) +\tilde R(a_1),
\ee
that is
\be \label{eq4.14}
\tilde W^2(x;a_1)+\tilde W'(x;a_1) = W^2(x;a_2)-W'(x;a_2)+\tilde
R(a_1).\label{twosteps}
\ee

We now show that when this condition holds, the energy eigenvalues and
eigenfunctions of the potential $V_1(x;a_1)$ can be obtained algebraically.
First of all, let us notice that unbroken SUSY implies zero energy ground
states
for the potentials $V_1(x;a_1)$ and $\tilde V_1(x;a_1)$:
\be \label{eq4.15}
E^{(1)}_0(a_1) = 0, \hspace{.5in} \tilde E^{(1)}_0(a_1)=0~.
\ee
The degeneracy of the energy levels for the SUSY partner potentials yields
\be \label{eq4.16}
E^{(2)}_n(a_1)=E^{(1)}_{n+1}(a_1); ~~\tilde E^{(2)}_n(a_1)=\tilde
E^{(1)}_{n+1}(a_1).
\ee
{}From eq. (\ref{eq4.11}) it follows that
\be \label{eq4.17}
E^{(2)}_n(a_1) = \tilde E^{(1)}_n(a_1)+R(a_1),
\ee
so that for $n = 0$, these two equations yield
\be \label{eq4.18}
E^{(1)}_1(a_1) = R(a_1).
\ee
Also, the shape invariance condition (\ref{eq4.13}) yields
\be \label{eq4.19}
\tilde E^{(2)}_n(a_1) = E^{(1)}_n(a_2)+\tilde R(a_1).
\ee
{}From the above equations one can then show that
\be \label{eq4.20}
E^{(1)}_{n+1}(a_1) = E^{(1)}_{n-1}(a_2)+R(a_1)+\tilde R(a_1).
\ee
On solving these questions recursively we obtain $(n=0,1,2,...)$
\be \label{eq4.21}
E^{(1)}_{2n}=\sum^n_{k=1}[R(a_k) + \tilde R(a_k)],
\ee
\be \label{eq4.22}
E^{(1)}_{2n+1}=\sum^n_{k=1} [R(a_k)+ \tilde R(a_k)]+R(a_{n+1}).
\ee

We now show that, similar to the discussion of the last subsection,
the bound state wave functions $\psi^{(1)}
_n(x;a_1)$ can also be easily obtained in terms of the ground state wave
functions $\psi^{(1)}
_0(x;a_1)$ and $\tilde \psi^{(1)}
_0(x;a_1)$ which in turn are known in terms of the superpotentials $W$
and $\tilde W$. In particular from eq. (\ref{eq4.11}) it follows that

\be \label{eq4.23}
\psi^{(1)}_{n+1}(x;a_1) \propto
A^{\dag}(x;a_1) \psi_n^{(2)}(x;a_1) \propto A^{\dag}(x;a_1)
 \tilde\psi^{(1)}_n(x;a_1),
\ee
while from eq. (\ref{eq4.13}) we have
\be \label{eq4.24}
\tilde \psi^{(1)}_{n+1}(x;a_1) \propto \tilde
A^{\dag}(x;a_1) \tilde \psi_n^{(2)}(x;a_1) \propto \tilde A^{\dag}(x;a_1)
 \psi^{(1)}_n(x;a_2).
\ee
Hence on combining the two equations we have the identity
\be \label{eq4.25}
\psi^{(1)}_{n+2}(x;a_1) \propto
A^{\dag}(x;a_1) \tilde A^{\dag}(x;a_1) \psi^{(1)}_n(x;a_2).
\ee
Recursive application of the above identity yields
\be \label{eq4.26}
\psi^{(1)}_{2n}(x;a_1) \propto [A^{\dag}(x;a_1) \tilde
A^{\dag}(x;a_1)]...[A^{\dag}(x;a_n) \tilde A^{\dag}(x;a_n)]
 \psi^{(1)}_0(x;a_{n+1}),
\ee
\be \label{eq4.27}
\psi^{(1)}_{2n+1}(x;a_1) \propto [A^{\dag}(x;a_1) \tilde
A^{\dag}(x;a_1)]...[A^{\dag}(x;a_n) \tilde A^{\dag}(x;a_n)] A^{\dag}(x;a_{n+1})
\tilde \psi^{(1)}_0(x;a_{n+1}),
\ee
where we have used the fact that
\be \label{eq4.28}
\psi^{(1)}_1(x;a_1) \propto A^{\dag}(x;a_1) \tilde \psi^{(1)}_0(x;a_1).
\ee

Finally, it is easily shown that the relation (\ref{rt}) between
the scattering amplitudes takes a particularly simple form
\be \label{eq4.29}
R_1(k;a_1) =\left({W_-(a_1)+ik\over W_-(a_1)-ik}\right)\left({\tilde
W_-(a_1)+ik\over \tilde W_-(a_1)-ik}\right) R_1(k;a_2) ,
\ee
\be \label{eq4.30}
T_1(k;a_1)=\left({W_+(a_1)-ik'\over W_-(a_1)-ik}\right)\left({\tilde
W_+(a_1)-ik'\over \tilde W_-(a_1)-ik}\right) T_1(k;a_2),
\ee
thereby relating the reflection and transmission coefficients of the
same Hamiltonian at $a_1$ and $a_2$.

It is clear that this procedure can be easily generalized and one can
consider multi-step shape invariant potentials and in these cases too
the spectrum, the eigenfunctions and the scattering matrix can be
obtained algebraically.

\subsection{Strategies For Categorizing Shape Invariant Potentials}
\label{sec4.3}

Let us now discuss the interesting question of the
classification of various solutions to the shape invariance condition
(\ref{eq4.1}). This is clearly an important problem because once such
a classification is available, then one discovers new
SIPs which are solvable by purely algebraic
methods. Although the general problem is still unsolved, two classes of
solutions have been found and discussed. In the first class, the parameters
$a_1
 $
and $a_2$ are related to each other by translation $(a_2=a_1+\alpha)$
\cite{Cooper87,Chuan91}. Remarkably enough, all well known analytically
solvable potentials found in most text books on nonrelativistic quantum
mechanic
 s
belong to this class. Last year, a second class of solutions was discovered in
which the parameters $a_1$
and $a_2$ are related by scaling $(a_2=qa_1)$ \cite{Khare93a,Barclay93}.

\subsubsection{Solutions Involving Translation}
\label{sec4.3.1}

We shall now point out the key steps that go into the classification
of SIPs in case  $a_2=a_1+\alpha$ \cite{Cooper87}.
Firstly one notices the fact that the eigenvalue spectrum of the
Schr\"{o}dinger equation is always such that the $n$'th eigenvalue $E_n$
for large $n$ obeys the constraint \cite{Nieto79}
\be \label{eq4.31}
1/n^2 \leq E_n \leq n^2,
\ee
where the upper bound is saturated by the square well potential and
the lower bound is saturated by the Coulomb potential. Thus, for any
SIP, the structure of $E_n$ for large $n$ is
expected to be of the form
\be \label{eq4.32}
E_n \sim \sum_{\alpha}C_{\alpha} n^{\alpha},~~ -2\leq \alpha \leq 2.
\ee
Now, since for any SIP, $E_n$ is given by eq. (\ref{eq4.6}), it follows
that if
\be \label{eq4.33}
R(a_k) \sim \sum_{\gamma} k^{\gamma}
\ee
then
\be \label{eq4.34}~~
-3 \leq \gamma \leq 1
\ee

How does one implement this constraint on $R(a_k)$? While one has no
rigorous answer to this question, it is easily seen that a fairly
general factorizable form of $W(x;a_1)$ which produces the above
$k$-dependence in $R(a_k)$ is given by
\be \label{eq4.35}
W(x;a_1)=\sum^m_{i=1}(k_i+c_i)g_i(x)+h_i(x)/(k_i+c_i)+f_i(x)
\ee
where
\be \label{eq4.36}
a_1=(k_1,k_2...),~~ a_2 = (k_1+ \alpha, k_2+\beta...)
\ee
with $c_i, \alpha, \beta$ being constants. Note that this ansatz
excludes all potentials leading to $E_n$ which contain fractional
powers of $n$. On using the above ansatz for $W$ in the shape invariance
condition (\ref{eq4.1}) one can obtain the conditions to be satisfied
by the functions $g_i(x), h_i(x), f_i(x)$. One important condition is
of course that only those superpotentials $W$ are admissible which give a
square
integrable ground state wave function. It turns out that there are
no solutions in case $m \geq 3$ in eq. (\ref{eq4.35}), while there are
only two solutions in case $m = 2$ i.e. when
\be \label{eq4.37}
W(x;a_1)=(k_1+c_1)g_1(x)+(k_2+c_2)g_2(x)+f_1(x),
\ee
which are given by
\be \label{eq4.38}
W(r;A,B) = A \tanh \alpha r - B \coth \alpha r , \hspace{.2in} A > B > 0,
\ee
and
\be \label{eq4.39}
W(x;A,B) = A \tan \alpha x - B \cot \alpha x ; \hspace{.2in} A,B > 0,
\ee
where $0 \leq x \leq \pi/{2 \alpha}$ and $\psi(x=0)= \psi(x=\pi/{2 \alpha})=0$.
For the simplest possibility of $m = 1$, one has a number of solutions to the
shape invariance condition (\ref{eq4.1}). In Table 4.1, we give
expressions for the various shape invariant potentials $V_1(x)$,
superpotentials $W(x)$, parameters $a_1$ and $a_2$ and the corresponding
energy eigenvalues $E^{(1)}_n$ \cite{Dutt88,Levai89}.

As an illustration, let us consider the superpotential given in
eq. (\ref{eq4.39}). The corresponding partner potentials are
\[
V_1(x;A,B)= - (A+B)^2 + A(A-\alpha) \sec^2 \alpha x + B(B-\alpha)
{\rm cosec}^2 \alpha x\]
\be \label{eq4.40}
V_2(x;A,B)= - (A+B)^2 + A(A+\alpha) \sec^2 \alpha x + B(B+\alpha)
{\rm cosec}^2 \alpha x
\ee
$V_1$ and $V_2$ are often called P\"{o}schl-Teller I potentials in the
literature. They are shape invariant  partner potentials since
\be \label{eq4.41}
V_2(x;A,B) = V_1(x;A +\alpha,B +\alpha)+(A+B+2\alpha)^2-(A+B)^2
\ee
and in this case
\be \label{4.42}
\{a_1\} = (A,B); \{a_2\} = (A+\alpha,B+\alpha), R(a_1)=(A+B+2\alpha)^2-(A+B)^2.
\ee
In view of eq. (\ref{eq4.6}), the bound state energy eigenvalues of
the potential $V_1(x;A,B)$ are then given by
\be \label{4.43}
E^{(1)}_n=\sum^n_{k=1}R(a_k)
 = (A+B+2n \alpha)^2 - (A+B)^2.
\ee
The ground state wave function of $V_1(x;A,B)$ is calculated from the
superpotential $W$ as given by eq. (\ref{eq4.39}). We find
\be \label{eq4.44}
\psi^{(1)}_0 (x;A,B) \propto (\cos \alpha x)^s (\sin \alpha x)^{\lambda}
\ee
where
\be \label{eq4.45}
s = A/{\alpha} ; \hspace{.2in} \lambda = B/{\alpha}.
\ee

The requirement of $A,B > 0$ that we have assumed in eq. (\ref{eq4.39})
guarantees that $\psi^{(1)}_0 (x;A,B)$ is well behaved and hence
acceptable as $x \longrightarrow 0, \pi/{2 \alpha}$. Using this expression for
the ground state wave function and  eq. (\ref{eq4.8}) one can also
obtain explicit expressions for the bound state eigenfunctions
$\psi^{(1)}_n (x;A,B)$. In particular, in this case, eq. (\ref{eq4.8})
takes the form
\be \label{eq4.46}
\psi_n(x;\{a_1\}) = (-{d\over dx} +A \tan \alpha x
- B \cot \alpha x) \psi_{n-1}(x;\{a_2\}).
\ee
On defining a new variable
\be \label{eq4.47}
 y = 1 - 2 \sin^2 \alpha x
\ee
and factoring out the ground state state wave function
\be \label{eq4.48}
\psi_n(y;\{a_1\}) = \psi_0(y;\{a_1\}) R_n(y;\{a_1\})
\ee
with $\psi_0$ being given by eq. (\ref{eq4.44}), we obtain
\bea \label{eq4.49}
R_n(y;A.B) & = & \alpha (1-y^2){d\over dy}R_{n-1}(y;A+\alpha,B+
\alpha)\nonumber \\
& & +[(A-B)-(a+B+\alpha) y] R_{n-1}(y;A+\alpha,B+\alpha).
\eea

It is then clear \cite{Dabrowska88} that $R_n(y;A,B)$ is proportional
to the Jacobi polynomial so that the unnormalized bound state energy
eigenfuncti
 ons for this
potential are
\be \label{eq4.50}
\psi_n(y;A,B) = (1-y)^{\lambda/2} (1+y)^{s/2} P_n^{\lambda-1/2,s-1/2}(y).
\ee
The procedure outlined above has been applied to all known SIPs
 \cite{Dabrowska88,Khare91} and the energy eigenfunctions
$\psi^{(1)}_n(y)$  have been obtained in Table 4.1, where we also give the
variable $y$ for each case.

Several remarks are in order at this time.

\begin{enumerate}
\item The P\"{o}schl-Teller I and II superpotentials as given by eqs.
(\ref{eq4.39}) and (\ref{eq4.38}) respectively have not been included in
Table 4.1 since they are equivalent to the Scarf I (trigonometric) and
generalized P\"{o}schl-Teller superpotentials
\bea \label{eq4.51}
W_1 & = & - A \tan \alpha x + B {\rm sec}~ \alpha x,\nonumber \\
W_2 & = &   A \coth \alpha r - B {\rm cosech}~ \alpha r,
\eea
by appropriate
redefinition of the parameters \cite{Dutt93a}. For example, one
can write
\be \label{eq4.52}
W_2  = ({A+B \over 2}) \tanh ({\alpha r\over 2}) - ({B-A \over 2})
\coth ({\alpha r\over 2}),
\ee
which is just the P\"{o}schl-Teller II superpotential of eq. (\ref{eq4.38})
with redefined parameters.
\item Throughout this section we have used the convention of
$\hbar = 2m=1$. It would naively appear that if we had not put $\hbar
=1$, then the shape invariant potentials as given in Table 4.1 would all
be $\hbar$-dependent. However, it is worth noting that in each and
every case, the $\hbar$-dependence is only in the constant multiplying the
$x$-dependent function so that in each case we can always redefine the
constant multiplying the function and obtain an $\hbar$-independent
potential. For example, corresponding to the superpotential given by
eq. (\ref{eq4.39}), the $\hbar$-dependent potential is given by $(2m = 1)$\\
\bea  \label{eq4.53}
V_1(x;A,B)= W^2 -\hbar W' &=& - (A+B)^2+A(A+\hbar \alpha) \sec^2 \alpha x
\nonumber  \\
 &+& B(B+\hbar \alpha) {\rm cosec}^2 \alpha x.
\eea
On redefining
\be \label{eq4.54}
A(A+\hbar\alpha) = a ; \hspace{.2in}  B(B+\hbar\alpha) = b,
\ee
where $a,b$ are $\hbar$-independent parameters, we then have
a $\hbar$-independent potential.
\item  In Table 4.1, we have given conditions (like $A > 0, B >0$)
 for the superpotential (\ref{eq4.39}), so that $\psi^{(1)}_0 = N
\exp(-\int^x W(y)dy)$ is an acceptable ground state energy
eigenfunction. Instead one can also write down conditions for
$\psi^{(2)}_0 = N \exp (\int^x W(y) dy)$ to be an acceptable ground
state energy eigenfunction.
\item It may be noted that the Coulomb as well as the harmonic oscillator
potentials in $n$-dimensions are also shape invariant potentials.
\item Does this classification exhaust all shape invariant
potentials? It was believed that the answer to the question
is yes \cite{Barclay91,Montemayor89} but as we shall see in
the next subsection, the answer to the question is in fact negative.
However, it appears that this classification has perhaps exhausted
all SIPs where $a_2$ and $a_1$ are related by translation.
\item No new solutions (apart from those in Table 4.1) have been
obtained so far in the case of multi-step shape invariance and when
$a_2$ and $a_1$ are related by translation.
\item What we have shown here is that shape invariance is a sufficient
condition for exact solvability. But is it also a necessary condition?
 This question has been discussed in detail in ref. \cite{Cooper87} where it
has
been shown that the solvable Natanzon potentials
\cite{Natanzon71,Ginocchio84}
are in general not shape invariant.
\end{enumerate}

Before ending this subsection, we would like to remark that for the
SIPs (with translation) given in Table 4.1,
the reflection and transmission amplitudes $R_1(k)$ and $T_1(k)$ (or
phase shift $\delta_1(k)$ for the three-dimensional case) can also be
calculated by operator methods. Let us first notice that since for
all the cases $a_2=a_1+\alpha$, hence $R_1(k;a_1)$ and $T_1(k;a_1)$
are determined for all values of $a_1$ from eqs. (\ref{eq4.9})
and (\ref{eq4.10}) provided they are known in a finite strip.
For example, let us consider the shape invariant superpotential
\be \label{eq4.55}
W = n \tanh x,
\ee
where $n$ is positive integer (1,2,3,...). The two partner potentials
\bea \label{eq4.56}
V_1(x;n) & = & n^2 - n (n+1) {\rm sech}^2 x,\nonumber \\
V_2(x;n) & = & n^2 - n (n-1) {\rm sech}^2 x,
\eea
are clearly shape invariant with
\be \label{eq4.57}
a_1 = n ~,~ \hspace{.2in} a_2 = n-1 ~.
\ee
On going from $V_1$ to $V_2$ to $V_3$ etc., we will finally
reach the free particle potential which is reflectionless and for which
$T = 1$. Thus we immediately conclude that the series of potentials
$V_1,V_2,...$ are all reflectionless and the transmission
coefficient of the reflectionless potential $V_1(x;n)$ is given by
\bea \label{eq4.58}
T_1(k,n) & = & {(n-ik)(n-1-ik)...(1-ik)\over (-n-ik)(-n+1-ik)...(-1-ik)}
\nonumber \\
         & = & {\Gamma(-n-ik) \Gamma(n+1-ik)}\over {\Gamma(-ik) \Gamma(1-ik)}.
\eea
The scattering amplitudes of the Coulomb \cite{Cooper88a} and the potential
corresponding to $W = A \tanh x + B{\rm sech}~x$ \cite{Khare88} have
also been obtained in this way.

There is, however, a straightforward method \cite{Khare88} for
calculating the scattering amplitudes by making use of the $n$'th state
wave functions as given in Table 4.1. In order to impose boundary
conditions appropriate to the scattering problem, two modifications
of the bound state wave functions have to be made: (i) instead of the
parameter $n$ labelling the number of nodes, one must use the wave
number $k'$ so that the asymptotic behaviour is
$\exp(ik'x)$ as $x\rightarrow \infty$. (ii) the second solution of the
Schr\"{o}dinger equation must be kept (it had been discarded for bound
state problems since it diverged asymptotically). In this way the
scattering amplitude of all the SIPs of Table 4.1 have been calculated
in ref. \cite{Khare88}.

\subsubsection{Solutions Involving Scaling}
\label{sec4.3.2}

For almost nine years, it was believed that the only shape invariant
potentials are those given in Table 4.1 and that there were no more shape
invariant potentials. However, very recently we have been able to discover a
huge class of new shape invariant potentials \cite{Khare93a,Barclay93}. It
turns out that for many of these new shape
invariant potentials, the parameters $a_2$ and $a_1$ are related by
scaling $(a_2=qa_1, 0 < q < 1)$ rather than by translation, a
choice motivated by the recent interest in $q$-deformed Lie algebras. We
shall see that many of these potentials are reflectionless and have an
infinite number of bound states. So far none of these potentials have
been obtained in a closed form but are obtained only in a series form.

Let us consider an expansion of the superpotential of the from
\be \label{eq4.59}
W(x;a_1) = \sum^{\infty}_{j=0}g_j(x)a^j_1
\ee
and further let
\be \label{eq4.60}
a_2 = qa_1 , \hspace{.2in} 0 < q < 1 ~~.
\ee
This is slightly misleading in that a reparameterization of the form
$a_2 =qa_1$, can be recast as $a'_2 = a'_1+\alpha$ merely by taking
logarithms. However, since the choice of parameter is usually an
integral part of constructing a SIP, it is in
practice part of the ansatz. For example, we will construct below
potentials by expanding in $a_1$, a procedure whose legitimacy and
outcome are clearly dependent on our choice of parameter and hence
reparameterization. We shall see that, even though the construction
is non-invariant, the resulting potentials will still be invariant
under redefinition of $a_1$. On using eqs. (\ref{eq4.59}) and
(\ref{eq4.60}) in the shape invariance condition (\ref{eq4.1}),
writting $R(a_1)$ in the form
\be \label{eq4.61}
R(a_1) = \sum^{\infty}_{j=0} R_j a^j_1~,
\ee
and equating powers of $a_1$ yields \cite{Barclay93,Khare93a}
\be \label{eq4.62}
2g'_0(x)=R_0 ;\hspace{.2in} g'_1(x)+2d_1 g_0(x) g_1(x)=r_1 d_1,
\ee
\be \label{eq4.63}
g'_n(x)+2d_n g_0(x) g_n(x) = r_n d_n-d_n\sum^{n-1}_{j=1} g_j(x) g_{n-j}(x),
\ee
where
\be \label{eq4.64}
r_n \equiv R_n/(1-q^n), \hspace{.2in} d_n = (1-q^n)/(1+q^n),
\hspace{.2in} \ n = 1,2,3,...
\ee
This set of linear differential equations is easily solvable in
succession to give a general solution of eq. (\ref{eq4.1}).
Let us first consider the special case $g_0(x) = 0$,
which corresponds to $R_0=0$. The general solution of eq. (\ref{eq4.63})
then turns out to be
\be \label{eq4.65}
g_n(x)=d_n \int dx[r_n- \sum^{n-1}_{j=1} g_j(x) g_{n-j}(x)],~~ n=1,2,...
\ee
where without loss of generality we have assumed the constants of
integration to be zero. We thus see that once a set of $r_n$
are chosen, then the shape invariance condition essentially
fixes the $g_n(x)$ (and hence $W(x;a_1))$ and determines the shape
invariant potential. Implicit constraints on this choice are that the
resulting ground state wave function be normalizable and the spectrum
be sensibly ordered which is ensured if $R(q^na_1)>0$.

The simplest case is $r_1>0$ and $r_n=0, n\geq 2$. In this case the
eq. (\ref{eq4.65}) takes a particularly simple form
\be \label{eq4.66}
g_n(x) = \beta_n x^{2n-1}~,
\ee
where
\be \label{eq4.67}
\beta_1 = d_1 r_1,\hspace{.5in} \beta_n =-{d_n\over {(2n-1)}}
\sum^{n-1}_{j=1}\beta_j \beta_{n-j}
\ee
and hence
\be \label{eq4.68}
W(x;a_1)=\sum^{\infty}_{j=1} \beta_ja^j_1 x^{2j-1} =\sqrt{a_1}
F(\sqrt{a_1} x)~.
\ee
For $a_2 = qa_1$, this gives
\be \label{eq4.69}
W(x;a_2) = \sqrt q W(\sqrt q x, a_1)~,
\ee
which corresponds to the self-similar $W$ of Shabat and Spiridonov
\cite{Shabat92,Spiridonov92a}. It is worth pointing out that these
self-similar potentials can be shown to satisfy $q$-supersymmetry
\cite{Spiridonov92b}. It may be noted here that instead of choosing $r_n=0,
n\geq 2,$ if any one $r_n$ (say $r_j$) is taken to be nonzero then one
again obtains self-similar potentials \cite{Barclay93,Khare93a} and in these
instances the results obtained from shape invariance and
self-similarity are entirely equivalent and the Shabat-Spiridonov
self-similarity condition turns out to be a special case of the shape
invariance condition.

It must be emphasized here that shape invariance is a much more
general concept than self-similarity. For example, if we choose more
than one $r_n$ to be nonzero, then SIP are
obtained which are not contained within the self-similar ansatz.
Consider for example, $r_n=0, n\geq 3$. Using eq. (\ref{eq4.65}) one can
readily calculate all the $g_n(x)$, of which the first three are
\bea \label{eq4.70}
g_1(x) & = & d_1 r_1 x ,\hspace{.2in} g_2(x) = d_2 r_2 x-{1\over 3}d^2_1
r^2_1 x^3,\nonumber \\
g_3(x) & = & -{2\over 3}d_1 r_1 d_2 r_2 d_3 x^3 + {2\over 15}d^3_1 r^3_1
d_2 d_3 x^5.
\eea
Notice that in this case $W(x)$ contains only odd powers of $x$. This
makes the potentials $V_{1,2}(x)$ symmetric in $x$ and also guarantees
unbroken SUSY. The energy eigenvalues follow immediately from
eqs. (\ref{eq4.6}) and (\ref{eq4.61}) and are given by $(0<q<1)$
\be \label{eq4.71}
E^{(1)}_n(a_1) = \Gamma_1{(1+q)(1-q^n)\over
(1-q)} + \Gamma_2{(1+q^2)(1-q^{2n})\over (1-q^2)}, n=0,1,2,..
\ee
where $\Gamma_1 = d_1r_1a_1, \Gamma_2 = d_2r_2a^2_1$ while the
unnormalized ground state wave function is
\be \label{eq4.72}
\psi^{(1)}_0(x;a_1) = \exp[-{x^2\over 2}(\Gamma_1+\Gamma_2)
+{x^4\over 4}(d_2\Gamma^2_1+2d_3\Gamma_1\Gamma_2+d_4\Gamma^2_2)+0(x^6)]
\ee
The wave functions for the excited states can be recursively calculated from
the
relation (\ref{eq4.8}).

We can also calculate the transmission coefficient of this symmetric
potential $(k=k')$ by using the relation (\ref{eq4.30}) and the fact that
for this SIP $a_2 = q a_1$. Repeated application of the relation
(\ref{eq4.30}) gives
\be \label{eq4.73}
T_1(k;a_1)={[ik-W(\infty,a_1)][ik-W(\infty,a_2)]...[ik-W(\infty,a_n)]\over
[ik+W(\infty,a_1)][ik+W(\infty,a_2)]...[ik+W(\infty,a_n)]}T_1(k;a_{n+1})
\ee
where
\be \label{eq4.74}
W(\infty,a_j) =\sqrt{E_{\infty}^{(1)}-E_j^{(1)}}~~.
\ee

Now, as $n\rightarrow\infty, a_{n+1}=q^na_1\rightarrow 0(0 < q < 1)$
and, since we have taken $g_0(x)=0,$ one gets
$W(x;a_{n+1})\rightarrow 0$. This corresponds to a free particle for
which the reflection coefficient $R_1(k;a_1)$ vanishes and the transmission
coefficient is given by
\be \label{eq4.75}
T_1(k;a_1)=\prod^{\infty}_{j=1}{[ik-W(\infty,a_j)]\over [ik+W(\infty,a_j)]}~.
\ee

The above discussion keeping only $r_1,r_2\not =0$ can be readily
generalized to an arbitrary number of nonzero $r_j$. The energy
eigenvalues for this case are given by $(\Gamma_j\equiv d_jr_ja^j_1)$
\be \label{eq4.76}
E^{(1)}_n(a_1) = \sum_j \Gamma_j{(1+q^j)(1-q^{jn})\over (1-q^j)}, ~~n =
0,1,2,..
 .
\ee

All these potentials are also symmetric and reflectionless with $T_1$
as given by eq. (\ref{eq4.75}). The limits $q\rightarrow 0$ and $q\rightarrow
1$ of all these potentials are simple and quite interesting. At $q = 1$, the
solution of the shape invariance condition (\ref{eq4.1}) is the standard
one dimensional harmonic oscillator with $W(x)=Rx/2$ while in the
limit $q\rightarrow 0$ the solution is the Rosen-Morse superpotential
corresponding to the one soliton solution given by
\be \label{eq4.77}
W(x) = \sqrt R \tanh (\sqrt R x).
\ee
Hence the general solution as obtained above with $0<q<1$ can be
regarded as the multi-parameter deformation of the hyperbolic tangent
function with $q$ acting as the deformation parameter. It is also worth
noting that the number of bound states increase discontinuously from
just one at $q = 0$ to infinity for $q > 0$. Further, whereas for $q = 1$ the
spectrum is purely discrete, for $q$ even slightly less than one, we have
the discrete as well as the continuous spectra.

Finally, let us consider the solution to the shape invariance
condition (\ref{eq4.1}) in the case when $R_0\not =0$. From eq.
(\ref{eq4.62}) it then follows that $g_0(x) = R_0 x/2$ rather than
being zero. One can again solve the set of linear differential equations
 (\ref{eq4.62}) and (\ref{eq4.63}) in succession yielding $g_1(x),
 g_2(x)$,...Further, the spectrum can be immediately obtained by using
eqs. (\ref{eq4.6}) and (\ref{eq4.61}). For example, in the case of an
arbitrary number of nonzero $R_j$ (in addition to $R_0$), it is given
by
\be \label{eq4.78}
E^{(1)}_n =nR_0+\sum_j \Gamma_j{(1+q^j)(1-q^{nj})\over (1-q^j)}
\ee
which is the spectrum of a $q$-deformed harmonic  oscillator
\cite{Biedenharn89,Macfarlane89}. It is worth pointing out here that, unlike
the
usual q-oscillator where the space is noncommutative but the
potential is normal $(\omega^2x^2)$, in our approach the space is
commutative, but the potential is deformed, giving rise to a
multi-parameter deformed oscillator spectrum.

An unfortunate feature of the new SIPs
obtained above is that they are not explicitly known in
terms of elementary functions but only as a Taylor series about
$x = 0$. Questions about series convergence naturally arise. Numerical
solutions pose no serious problems. As a consistency check, Barclay et al.
have checked numerically that the Schr\"{o}dinger equation solved with
numerically obtained potentials indeed has the analytical energy eigenvalues
given above. From numerical calculations  one finds that the superpotential and
the potential are as shown in Figs. 4.1 and 4.2 \cite{Barclay93}
corresponding to
the case when $r_1\not = 0, r_n=0, n\geq 2$. These authors, however, see no
evidence of the oscillations in $W$ and $V$ as reported by Degasperis and
Shabat
\cite{Degasperis92}. A very unusual new shape invariant potential has also been
obtained \cite{Barclay93} corresponding to $r_1 = 1, r_2 = - 1, r_n=0, n\geq 3$
(with $q = 0.3$ and $a = 0.75$) which is shown in Fig. 4.3. In this case,
whereas $V_1(x)$ is a double well potential, its shape invariant partner
potential $V_2(x)$ is a single well.

It is worth pointing out that even though the potentials are not known
in a closed form in terms of elementary functions, the fact that
these are reflectionless symmetric potentials can be used to
constrain them quite strongly. This is because, if we regard them
as a solution of the KdV equation at
time t = 0, then being reflectionless, it is well known that such
solutions as $t\rightarrow \pm\infty$ will break up into an infinite
number of solitons of the form $2k^2_i {\rm sech}^2k_i x$ \cite{Lamb80,Das89}.
On using the fact
that the KdV solitons obey an infinite number of conservation laws
corresponding to mass, momentum, energy ..., one can immediately
obtain constraints on the reflectionless SIPs obtained above \cite{Barclay93}.

\subsubsection{Solutions of Multi-Step Shape Invariance}
\label{sec4.3.3}

Having obtained potentials which are multi-parameter deformations of
the Rosen-Morse potential corresponding to the one soliton solution,
an obvious question to ask is if one can also obtain deformations of
the multi-soliton solutions. The answer is yes \cite{Barclay93} and as an
illustration we now explicitly obtain the multi-parameter deformations of
the two soliton case by using the formalism of two-step shape
invariance as developed earlier in this section.

Let us take the scaling ansatz $a_2=qa_1$ and expand the two
superpotentials $W$ and $\tilde W$ in powers of $a_1$
\be \label{eq4.79}
W(x;a_1)=\sum^{\infty}_{j=0}g_j(x)a^j_1 ; \hspace{.2in}
\tilde W(x;a_1)=\sum^{\infty}_{j=0}h_j(x)a^j_1
\ee
Further, we write $R$ and $\tilde R$ in the form
\be \label{eq4.80}
R(a_1) =\sum^{\infty}_{j=0}R_j a^j_1 ; \hspace{.2in}
\tilde R(a_1)= \sum^{\infty}_{j=0}\tilde R_ja^j_1
\ee

Using these in eqs. (\ref{eq4.12}) and (\ref{eq4.14}) and equating
powers of $a_1$ yields (n=0,1,2,...)
\be \label{eq4.81}
g'_n + \sum^n_{j=0}g_jg_{n-j} = \sum^n_{j=0} h_j h_{n-j}-h'_n +R_n
\ee
\be \label{eq4.82}
h'_n+\sum^n_{j=0}h_jh_{n-j}=q^n\sum^{\infty}_{j=0}g_jg_{n-j}-q^ng'_n+\tilde
R_n
\ee
This set of linear equations is easily solved in succession. For
example when $R_0=\tilde R_0=0$ (and hence $g_0(x)=h_0(x)=0)$ and
further $R_n = \tilde R_n=0, n\geq 3$ one can readily calculate all
the $g_n(x)$ and $h_n(x)$. The first two of each are
\bea \label{eq4.83}
g_1 & = & {(R_1-\tilde R_1)\over (1-q)} x,\nonumber \\
g_2 & = & {(R_2-\tilde R_2)\over
(1-q^2)} x + {x^3\over 3(1-q)^3}[(1-q)(\tilde
R^2_1-R^2_1)-2(1+q)R_1\tilde R_1],
\eea
\bea \label{eq4.84}
&&h_1  =  {(\tilde R_1-qR_1)\over (1-q)}x;\hspace{.5in} h_2 = {R_2x\over
(1-q^2)} \nonumber \\
    && - {x^3\over 3(1+q)(1-q)^2}[(1+q)\tilde R^2_1 +
(1+q)(1-q^2)R^2_1-2q(1-q)R_1\tilde R_1] \nonumber \\
\eea
The energy eigenvalues which follow from eqs. (\ref{eq4.22}) and
(\ref{eq4.23}) are \cite{Barclay93}
\bea \label{eq4.85}
E^{(1)}_{2n}(a_1) & = & \sum^2_{j=1}(R_j+\tilde R_j)a^j_1
{(1-q^{jn})\over (1-q^j)}\nonumber \\
E^{(1)}_{2n+1}(a_1) & = & \sum^2_{j=1} R_j a^j_1 {(1-q^{j(n+1)})\over
(1-q^j)}+\sum^2_{j=1}\tilde R^2_ja^j_1 {(1-q^{jn})\over (1-q^j)}
\eea

For the special case of $R_2=\tilde R_2=0$ the spectrum was obtained
previously by Spiridonov  from the considerations of the two-step
self-similar potentials \cite{Spiridonov92}.

The limit $q\rightarrow 0$ of the above equation is particularly
simple and yields the Rosen-Morse potential corresponding to the two
soliton case i.e.
\be \label{eq4.86}
W = 2\sqrt{\tilde R} \tanh \sqrt{\tilde R} x, \tilde W = \sqrt{\tilde
R} \tanh \sqrt{\tilde R} x
\ee
provided $R = 3\tilde R$. This procedure can be immediately
generalized and one can consider shape invariance with a scaling
ansatz in $3,4,.....p$ steps and thereby obtain multi-parameter
deformations of the $3,4,...p$ soliton Rosen-Morse potential.

\subsubsection{Other Solutions}
\label{sec4.3.4}

So far we have obtained solutions where $a_2$ and $a_1$ are related
either by scaling or by translation. Are there shape invariant
potential where $a_2$ and $a_1$ are neither related by scaling nor by
translation? It turns out that there are other possibilities for
obtaining new shape invariant potentials. Some of the other
possibilities are: $a_2=qa^p_1$ with $p$= 2,3,... and $a_2$ =
$qa_1 / (1+pa_1)$. Let us first consider the case when
\be \label{eq4.87}
a_2 = q a_1^2
\ee
i.e $p=2$. Generalization to arbitrary $p$ is straightforward \cite{Barclay93}.
On using eqs. (\ref{eq4.60}) and (\ref{eq4.61}) one
obtains the set of equations
\be \label{eq4.88}
g'_{2m}(x) + \sum^{2m}_{j=0}g_j(x) g_{2m-j}(x)
=q^m \sum^m_{j=0}g_j(x)g_{m-j}(x)-q^mg'_m(x)+R_{2m},\\
\ee
\be \label{eq4.89}
g'_{2m+1}(x) + \sum^{2m+1}_{j=0} g_j(x) g_{2m+1-j}(x) = R_{2m+1},
\ee
which can be solved in succession and one can readily calculate all
the $g_n(x)$. For example, when only $R_1$ and $R_2$ are nonzero, the
first three $g's$ are
\bea \label{eq4.90}
g_1(x) & = & R_1x, \hspace{.2in} g_2(x) = (R_2-qR_1)x-{1\over
3}R^3_1x^3,\nonumber \\
g_3(x) & = & {2\over 3}R_1(qR_1-R_2)x^3 + {2\over 15} R^3_1x^5.
\eea
The corresponding spectrum turns out to be $(E^{(1)}_0{(a_1)}=0)$
\be \label{eq4.91}
E^{(1)}_n={R_1\over q}\sum^n_{j=1}(a_1q)^{2^{j-1}}+{R_2\over q^2}
\sum^n_{j=1}(a_1q)^{2^j},n=1,2,...
\ee
The $q\rightarrow 0$ limit of these  equations again correspond to
the Rosen-Morse potential corresponding to the one soliton solution.
One can also consider shape invariance in multi-steps along with this
ansatz thereby obtaining deformations of the multi-soliton
Rosen-Morse potential.

One can similarly consider solutions to the shape invariance
condition (\ref{eq4.1}) in case
\be \label{eq4.92}
a_1= {qa_1\over 1+pa_1}
\ee
when $0<q<1$ and $pa_1<<1$ so that one can expand $(1+pa_1)^{-1}$ in
powers of $a_1$. For example, when only $R_1$ and $R_2$ are nonzero, then
one can show that the first two nonzero $g_n$ are
\be \label{eq4.93}
g_1(x)={R_1x\over (1+q)},\hspace{.2in} g_2(x) = (R_2+{pqR_1\over (1+q)})x
-{(1-q)\over (1+q)^2(1+q^2)}{x^3\over 3}
\ee
and the energy eigenvalue spectrum is $(E^{(1)}_0=0$)
\be \label{eq4.94}
E^{(1)}_n = R_1 \sum^n_{j=1}{q^{j-1}a_1\over [1+pa_1({1-q^{j-1}\over
1-q})]}+R_2\sum^n_{j=1} {(q^{j-1}a_1)^2\over [1+pa_1({1-q^{j-1}\over
1-q})]^2}
\ee

Generalization to the case when several $R_j$ are nonzero as well as
shape invariance in multi-steps is straight forward.

We would like to close this subsection with several comments.

\begin{enumerate}
\item Just as we have obtained $q$-deformations of the
reflectionless Rosen-Morse and harmonic oscillator potentials, can
one also obtain deformations of the other SIPs given in Table 4.1?
\item Have we exhausted the list of SIP?
We now have a significantly expanded list but it is
clear that the possibilities are far from exhausted. In fact it
appears that there are an unusually large number of shape invariant
potentials, for all of which the whole spectrum can be obtained
algebraically. How does one classify all these potentials? Do these
potentials include all solvable potentials \cite{Cooper87,Chuan91}?
\item For those SIP where $a_2$ and
$a_1$ are not related by translation, the spectrum has so far only
been obtained algebrically. Can one solve the Schr\"{o}dinger equation
for these potentials directly?
\item There is a fundamental difference between those shape
invariant potentials for which $a_2$ and $a_1$ are related by
translation and other choices (like $a_2=qa_1$). In particular,
whereas in the former case the potentials are explicitly known in a
closed form in terms of simple functions, in the other cases they are
only known formally as a Taylor series. Secondly, whereas in the
latter case, all the SIP obtained so far have
infinite number of bound states and are either reflectionless (or
have no scattering), in the former case one has also many SIP
with nonzero reflection coefficients.
\end{enumerate}

\subsection{Shape Invariance and Noncentral Solvable Potentials}
\label{sec4.4}

We have seen that using the ideas of SUSY and shape invariance, a number of
potential problems can be solved algebraically. Most of these
potentials are either one dimensional or are central potentials which are again
essentially one dimensional but on the half line. It may be worthwhile to
enquire if one can also algebraically solve some noncentral but separable
potential problems.  As has been shown
recently \cite{Khare93f}, the answer to the question is yes. It turns out that
the problem is algebraically solvable so long as the separated problems for
each of the coordinates belong to the class of SIP.
As an illustration, let us discuss noncentral separable potentials in spherical
polar coordinates.

In spherical polar coordinates the most general
potential for which the Schr\"{o}dinger equation is separable is given by
\cite{Morse53}
\be \label{eq4.97}
V(r,\theta,\phi) = V(r) +{V(\theta)\over r^2} + {V(\phi)\over r^2
\sin^2\theta}
\ee
where  $V(r), V(\theta)$ and $V(\phi)$ are arbitrary functions of their
argument.
First, let us see why the Schr\"{o}dinger equation with a potential of the
form given by  eq. (\ref{eq4.97}) is   separable in the $(r,\theta,\phi)$
coordinates. The equation for the wave function $\psi(r ,\theta, \phi )$ is
\be \label{eq4.98}
\bigg [ -({\partial^2\psi\over\partial^2r} + {2\over r}
{\partial\psi\over \partial r })-{1\over
r^2} ({\partial^2\psi\over \partial\theta^2}+{\rm cot} \theta
{\partial\psi\over \partial\theta}) - {1\over r^2 {\rm  sin}^2\theta}
{d^2\psi\over \partial\phi^2}\bigg ] = (E - V) \psi
\ee
It is convenient to write  $\psi(r,\theta,\phi)$  as
\be \label{eq4.99}
\psi(r,\theta,\phi) = {R(r)\over r} {H(\theta)\over (\sin
\theta)^{1/2}} K(\phi).
\ee
Substituting eq. (\ref{eq4.99}) in eq. (\ref{eq4.98}) and using the standard
separation of variables procedure, one obtains the following equations for the
functions $K(\phi),H(\theta)$ and $R(r)$ :

\be \label{eq4.101}
- {d^2 K\over d \phi^2} + V(\phi) K(\phi) = m^2 K(\phi),
\ee
\be \label{eq4.103}
- {d^2 H\over d \theta^2} + [ V(\theta) + ( m^2 - {1\over 4} )\,{\rm cosec}^2
 \theta]\, H(\theta) = l^2 H( \theta ),
\ee
\be \label{eq4.104}
- {d^2 R\over d r^2} + [ V(r) +{ ( l^2 - {1\over 4} )\over r^2}]\, R(r)
 = E\, R(r),
\ee
where $m^2$ and $l^2$ are separation constants.

As has been shown in ref. \cite{Khare93f}, the three Schr\"{o}dinger equations
given by (\ref{eq4.101}), (\ref{eq4.103}) and (\ref{eq4.104})
may be solved algebraically
by choosing appropriate SIPs for $V(\phi),V(\theta)$ and $V(r)$.
Details can be found in ref. \cite{Khare93f}.

Generalization of this technique to noncentral but separable potentials in
other orthogonal curvilinear coordinate systems as well as in other dimensions
is quite straightforward. Further, as we show below, one could use this
trick to discover
a number of new exactly solvable three-body potentials in one dimension.

\subsection{Shape Invariance and 3-Body Solvable Potentials}
\label{sec4.5}

Many years ago, in a classic paper, Calogero \cite{Calogero69} gave the
complete solution of the Schr\"{o}dinger equation for three particles in
one dimension, interacting via two-body harmonic and
inverse-square potentials. Later, Wolfes \cite{Wolfes74} used Calogero's
method to obtain analytical solutions of the same problem in the
presence of an added three-body potential of a special form.
Attention then shifted to the exact solutions of the many-body problem
and the general question of integrability
\cite{Sutherland71,Calogero71,Olshanetsky81}.
Recently, there has been renewed interest in the one-dimensional many-body
problem in connection with the physics of spin chains
\cite{Haldane88,Shastry88,Frahm93,Polychronakos92}. Also, there has been
a recent generalization of Calogero's potential for N-particles
to SUSY QM \cite{Freedman90}.

The purpose of this subsection is to show that using the results for SIPs
derived above, one can discover a number of new 3-body potentials for which
the 3-body problem in one dimension can be solved exactly \cite{Khare93g}.
There is a rule of
thumb that if one can solve a 3-body problem then one can also solve the
corresponding $n$-body problem. Thus hopefully one can also solve the
corresponding statistical mechanics problem.

The important point to note is that three particles in one dimension,
after the center of mass
motion is eliminated, have two degrees of freedom. This may therefore be mapped
on to a one-body problem in  two dimensions. Calogero \cite{Calogero69}
considered
the case where the two dimensional potential is noncentral but separable in
polar coordinates $r,\phi$. From the above discussion, it is
clear that if the potentials in each of the coordinates $r$ and $\phi$ are
chosen to be shape invariant, then the whole problem can be solved exactly.

Calogero's \cite{Calogero69} solution of the three-body problem is
for  the potential
\be \label{eq4.126}
V_{C}= \omega^2/8 \sum_{i<j} (x_{i}-x_{j})^2 + g \sum_{i<j} (x_{i}-x_{j})^{-2},
\ee
where $g>-1/2$ to avoid a collapse of the system. Wolfes \cite{Wolfes74}
showed that a three-body potential
\be \label{eq4.127}
V_{W}= f[(x_{1}+x_{2}-2x_{3})^{-2}+(x_{2}+x_{3}-2x_{1})^{-2}
       +(x_{3}+x_{1}-2x_{2})^{-2}]
\ee
is also solvable when it is added to $V_{C}$, with or without the
pairwise centrifugal term. The last two terms on the right-hand side
of eq. (\ref{eq4.127})
are just cyclic permutations of the first. Henceforth, such terms
occuring in  any potential are referred to as ``cyclic terms".
In this subsection, we give more examples
of three-body potentials that can be solved exactly. Our first solvable
example is the three-body potential of the form
\be \label{eq4.128}
V_{1}={{\sqrt{3}f_{1}}\over{2 r^{2}}}\left[{{(x_{1}+x_{2}-2x_{3})}
\over{(x_{1}-x_{2})}}+\; \rm {cyclic\;
terms}\;\right]\;,
\ee
which is added to the Calogero potential $V_{C}$ of eq. (\ref{eq4.126}).
In the above equation,
\be \label{eq4.129}
    r^2={{1}\over{3}} [(x_{1}-x_{2})^2+(x_{2}-x_{3})^2+(x_{3}-x_{1})^2].
\ee
To see why potentials given by eqs. (\ref{eq4.126}) to (\ref{eq4.128}) are
solvable, define the Jacobi coordinates
\be \label{eq4.130}
    R={{1}\over{3}}(x_{1}+x_{2}+x_{3}),
\ee
\be \label{eq4.131}
x={{(x_{1}-x_{2})}\over{\sqrt{2}}},\;\;
y={{(x_{1}+x_{2}-2x_{3})}\over{\sqrt{6}}}.
\ee
After elimination of the center of mass part from the Hamiltonian, only
the $x$- and  $y$-degrees of freedom remain, which may be mapped into
polar coordinates
\be \label{eq4.132}
    x=r~\sin~\phi~~,~~y=r~\cos~\phi~.
\ee
Obviously, the variables $r$, $\phi$ have ranges  $0\leq r\leq\infty$ and
$0\leq \phi\leq2\pi$. It is straightforward to show that
\bea \label{eq4.133}
 (x_{1}-x_{2})&=&\sqrt{2}\;r\;\rm{sin}\,\phi,\nonumber \\
 (x_{2}-x_{3})&=&\sqrt{2}\;r\;\rm{sin}(\phi+
 2\pi/3),\nonumber \\
 (x_{3}-x_{1})&=&\sqrt{2}\;r\;\;\rm{ sin}(\phi+4\pi/3).
\eea
It turns out that $V_C, V_W$ as well as $V_1$ are all noncentral but separable
potentials in the polar coordinates $r$,$\phi$. As a result,
the Schr\"{o}dinger equation separates cleanly in the radial and angular
variables, and the wave function can be written as
\be \label{eq4.134}
        \psi_{nl}(r,\phi)= {R_{nl}(r)\over \sqrt{r}}F_{l}(\phi).
\ee
In all three cases, the radial wave function obeys the equation
\be \label{eq4.135}
 \left\{-{{d^2}\over{dr^2}}+{{3}\over{8}} \omega^2r^2+{(B_{l}^2-{1\over 4})
\over r^2}\right\}R_{nl}=E_{nl}R_{nl}(r),
\ee
where $B_{l}^2$ is the eigenvalue of the Schr\"{o}dinger equation in the
angular
variable. Eq. (\ref{eq4.135}) corresponds to a SIP and the eigenvalues $E_{nl}$
and the eigenfunctions $R_{nl}$ which follow from Table 4.1 are
\be \label{eq4.136}
 E_{nl}=\sqrt{3/2}\;\omega(2n+B_{l}+1)~,~~~n,l=0,1,2...
\ee
\be \label{eq4.137}
 R_{nl}=r^{B_{l}}
\exp[{{-1}\over{4}}(\sqrt{3/2})\omega r^2]L_{n}^{B_{l}}[{{1}\over{2}}
(\sqrt{3/2})\omega r^2],
\ee
where $ B_{l}>0$. To examine the angular part of the eigenfunction
$F_{l}(\phi)$, take the potential $V_{C}+V_{1}$. Then, in the variable $\phi$,
the Schr\"{o}dinger equation is
\[
\left\{-{{d^2}\over{d\phi^2}}+{{g}\over{2}}\sum_{m=1}^{3}{\rm
cosec}^2[\phi+2(m-1) {\pi\over{3}}]+{{3}\over{2}}\,f_{1}\sum_{m=1}^{3}{\rm cot}
[\phi+2(m-1){\pi\over{3}}] \right\}F_{l}(\phi) \]
\be \label{eq4.138}
\hspace{3in} =B_{l}^2\,F_{l}(\phi).
\ee
On using the identities
\be \label{eq4.139}
\sum_{m=1}^{3}{\rm cosec}^2[\phi+2(m-1)\pi/3]=9{\rm cosec}^2 3\phi
\ee
\be \label{eq4.140}
\sum_{m=1}^{3}{\rm cot}[\phi+2(m-1)\pi/3]=3{\rm cot}~3\phi
\ee
eq. (\ref{eq4.138}) reduces to
\be \label{eq4.141}
\left\{-{{d^2}\over{d\phi^2}}+{{9}\over{2}}\,g\,{\rm
cosec}^2 3\phi+{{9}\over{2}} \, f_{1}\,{\rm cot}~3\phi\right\}\,F_{l}=
 B_{l}^2\,F_{l}(\phi).
\ee
Now this is again a SIP and hence its eigenvalues and eigenfunctions are
\be \label{eq4.142}
B_{l}^2= 9(l+a+1/2)^2-{{9}\over{16}} f_{1}^2/(l+a+1/2)^2,
\ee
\be \label{eq4.143}
F_{l}(\phi)=\exp(-i{\pi\over 2}{\tilde l})(\sin\,3\phi)^{\tilde l}
 \exp[{3\over4}{{ f_{1}\phi}\over{\tilde l}}] P_{l}^{-{\tilde l}
-{{if_{1}}\over{4{\tilde l}}},-{\tilde l}+{{if_{1}}\over{4  {\tilde l}}}}
(i\;\rm{cot}\,3\phi)
\ee
where $ P^{\alpha,\beta}_{n}$ is the Jacobi polynomial and
\be \label{eq4.144}
a= 1/2(1+2g)^{1/2}\;.
\ee
It must be noted that $g>-1/2$ for meaningful solutions. As expected, we
recover the results for the Calogero potential $V_C$ in the limit $f_1=0$.

For distinguishable particles, a given value of $\phi$ defines a specific
ordering. For ${0\leq\phi\leq\pi/3}$, eq. (\ref{eq4.133}) implies
$x_{1}\geq x_{2}\geq x_{3}$, and other ranges of $\phi$ correspond to
different orderings \cite{Calogero69,Wolfes74}. For singular repulsive
potentials, crossing is not allowed, and $F_{l}(\phi)$ of eq. (\ref{eq4.143})
is zero outside $0\leq\phi\leq\pi/3$. Following
Calogero, the wave function for the other ranges may be constructed.
Similarly, for indistiguishable particles, symmetrized or antisymmetrized
wave functions may be constructed.

Proceeding in the same way and using the results of Table 4.1, it is easily
show
 n
that the other exactly solvable potentials are \cite{Khare93g}
\bea \label{eq4.145}
V_{2}={{1}\over{8}} \omega^{2}\sum_{i<j}(x_{i}-x_{j})^2
+3 g\; [(x_{1}+x_{2}-2x_{3})^{-2}+\;\rm{cyclic\;terms}\;]\nonumber \\
-{{3\sqrt{3}}\over{2}}\;{{ f_{1}}\over{r^2}}\;\left[{{(x_{1}-x_{2})}
\over{(x_{1}+x_{2}-2x_{3})}}+\;\rm{cyclic\;terms}\,\right],
\eea
\be \label{eq4.146}
V_{3}={{1}\over{8}} \omega^2\sum_{i<j}
(x_{i}-x_{j})^2 +g\sum_{i<j}(x_{i}-x_{j})^{-2}-{{f_{3}}\over{\sqrt{6} r}}
\left[{{(x_{1}+x_{2}-2 x_{3})}\over{(x_{1}-x_{2})^2\,}}+
\,\rm{cyclic\;terms}\,\right],
\ee
\bea \label{eq4.147}
V_{4}=\,{{1}\over{8}}\omega^2\sum_{i<j}(x_{i}-x_{j})^2\,+3\,g\,
[\,(x_{1}+x_{2}-x_{3})^{-2}\,+\;\rm{cyclic\;terms}\,]\nonumber \\
+{1\over{3\sqrt{2}}}{f_{3}\over r}\,\left[\,{{(x_{1}-
x_{2})}\over{(x_{1}+x_{2}-2x_{3})^2}}+\;\rm{ cyclic\;terms}\,\right].
\eea
Notice that in all these cases one has combined the SIP in the angular variable
with the harmonic confinement. The three body scattering problem has also
been studied in these cases after droping the harmonic term \cite{Khare93g}.
The possibility of replacing the harmonic confinement term with the attractive
$1\over r$-type interaction has also been considered. Note that in this case
one has both discrete and continuous spectra. Further, in this case one can
obtain exact
solutions of three-body problems for all the SIPs discussed above
(i.e.$V_1$,...
$V_4, V_C, V_W, V_C+V_W$) along with the attractive $1\over r$
potential \cite{Khare93g}.

\section{Operator Transforms -- New Solvable Potentials from Old}
\label{sec5}

   In 1971, Natanzon \cite{Natanzon71} wrote down (what he thought at that time
 to be)
the most general solvable potentials i.e. for which the Schr\"{o}dinger
equation
can be reduced to either the hypergeometric or confluent hypergeometric
equation
 .
It turns out that most of these potentials are not shape invariant. Further,
for
most of them, the energy eigenvalues and eigenfunctions are known implicitly
rather than explicitly as in the shape invariant case.  One might ask if one
can
obtain these solutions from the explicitly solvable shape invariant ones.  One
strategy for doing this is to start with a Schr\"{o}dinger equation which is
exactly solvable (for example one having a SIP) and to see what happens to this
equation under a point canonical coordinate transformation.  In order for the
Schr\"{o}dinger equation to be mapped into another Schr\"{o}dinger equation,
there are severe restrictions on the nature of the coordinate transformation.
Coordinate transformations which satisfy these restrictions give rise to
new solvable problems.  When the relationship between  coordinates is implicit,
then the new solutions are only implicitly determined, while if the
relationship
is explicit then the newly found solvable potentials are also shape invariant
\cite{Junker90,De92,Gangopadhyaya94}. In a more specific special
application of these ideas,
Kostelecky et al. \cite{Kost85} were able to relate, using an explicit
coordinate transformation, the Coulomb problem in d dimensions with the
d-dimensional harmonic oscillator. Other explicit applications of the
coordinate
transformation idea are found in the review article of Haymaker and Rau
\cite{Haymaker86}.

 Let us see how this works.  We start from the one-dimensional Schr\"{o}dinger
 equation
\be
\left \{ -{d^2 \over dx^2} + [V(x)- E_n] \right \} \psi_n (x) = 0.
\ee
Consider the coordinate transformation from $x$ to $z$ defined by
\be
f(z)= {dz \over dx},
\ee
so that
\be
{d \over dx} = f {d \over dz}.
\ee
The first step in obtaining a new  Schr\"{o}dinger equation is to change
coordin
 ates and
divide by $f^2$ so that we have:
\be
 \left \{ -{d^2 \over dz^2}- { f^{\prime} \over f} {d \over dz}  + {[V- E_n]
\over f^2} \right \} \psi_n = 0
\ee
To eliminate the first derivative term, one next rescales the wave function:
\be
\psi = f^{-1/2} \bar{\psi}.
\ee
Adding a term
$\epsilon_n \psib$ to both sides of the equation  yields
\be
-\frac{d^2\psib_n}{dz^2} + [\bar{V} (E_n) + \epsilon_n] \psib_n = \epsilon_n
\psib_n,
\ee
where
\be
\bar{V} (E_n) = { V-E_n \over f^2} -\left[ {f^{\prime 2} \over 4f^2}
-{f^{\prime
\prime} \over 2f} \right].
\ee

In order for this to be a legitimate Schr\"{o}dinger equation, the potential
$\bar{V} (E_n) + \epsilon_n$  must be independent of $n$.  This can be
achieved if the quantity  $G$ defined by
\be
  G= V-E_n + \epsilon_n f^2
\ee
is independent of $n$.  How can one satisfy this condition?  One way is
to have $f^2$ and $G$ to have the same functional dependence on $x(z)$ as the
original potential $V$.  This  further requires that in order for $\bar{V}$ to
b
 e
independent of $n$, the parameters of $V$ must change with $n$ so that
the  wave function corresponding to the $n$'th energy level of the new
Hamiltoni
 an
is related to a wave function of the old Hamiltonian with parameters which
depen
 d on
$n$. This can be made clear by a simple example.

Let us consider an exactly solvable problem - the three dimensional harmonic
oscillator in a given angular momentum state with angular momentum $\beta$.
The reduced ground state wave function for that angular momentum is
\be
\psi_0(r) = r^{\beta+1} e^{-\alpha r^2/2},
\ee
so that the superpotential is given by
\be
W(r) = \alpha r - (\beta+1)/r,
\ee
and $H_1$ is given by:
\be
H_1 = - { d^2 \over dr^2} + { \beta (\beta+1) \over r^2} + \alpha^2 r^2 - 2
\alpha (\beta+3/2).
\ee
By our previous argument we must choose $f={dz \over dr}$
 to be of the form:
\be
f^2 = ({dz \over dr})^2 = { A \over r^2} + B r^2 + C.
\ee
The solution of this equation gives $z= z(r)$ which in general is not
invertible so that one
knows $r=r(z) $ only implicitly as discussed before.  However for special
cases one has an invertible function.  Let us for simplicity now choose
\be
f=r , \hspace{.2 in}   z=r^2/2~~.
\ee
As discussed earlier, the energy eigenvalues of the three dimensional harmonic
oscillator are give by
\be
E_n = 4 \alpha n~,
\ee
so that the condition we want to satify is
\be
V - 4 \alpha n + \epsilon_n f^2 = G = {D \over r^2} +E r^2 + F.
\ee
Equating coefficients we obtain
\bea
D &=& \beta (\beta+1)~, \nonumber \\
E &=& \epsilon_n + \alpha^2 \nonumber = \gamma ~, \\
F &=& - 4 \alpha n - 2 \alpha (\beta + 3/2) = -2Ze^2.
\eea

We see that for the quantities $D,E,F$ to be independent of $n$ one needs to
hav
 e
that $\alpha$, which describes the strength of the oscillator potential,
be dependent on $n$.  Explicitly, solving the above three equations and
choosing $\beta = 2 l + 1/2$,  we obtain the relations:
\be
\alpha(l,n) =  {Z e^2 \over 2(l+n+1)}~,~
\epsilon_n =  \gamma - {Z^2 e^4 \over 4(l+1+n)^2}~~.
\ee
We now choose $\gamma = \alpha^2(l,n=0)$ so that the ground state energy is
zero. These energy levels are those of the hydrogen atom. In fact, the new
Hamiltonian written in terms of $z$ is now
\be
\bar{H} = -{d^2 \over dz^2} + {l(l+1) \over z^2} - {Ze^2 \over z} +{Z^2 e^4
\over 4(l+1)^2 }
\ee
and the ground state wave function of the hydrogen atom is obtained from
the ground state wave function of the harmonic oscillator via
\be
\psib_0 = f^{1/2} \psi_0 = x^{l+1} e^{-\alpha(l,n=0) x}.
\ee
Higher wave functions will have values of $\alpha$ which depend on $n$ so that
the different wave functions correspond to harmonic
oscillator solutions with different strengths.

All the exactly solvable shape invariant potentials of Table 4.1 can be
inter-related by point canonical coordinate transformations
\cite{De92,Gangopadhyaya94}. This
is nicely illustrated in Fig. 5.1 .
In general, $r$ cannot be explicitly found in terms of $z$, and one has
\be
dz/dr = f = \sqrt{A/r^2 + Br^2 + C}~~,
\ee
whose solution is:

\bea
z &=& {{{\sqrt{A + {\rm C}\,{r^2} + B\,{r^4}}}}\over 2} +
   {{{\sqrt{A}}\,\log ({r^2})}\over 2} \nonumber \\
& -&
   {{{\sqrt{A}}\,\log (2\,A + {\rm C}\,{r^2} +
         2\,{\sqrt{A}}\,{\sqrt{A + {\rm C}\,{r^2} + B\,{r^4}}})}\over 2}
\nonumber \\
 &+&
   {{{\rm C}\,\log ({\rm C} + 2\,B\,{r^2} +
         2\,{\sqrt{B}}\,{\sqrt{A + {\rm C}\,{r^2} + B\,{r^4}}})}\over
     {4\,{\sqrt{B}}}}~~.
\eea
This clearly is not invertible in general. If we choose this general coordinate
transformation, then the potential that one obtains is the particular class of
Natanzon potentials whose wave functions are confluent hypergeometric functions
in the variable $r$ and are thus only implicitly known in terms of the true
coordinate $z$. In fact, even the expression for the transformed potential is
only known in terms of $r$:

\be
\bar{V}(z,D,E) = {1/f^2} [{D/r^2}+E {r^2}+F-{f'^{2}/4}+ff''/2]
\ee
and thus only implicitly in terms of $z$. Equating coefficients, we get an
implicit expression for the eigenvalues:
\be
{C \epsilon_n -F\over {2\sqrt{E-B\epsilon_n}}} - \sqrt{D+{1/4} -A \epsilon_n}
=2n+1
\ee
as well as the state dependence on $\alpha$ and $\beta$ necessary for the new
Hamiltonian to be energy independent:
\be
\alpha_n = \sqrt{E-B\epsilon_n} ,
\hspace{.2in} \beta_n = -1/2 + \sqrt{D +{1/4}-A \epsilon_n}.
\ee

\subsection{Natanzon Potentials}
\label{sec5.1}

   The more general class of Natanzon potentials whose
wave functions are hypergeometric functions can be obtained by making
an operator transformation of the generalized P\"{o}schl-Teller potential
whose Hamiltonian is:

\be
H_1 = A_1^{\dag} A_1 = -{d^2 \over dr^2} + { \beta (\beta-1) \over
{\sinh^2 r}} -{ \alpha (\alpha+1) \over
{\cosh^2 r}} +(\alpha-\beta)^2
\ee
This corresponds to a superpotential
\be
W= \alpha \tanh r - \beta {\rm coth}~r
\ee
and a ground state wave function given by:
\be
\psi_0 =  \sinh^{\beta} r \hspace{.1in}  \cosh^{-\alpha} r~.
\ee
The energy eigenvalues were discussed earlier and are
\be
E_n = (\alpha-\beta)^2 -(\alpha-\beta-2n)^2.
\ee

The most general transformation of coordinates from
$r$ to $z $ which preserves the Schr\"{o}dinger equation is described by:
\be
f^2 = { B \over
{\sinh^2 r}} -{A \over
{\cosh^2 r}} + C  = ({dz \over dr})^2
\ee
{}From this we obtain an explicit expression for $z$ in terms of $r$.
\bea
z&=&
{\sqrt{A}}\,\tan^{-1}\{{{-3\,A + B - C + (A+B-C)\,\cosh ~2r}\over
       {2\,{\sqrt{A}}\,{\sqrt{-2\,A + 2\,B -  C + 2\,(A+B)\,\cosh ~2r +
       C \,{{\cosh^2 2r}}}}}}\} \nonumber \\
& -&
   {\sqrt{B}}\,\log \{-A + 3\,B - C  + (A+B+C)\,\cosh ~2r  \nonumber \\
&+&
      2\,{\sqrt{B}}\,{\sqrt{-2\,A + 2\,B - C  + 2\,(A+B)\,\cosh ~2r +
 C\,{{\cosh^2 2r}}}}\} \nonumber \\ & +&
   {\sqrt{C }}\,\log \{A + B + C \,\cosh ~2r \nonumber \\
 &+&
      {\sqrt{C }}\,{\sqrt{-2\,A + 2\,B - C  + 2\,(A+B)\,\cosh ~2r +
            C \,{{\cosh^2 2r}}}} \} \nonumber \\
 &+&
   {\sqrt{B}}\,\log \{2\,{{{\sinh}^2~r}} \}
\eea
However the expression for the transformed potential is only known in
terms of $r$:
\be
\bar{V}(z,\gamma,\delta)  = {1 \over f^2} ( {\delta (\delta -1) \over
{\sinh^2 r}} -{\gamma (\gamma+1) \over
{\cosh^2 r}} + \sigma  - {f^{\prime 2} \over 4f^2} +{f^{\prime
\prime} \over 2f}  )
\ee
and thus only implicitly in terms of $z$.   Equating coefficients,
we get an implicit expression for the eigenvalues:
\be
[(\gamma+1/2)^2 - A \epsilon_n]^{1/2} -[(\delta-1/2)^2 - B \epsilon_n]^{1/2}
-(\sigma - C \epsilon_n)^{1/2} = 2n+1
\ee
as well as the state dependence on $\alpha$ and $\beta$ necessary for the new
Hamiltonian to be energy independent:
\bea
\alpha_n &=& [(\gamma+1/2)^2 - A \epsilon_n]^{1/2}-1/2~, \nonumber \\
\beta_n &=&  [(\delta-1/2)^2 - B \epsilon_n]^{1/2}+1/2~.
\eea

\subsection{Generalizations of Ginocchio and Natanzon Potentials}
\label{sec5.2}

In ref. \cite{Cooper87}  it was explicitly shown that the Ginocchio and
Natanzon
potentials, whose wave functions are hypergeometric functions
of  an implicitly determined variable are not shape invariant and
thus one could construct new exactly solvable potentials from these using the
factorization method.  It is convenient to use the original
approach  \cite{Ginocchio84,Natanzon71} to define the
Ginocchio and Natanzon potentials and then obtain their generalization to more
complicated solvable potentials whose wave functions are sums of hypergeometric
functions. These generalized potentials have coordinates which are
only implicitly
known. The operator which allows one to construct the
eigenfunctions of $H_2$ from
those of $H_1$ converts single hypergeometric functions of the implicitly known
coordinate to sums of hypergeometric functions. This process yields totally new
solvable potentials. The resulting potentials are ratios of polynomials in the
transformed  coordinate in which the wave functions are sums of hypergeometric
functions. The transformed coordinate is only implicitly known in terms of the
coordinate system for the Schr\"{o}dinger equation.

The original method of obtaining the Natanzon potentials was
to find a coordinate transformation that mapped the Schr\"{o}dinger
equation onto the equation for the hypergeometric functions.

If we denote by $r$ the coordinate appearing in the Schr\"{o}dinger equation :
\be
[-{ d^2 \over dr^2} + V(r)] \psi(r)  = 0
\ee
where $ -\infty \leq r \leq \infty$
and
$z$ the coordinate describing the hypergeometric function
$F(\alpha,\beta;\gamma;z)$ appearing in the differential equation:
\be
 z(1-z) {d^2 F \over dz^2} + [\gamma-(\alpha+\beta+1)z] {dF \over dz} -\alpha
\beta  F=0,
\ee
where $ 0 \leq z \leq 1$,
then the transformation of coordinates is given by \cite{Natanzon71}
\be
{dz \over dr} = {2z (1-z)  \over R^{1/2}} ,  \hspace{.5in} R= a z^2 + (c_1 -
c_0
  -a) z + c_0
\ee

In terms of the coordinate $z$ the most general potential $V(r)$ which
correspon
 ds to a Schr\"{o}dinger equation
getting  mapped into a hypergeometric function equation is given by:
\bea \label{eq5.4}
V(r) &= &[fz(z-1) + h_0(1-z) +h_1 z+1]/R +  \nonumber \\
\{a &+& { a + (c_1-c_0)(2z-1) \over z (z-1) } - {5 \triangle \over 4R} \}
{z^2 (1-z)^2 \over R^2}
\eea
where
\[ \triangle = (a-c_0-c_1)^2 - 4 c_0 c_1. \]
We note that $z$ is only implicitly known in terms of $r$.
The potential $V(r)$ is a function of six dimensionless parameters  $f,
h_0,h_1,
a, c_0$ and $c_1$. These six parameters will be  related to the energy
$\epsilon_n$ and the parameters $(\alpha,\beta,\gamma)$ of the hypergeometric
function below.

The class of potentials called the Ginocchio potentials
is a subclass of the Natanzon class which has only two independent variables $
\
 nu$ and $\lambda$ where $ c_0=0$, $c_1=1/\lambda^4 $, $a= c_1- 1/\lambda^2$,
$h
 _0 = -3/4$, $h_1 = -1$, and $f=(\nu+1/2)^2 -1$.

The Ginocchio potentials are more easily discussed, however by considering
them as derived from coordinate transformations which map the
Schr\"{o}dinger equation into the differential equation for the Gegenbauer
polynomials.

The transformation is given by
\be
{ dy \over dr} = (1-y^2) [ 1+ (\lambda^2-1) y^2 ]
\ee
where now $ -1 \leq y \leq 1$ and the potential is given by:
\be
V(r) =\{-\lambda^2 \nu(\nu+1) + {1 \over 4} (1-\lambda^2) [5(1-\lambda^2) y^4
-(
 7-\lambda^2) y^2 +2]\}(1-y^2).
\ee
Although $y$ is not known explictly in terms of $r$, it is implicitly known
via:

\[
r \lambda^2 = {\rm tanh}^{-1}(y) -(1-\lambda^2)^{1/2} {\rm tanh}^{-1}
[(1-\lambda^2)^{1/2}y], \hspace{.2in} \lambda <1 ,
 \]
\be
r \lambda^2 = {\rm tanh}^{-1}(y) +(\lambda^2-1)^{1/2} {\rm tan}^{-1}
[(\lambda^2-1)^{1/2}y], \hspace{.2in} \lambda >1 . \ee
By introducing the variables: $x=\lambda y/[g(y)]^{1/2}$  where
\[
g(y) = 1+(\lambda^2-1) y^2,
\]
and changing the variable $x$ to an angle $ x=~\cos \theta$, one
obtains from the Schr\"{o}dinger equation, the equation for the Gegenbauer
polynomials:

\be [ {d^2 \over d \theta ^2} + (n+ \mu + 1/2)^2 - {\mu^2-1/4 \over \sin^2
\theta} ] Z^{1/2} \psi_n =0
\ee
where
\[Z= { |1-y^2|^{1/2} \over \lambda} \]
\be
\epsilon_n = - \mu_{n} ^2 \lambda^4.
\ee
The corresponding energy eigenvalues and eigenfuctions are

\be
\mu_n \lambda^2= [\lambda^2(\nu+1/2)^2 +(1-\lambda^2)(n+1/2)^2]^{1/2} -
(n+1/2),
\ee

\be
\psi_n  = (1-y^2)^{\mu_n/2} [g(y)]^{-(2\mu_n+1)/4} C_{n} ^ {(\mu_n+1/2)}(x).
\ee
Thus the superpotential is :
\be
W(r) = -{\psi'(r)\over {\psi(r)}} = {1 \over 2} ( 1-\lambda^2) y (y^2-1) +
\mu_0
  y \lambda^2
\ee
For the Natanzon potential (\ref{eq5.4}) we have that the energy eigenvalues
are given by:
\bea
(2n+1) &= &(1-a \epsilon_n +f)^{1/2} - (1-c_0\epsilon_n +h_0)^{1/2} \nonumber
\\
&-& (1-c_1 \epsilon_n+h_1)^{1/2} \equiv \alpha_n -\beta_n-\delta_n.
\eea
and the corresponding (unnormalized) energy eigenfunctions are given by :
\be
\psi_n = R^{1/4} z^{\beta_n/2} (1-z)^{\delta_n/2}\hspace{.1cm}
F(-n,\alpha_n-n;1
 +\beta_n;z).
\ee

It is easy to determine the superpotential from the ground state wave
function:
\bea
W(r) &= & [\delta_0 z - (1+ \beta_0)(1-z)]/R^{1/2} \nonumber \\
&+& [(c_1-c_0-a)z + 2 c_0](1-z)/(2R^{3/2}).
\eea
Once we have the superpotential and the explicit expression for the wave
functions we can determine all the wave functions of the partner potential
using the operator:

 \[ A_1 ={d \over dr} + W(r) = {dz \over dr} {d \over dz} + W(z) =
 {2z (1-z)  \over R^{1/2}}{d \over dz} + W(z)
\]
for the case of the Natanzon potential and the operator:
\[ A_1 ={d \over dr} + W(r) = {dy \over dr} {d \over dy} + W(y) =
 (1-y^2) [ 1+ (\lambda^2-1) y^2 ]{d \over dy} + W(y)
\]
for the case of the Ginocchio potentials.
In general, these operators take a single hypergeometric function into a
sum of two hypergeometric functions which cannot be reexpressed as
a single hypergeometric function except in degenerate cases where one
obtains the shape invariant special cases discussed earlier. Explicitly,
one has for the unnormalized eigenfunctions of $H_2$, the partner to the
origina
 l
Natanzon Hamiltonian:
\[
\psi_{n-1}^{(2)} = R^{-1/4} z^{\beta_n/{2}} (1-z)^{\delta_n/{2}}
[(1-z)[\beta_n-\beta_0-z(\delta_n- \delta_0)]
F(-n,\alpha_n-n;1+\beta_n;z)
\]
\be
- 2z (1-z)n (n+1+ \beta_n + \delta_n) F(-n+1,\alpha_n-n+1 ;2 + \beta_n;z)]
\ee
and for the partner potential $V_2 = W^2 + W'$ one has
\[
V_2 = \epsilon_0 + [(\beta_0+ \delta_0)(\beta_0+ \delta_0+2) z (z-1) +
(\beta_0^2-1) (1-z) + (\delta_{0}^{2} -1)z +1] /R
\]
\[
+ [a-[c_1(3 \beta_0 + \delta_0) + c_0(\beta_0+3 \delta_0) a(
1-\beta_0-\delta_0)] / [z(z-1)]
\]
\[
- (2z-1)[(c_1-c_0) (\beta_0 + \delta_0+1) + a (\beta_0 - \delta_0)] /[z(z-1)]
\]
\be
+ 7 \triangle /(4R)] z^2 (1-z^2) /R^2.
\ee
By using the hierarchy of Hamiltonians one can construct in the usual manner
more and more complicated potentials which are ratios of higher
and higher order polynomials in $z$ (as well as $R$) which are isospectral
to the original Natanzon potential except for the usual missing states.
The wave functions will be sums of hypergeometric functions.  The arguments
are identical for the Ginocchio class with the hypergeometric functions
now being restricted to being Gegenbauer polynomials.  The potentials
one obtains can have multiple local minima and several of these are
displayed in  \cite{Cooper87}. In \cite{Cooper87} it was also shown that the
ser
 ies
of Hamiltonians and SUSY charges form the graded Lie algebra
$sl(1/1)\otimes SU(2)$.  However this algebra did not lead to any new insights.
We shall also see in Sec.13 that the series of Hamiltonians form a
parasupersymmetry of order $p$ if the original Hamiltonian has $p$ bound
states.

\section{Supersymmetric WKB Approximation}
\label{sec6}

The semiclassical WKB method \cite{Froman65} is one of the most useful
approximations for computing the energy eigenvalues of the
Schr\"{o}dinger equation.
It has a wider range of applicability than standard perturbation theory
which is restricted to perturbing potentials with small coupling
constants.
The purpose of this section is to describe and give applications of the
supersymmetric WKB method (henceforth called SWKB) \cite {Comtet85,Khare85}
which has been inspired by supersymmetric quantum mechanics.

\subsection{SWKB Quantization Condition for Unbroken Supersymmetry}
\label{sec6.1}

As we have seen in previous sections, for quantum mechanical problems,
the main implication of SUSY is that it relates the energy
eigenvalues, eigenfunctions and phase shifts of two supersymmetric
partner potentials $V_1(x)$ and $V_2(x)$. Combining the ideas of SUSY
with the lowest order WKB method, Comtet, Bandrauk and Campbell
\cite{Comtet85} obtained the lowest order SWKB quantization condition
in case SUSY is unbroken and showed that it yields energy
eigenvalues which are not only
exact for large quantum numbers $n$ (as any WKB approximation scheme
should in the semiclassical limit) but which are also exact
for the ground state $(n=0)$. We shall now show this in detail.

In lowest order, the WKB quantization condition for the one dimensional
potential $V(x)$ is \cite{Froman65}
\be \label{3.1}
\int_{x_1}^{x_2}\sqrt{2m[E_n -V(x)]}~ dx =(n+1/2) \hbar\pi,
{}~~ n=0,1,2,... ~~,
\ee
where $x_1$ and $x_2$ are the classical turning points defined by
$E_n=V(x_1)=V(x_2)$. For the potential $V_1(x)$ corresponding to the
superpotential $W(x)$, the quantization condition
(\ref{3.1}) takes the form
\be \label{3.2}
\int_{x_1}^{x_2} \sqrt{2m[E_n^{(1)}-W^2(x)+\frac{\hbar}{\sqrt{2m}}W'(x)]}~dx
        =(n+1/2)\hbar\pi.
\ee
Let us assume that the superpotential $W(x)$ is formally $O(\hbar^0)$.
Then, the $W'$ term is clearly
$O(\hbar)$. Therefore, expanding the left hand side in powers of $\hbar$ gives
\be  \label{3.3}
\int_{a}^{b} \sqrt{2m[E_n^{(1)}-W^2(x)]}~dx+\frac {\hbar}{2}
\int_{a}^{b} \frac{W'(x) ~dx}{\sqrt{E_n^{(1)}-W^2(x)}}+...
=(n+1/2)\hbar\pi,
\ee
where $a$ and $b$ are the turning points defined by $E_n^{(1)}=W^2(a)
=W^2(b)$. The $O(\hbar)$ term in eq. (\ref{3.3}) can be integrated
easily to yield
\be \label{term}
\frac{\hbar}{2} \sin^{-1}\left[\frac{W(x)}{\sqrt{E_n^{(1)}}}\right]_{a}^{b}.
\ee
In the case of unbroken SUSY, the superpotential $W(x)$ has
opposite signs at the two turning points, that is
\be \label{oppsign}
-W(a)=W(b)=\sqrt{E_n^{(1)}}~.
\ee
For this case, the $O(\hbar)$ term in (\ref{term}) exactly gives
$\hbar\pi/2$, so that to leading order in $\hbar$ the SWKB quantization
condition when SUSY is unbroken is \cite{Comtet85,Khare85}
\be  \label{3.4}
\int_{a}^{b} \sqrt{2m[E_n^{(1)}-W^2(x)]}~dx
=n\hbar\pi,~~n=0,1,2,...
\ee
Proceeding in the same way, the SWKB quantization condition
for the potential $V_2(x)$ turns out to be
\be  \label{3.5}
\int_{a}^{b} \sqrt{2m[E_n^{(2)}-W^2(x)]}~dx =(n+1)\hbar\pi,~~n=0,1,2,...
\ee

Some remarks are in order at this stage.

(i) For $n=0$, the turning points $a$ and $b$ in eq. (\ref{3.4}) are
coincident and $E_0^{(1)} = 0$. Hence SWKB is exact by
construction for the ground state energy of the
Hamiltonian $H_1=(-\hbar^2/{2m})d^2/dx^2+V_1(x)$.

(ii) On comparing eqs. (\ref{3.4}) and (\ref{3.5}), it follows that the
lowest order SWKB quantization condition preserves the SUSY
level degeneracy i.e. the approximate energy eigenvalues computed from the
SWKB quantization conditions for $V_1(x)$ and $V_2(x)$ satisfy the exact
degeneracy relation $E_{n+1}^{(1)}=E_n^{(2)}$.

(iii) Since the lowest order SWKB approximation is not only exact
as expected for large $n$,
but is also exact by construction for $n=0$, hence, unlike the ordinary
WKB approach, the SWKB eigenvalues are constrained to be accurate at both
ends of allowed values of $n$ at least when the spectrum is purely discrete.
One can thus reasonably expect better results than the WKB scheme.

\subsection{Exactness of the SWKB Condition for Shape Invariant Potentials}
\label{sec6.2}

How good is the SWKB quantization condition [eq. (\ref{3.4})]?
To study this question, the obvious first attempts consisted of
obtaining the SWKB bound state spectra of several analytically solvable
potentials like Coulomb, harmonic oscillator,
Morse, etc. \cite{Comtet85,Khare85}. In fact, it
was soon shown that the lowest order SWKB condition gives the
exact eigenvalues for all SIPs
\cite{Dutt86}! The proof of this statement follows from the facts that
the SWKB condition preserves the level
degeneracy and a vanishing ground state energy eigenvalue.
For the hierarchy of Hamiltonians $H^{(s)}$ discussed in Sec. 3, the SWKB
quantization condition takes the form
\be \label{3.10}
\int\sqrt{2m[E_n^{(s)}-\sum_{k=1}^{s}R(a_k)-W^2(a_s;x)]}~dx=n\hbar\pi~.
\ee

Now, since the SWKB quantization condition is exact
for the ground state energy when SUSY is unbroken, hence
\be
E_0^{(s)}=\sum_{k=1}^{s}R(a_k)
\ee
as given by eq. (\ref {3.10}), must be exact for Hamiltonian $H^{(s)}$.
One can now go back in sequential manner from $H^{(s)}$ to $H^{(s-1)}$ to
$H^{(2)}$ and $H^{(1)}$ and use the fact that the SWKB
method preserves the level degeneracy $E_{n+1}^{(1)}=E_n^{(2)}$.
On using this relation $n$ times , we find that for all SIPs,
the lowest order SWKB condition
gives the exact energy eigenvalues \cite{Dutt86}.
This is a very substantial improvement over the usual WKB formula
eq. (\ref{3.1}) which is not exact for most SIPs. Of
course, one can artificially restore exactness by ad hoc Langer-like
corrections \cite {Krieger67}. However, such modifications are unmotivated
and have different forms for different potentials.
Besides, even with such corrections,
the higher order WKB contributions are non-zero for most of these
potentials\cite{Krieger67,Seetharaman84}.

What about higher order SWKB contributions? Since the lowest order SWKB
energies are exact for shape invariant potentials,
it would be nice to check that higher
order corrections vanish order by order in $\hbar$. By starting from the
higher order WKB formalism, one can readily develop the higher order
SWKB formalism \cite{Adhikari88}. It has been explicitly checked for all
known SIPs that up to $O(\hbar^6)$
there are indeed no corrections.. This
result can be extended to all orders in $\hbar$ \cite{Raghunathan87,Barclay91}.
Conditions on the superpotential which ensure exactness of the
lowest order SWKB condition have been given in ref. \cite{Barclay91}.

It has been proved above that the lowest order SWKB approximation
reproduces the exact bound state spectrum of any SIP. This
statement has indeed been explicitly checked for all SIPs
known until last year i.e. solutions of the shape invariance
condition involving a translation of parameters $a_2=a_1+$ constant. However,
it has recently been shown \cite{Barclay93b} that the above statement
is not true for the newly
discovered class \cite{Khare93a,Barclay93,Shabat92,Spiridonov92} of
SIPs discussed in Sec. 4.2.2, for which the parameters
$a_2$ and $a_1$ are related by scaling $a_2=qa_1$.
What is the special feature of these new potentials that interferes
with the proof that
eq. (\ref{3.4}) is exact for SIPs ?
To understand this, let
us look again at the derivation of the lowest
order SWKB quantization condition. In the derivation,
$W^2$ is taken as $O(\hbar^0)$ while $\hbar W'$ is $O(\hbar)$ and
hence one can expand the integrand on the left hand side
in powers of $\hbar$. This assumption is justified for all the
standard SIPs \cite{Infeld51,Dutt88,Levai89}
since $W^2$ is indeed of $O(\hbar^0)$ while $\hbar W'$ is indeed of
$O(\hbar)$. One might object to this procedure since the resulting potential
$V_1$  is then $\hbar$-dependent. However, in all
cases, this $\hbar$-dependence can be absorbed into
some dimensionful parameters in the problem.
For example, consider
\be
W(x) = A \tanh x  ~~,~~V_1(x) = A^2 - A(A+\hbar) {\rm sech}^2x.
\ee
Taking $A$ such that $A(A+\hbar)$ is independent of $\hbar$ gives the desired
$\hbar$-independent potential (the additive constant is irrelevant and
so can contain $\hbar$).
Such a move may appear to be of limited value since
one cannot apply SWKB directly to a superpotential $W$ which is
now $\hbar$-dependent.
However, because $A$ is a free parameter, one can
continue the SWKB results obtained
for $A$ (and hence $W$) independent of $\hbar$ over to this superpotential and
so obtain an SWKB approximation for a $\hbar$-independent potential.

What about the new potentials~?
In the simplest of these cases, the only free parameter in the problem (apart
from $q$) is the combination $R_1 a_1$, on which $W$ depends as $W(x, R_1 a_1)
=\sqrt{R_1 a_1} F(\sqrt{R_1 a_1} x/\hbar)$.
Incorporating different dependences on $\hbar$ in $R_1 a_1$ will give
different ones in $W$, $V_1$ and $E_n$, but $F$ is a sufficiently
complicated function that there is no way of eliminating $\hbar$ from $W^2$.
This is a direct consequence of the  scaling reparameterisation $a_2=q a_1$
not involving $\hbar$. If $W^2(x;a_1)$ were independent of $\hbar$, so would
$W^2(x;a_2)$ be and in taking the lowest order of the
shape invariance condition one would get
$W^2(x;a_1)=W^2(x;a_2)$, which corresponds to the harmonic oscillator.
Furthermore, with $a_2=q a_1$, $W^2$ and $\hbar W'$ are now of a similar
order in $\hbar$. The basic distinction between them involved
in deriving eq. (\ref{3.4}) is thus
no longer valid and we are prevented from deriving the SWKB condition for
these new potentials.

We thus see that the SWKB quantization condition is
not the correct lowest order formula in the case of the new SIPs and hence
it is not really surprising that eq. (\ref{3.4}) does not give
the exact eigenvalues for these potentials.
In other words, it remains true that the lowest order SWKB quantization
condition is
exact for SIPs (if the SUSY is unbroken), but only in those cases
for which the formula is applicable in the first place.
It is thus still the case that the SIPs given in refs. \cite{Dutt88,Levai89}
are the only known ones for which the lowest order SWKB formula is
exact and the higher order corrections are all zero.

\subsection{Comparison of the SWKB and WKB Approaches}
\label{sec6.3}

Let us now compare the merits of the two schemes [WKB and SWKB]. For
potentials for which the ground state wave function (and hence the
superpotential $W$) is not known, clearly the WKB approach is preferable, since
one cannot directly make use of the SWKB quantization condition eq.
(\ref{3.4}).
On the other hand, we have already seen that for shape invariant potentials,
SWKB is clearly superior. An obvious interesting question is to compare
WKB and SWKB for
potentials which are not shape invariant but for whom the ground state
wave function is known. One choice which readily springs to mind is
the Ginocchio potential given by \cite{Ginocchio84}
\be \label{4.1}
V(x)=(1-y^2)\left\{-\lambda^2\nu(\nu+1)+\frac{(1-\lambda^2)}{4}[2-
(7-\lambda^2)y^2+5(1-\lambda^2)y^4]\right\}
\ee
where $y$ is related to the independent variable $x$ by
\be
\frac{dy}{dx}=(1-y^2)[1-(1-\lambda^2)y^2]~.
\ee
Here the parameters $\nu$ and $\lambda$ measure the depth and shape of
the potential respectively. The corresponding
superpotential is \cite{Cooper87}
\be
W(x)=(1-\lambda^2)y(y^2-1)/2+\mu_0\lambda^2y
\ee
where $\mu_n$ is given by \cite{Ginocchio84}
\be \label{4.4}
\mu_n\lambda^2=\sqrt{[\lambda^2{(\nu+1/2)}^2 + (1-\lambda^2){(n+1/2)}^2]}
        -(n+1/2)
\ee
and the bound state energies are
\be \label{4.5}
E_n=-\mu_n^2\lambda^4 ,~~ n=0,1,2,...
\ee
For the special case $\lambda=1$, one has the Rosen-Morse
potential, which is shape invariant.
The spectra of the Ginocchio potential using both the WKB and SWKB
quantization conditions have been computed \cite{Khare89}. The results are
shown
  in
Table 6.1. In general, neither semiclassical method gives the exact
energy spectrum. The only exception is the shape invariant limit
$\lambda=1$, in which case the SWKB results are exact, as expected. Also,
for $n=0,1$ the SWKB values are consistently better, but there is no clear cut
indication that SWKB results are always better. This example, as well as
other potentials studied in ref.
\cite{Dutt87,Roy88,Khare89,DeLaney90,Varshni92}
 ,
support the conjecture
that shape invariance is perhaps a necessary condition for the lowest order
SWKB approximation to yield the exact spectrum \cite{Khare89}.

So far, we have concentrated our attention on the energy eigenvalues. For
completeness, let us note that several authors have also obtained the wave
functions in the SWKB
approximation \cite{Comtet85,Fricke88,Murayama89,Sukhatme90}.
As in the WKB method, the SWKB wave functions diverge at the
turning points $a$ and $b$. These divergences can be regularized either by
the uniform approximation \cite{Fricke88,Miller53} or by appropriate
retention of higher orders in $\hbar$ \cite{Murayama89,Sukhatme90}.

\subsection{SWKB Quantization Condition for Broken Supersymmetry}
\label{sec6.4}

The derivation of the lowest order SWKB quantization condition for the
case of unbroken SUSY  is given in Sec. \ref{sec6.1} [eqs. (\ref{3.1})
to (\ref{3.4})]. For the case of broken SUSY, the same derivation
applies until one examines the $O(\hbar)$ term in eq. (\ref{term}). Here, for
broken SUSY, one has
\be \label{samesign}
W(a)=W(b)=\sqrt{E_n^{(1)}}
\ee
and the $O(\hbar)$ term in (\ref{term}) exactly
vanishes. So, to leading order in $\hbar$ the SWKB quantization
condition for broken SUSY is \cite{Inomata92,Dutt93b}
\be  \label{bswkb}
\int_{a}^{b} \sqrt{2m[E_n^{(1)}-W^2(x)]}~dx
=(n+1/2)\hbar\pi,~~n=0,1,2,...
\ee
As before, it is easy to obtain the quantization condition which includes
higher orders in $\hbar$ \cite{Dutt93b} and to test how well the lowest
order broken SWKB condition works for various specific examples. As for the
case of unbroken SUSY, it is found that exact spectra are obtained for
shape invariant potentials with broken SUSY \cite{Dutt93a}. For
potentials which are not analytically solvable, the results using
eq. (\ref{bswkb}) are usually better than standard WKB computations.
Further discussion can be found in ref. \cite{Dutt93b}.

\section{Isospectral Hamiltonians}
\label{sec7}

In this section, we will describe how one can start from any given
one-dimensional potential $V_1(x)$ with $n$ bound states, and use
supersymmetric quantum mechanics to construct an $n$-parameter
family of strictly isospectral potentials $V_1(\lambda_1, \lambda_2,
\ldots, \lambda_n; x)$ i.e., potentials with eigenvalues, reflection and
transmission coefficients identical to those for $V_1(x)$.  The fact that
such families exist has been known for a long time from the inverse
scattering approach  \cite{Chadan77},
but the Gelfand-Levitan approach to finding them is
technically much more complicated than the supersymmetry approach described
here.  Indeed, the advent of SUSY QM has
produced a revival of interest in the determination of isospectral
potentials \cite{Abraham80,Nieto84,Sukumar85a,Chaturvedi86,Baye87,Amado88}
\cite{Pursey86,Khare89a,Keung89,Wang90}. In Sec.~\ref{sec7.1} we
describe how a one parameter isospectral family is obtained by first deleting
and then re-inserting the ground state of $V_1(x)$
using the Darboux procedure \cite{Darboux82,Khare89a}.
The generalization to obtain an $n$-parameter family is described in
Sec.~\ref{sec7.2} \cite{Keung89}. When applied to
a reflectionless potential (Sec. \ref{sec7.3}), the $n$-parameter
families provide surprisingly simple expressions for the
pure multi-soliton solutions \cite{Wang90} of the
Korteweg-de Vries (KdV) and other nonlinear evolution
equations \cite{Gardner74,Eckhaus81,Lamb80,Chern79,Calogero82}.

\subsection{One Parameter Family of Isospectral Potentials}
\label{sec7.1}

In this subsection, we describe two approaches of obtaining the
one-parameter family $V_1(\lambda_1; x)$ of potentials isospectral to a given
potential $V_1(x)$.

The first approach follows from asking the following question:  Suppose
$V_2(x)$ is the SUSY partner potential of the original potential $V_1(x)$,
and let $W(x)$ be the superpotential such that $V_2(x) = W^2 + W^\prime$
and $V_1(x) = W^2 - W^\prime$.  Then, given $V_2(x)$, is the original
potential $V_1(x)$ unique i.e., for a given $V_2(x)$, what are the various
possible superpotentials $\hat{W}(x)$ and corresponding potentials
$\hat{V}_1(x) = \hat{W}^2 - \hat{W}^\prime$? Let us
assume \cite{Nieto84} that there exists a more general
superpotential which satisfies

\be \label{5.1}
V_2(x) = \hat{W}^2(x) + \hat{W}^\prime(x).
\ee
Clearly, $\hat{W} = W$ is one of the solutions to eq. (\ref{5.1}).  To find the
most general solution, let us set

\be
\hat{W}(x) = W(x) + \phi(x)
\ee
in eq. (\ref{5.1}).  We find then that $y(x) = \phi^{-1}(x)$ satisfies the
Bernoulli equation

\be
y^\prime(x) = 1 + 2Wy
\ee
whose solution is

\be
\frac{1}{y(x)} = \phi(x) = \frac{d}{dx}~\ln [{\cal I}_1(x) + \lambda_1].
\ee
Here
\be \label{5.5}
{\cal I}_1(x) \equiv \int_{-\infty}^{x} \psi_1^2 (x^\prime) dx^\prime.
\ee
$\lambda_1$ is a constant of integration and $\psi_1(x)$ is the
normalized ground state wave function of $V_1(x) = W^2(x) - W^\prime(x)$.
Thus the most general $\hat{W}(x)$ satisfying eq. (\ref{5.1}) is given by

\be
\hat{W}(x) = W(x) + \frac{d}{dx} \ln [{\cal I}_1(x) + \lambda_1],
\ee
so that the one parameter family of potentials

\be
\hat{V}_1(x) = \hat{W}^2(x) - \hat{W}^\prime(x) = V_1(x) - 2 \frac{d^2}
{dx^2} \ln [{\cal I}_1(x) + \lambda_1]
\ee
have the same SUSY partner $V_2(x)$.

In the second approach, we delete the ground state $\psi_1$ at energy $E_1$
for the potential $V_1(x)$.  This generates the SUSY partner potential $V_2
(x) = V_1 - 2 \frac{d^2}{dx^2} \ln \psi_1$, which is almost isospectral to
$V_1(x)$ i.e., it has the same eigenvalues as $V_1(x)$ except for the bound
state at energy $E_1$.  The next step is to reinstate a bound state at
energy $E_1$.

Although the potential $V_2$ does not have an eigenenergy $E_1$, the
function $1/\psi_1$ satifies the Schr\"{o}dinger equation with potential
$V_2$ and energy $E_1$.  The other linearly independent solution is
$\int_{-\infty}^{x} \psi_1^2(x^\prime) dx^\prime /\psi_1$.  Therefore, the
most general solution of the Schr\"{o}dinger equation for the potential
$V_2$ at energy $E_1$ is

\be
\Phi_1(\lambda_1) = ({\cal I}_1 + \lambda_1)/\psi_1~.
\ee

Now, starting with a potential $V_2$, we can again use the standard
SUSY (Darboux)
procedure to add a state at $E_1$ by using the general solution $\Phi_1
(\lambda_1)$,

\be
\hat{V}_1(\lambda_1) = V_2 - 2 \frac{d^2}{dx^2} \ln\Phi_1(\lambda_1)~~.
\ee
The function $1/\Phi_1(\lambda_1)$ is the normalizable ground state wave
function of $\hat{V}_1(\lambda_1)$, provided that $\lambda_1$ does not lie
in the interval $-1 \leq \lambda_1 \leq 0$.  Therefore, we find a
one-parameter family of potentials $\hat{V}_1(\lambda_1)$ isospectral to $V_1$
for $\lambda_1 > 0$ or $\lambda_1 < -1$.

\begin{eqnarray} \label{vhatt}
\hat{V}_1(\lambda_1) & = & V_1 - 2 \frac{d^2}{dx^2}~\ln
                           (\psi_1 \Phi_1(\lambda_1)) \nonumber \\
                     & = & V_1 -2 \frac{d^2}{dx^2} \ln ({\cal I}_1 +
                           \lambda_1)~~.
\end{eqnarray}
The corresponding ground state wave functions are

\be \label{psihat}
\hat{\psi}_1(\lambda_1;x)=1/\Phi_1(\lambda_1)~~.
\ee
Note that this family contains the original potential $V_1$.  This
corresponds to the choices $\lambda_1 \rightarrow \pm \infty$.

To elucidate this discussion, it may be worthwhile to explicitly construct
the one-parameter family of strictly isospectral potentials corresponding
to the one dimensional harmonic oscillator \cite{Khare89a}. In this case
\be
W(x) = \frac{\omega}{2}x
\ee
so that

\be
V_1(x) = \frac{\omega^2}{4} x^2 - \frac{\omega}{2}~~.
\ee
The normalized ground state eigenfunction of $V_1(x)$ is

\be
\psi_1(x) = \left(\frac{\omega}{2\pi}\right)^{1/4}\exp(-\omega{x^{2}}/4)
\ee
Using eq. (\ref{5.5}) it is now straightforward to compute the corresponding
${\cal I}_1(x)$.  We get

\be
{\cal I}_1(x)=1-\frac{1}{2}~{\rm erfc} \left(\frac{\sqrt{\omega}}{2}
x\right)~;~
{\rm erfc}(x) = \frac{2}{\sqrt{\pi}}~\int_x^\infty~e^{-t^2} dt~.
\ee

Using eqs. (\ref{vhatt}) and (\ref{psihat}),
one obtains the one parameter family of
isospectral potentials and the corresponding wave functions.  In Figs. 7.1
and 7.2, we have plotted some of the potentials and
wave functions for the case $\omega = 2$.
We see that as $\lambda_1$ decreases from $\infty$ to 0,
$\hat{V}_1$ starts developing a minimum which shifts towards $x = -\infty$.
Note that as $\lambda_1$ finally becomes zero this attractive
potential well is lost
and we lose a bound state. The remaining potential is called the Pursey
potential $V_P(x)$ \cite{Pursey86}. The general formula for $V_P(x)$
is obtained by
putting $\lambda_1=0$ in eq. (\ref{vhatt}). An analogous situation occurs in
the limit $\lambda_1=-1$, the remaining potential being the Abraham-Moses
potential \cite{Abraham80}.

\subsection{Generalization to $n$-Parameter Isospectral Family}
\label{sec7.2}

The second approach discussed in the previous subsection can be generalized
by first deleting all $n$ bound states of the original potential $V_1(x)$
and then reinstating them one at a time.  Since one parameter is generated
every time an eigenstate is reinstated, the final result is a
$n$-parameter isospectral family \cite{Keung89}.
Recall that deleting the eigenenergy
$E_1$ gave the potential $V_2(x)$.  The ground state $\psi_2$ for the potential
$V_2$ is located at energy $E_2$.  The procedure can be repeated ``upward",
producing potentials $V_3, V_4,\ldots$ with ground states $\psi_3, \psi_4,
\ldots$ at energies $E_3, E_4,\ldots,$ until the top
potential $V_{n+1}(x)$ holds
no bound state (see Fig. 7.3, which corresponds to $n = 2$).

In order to produce a two-parameter family of isospectral potentials, we go
from $V_1$ to $V_2$ to $V_3$ by successively deleting the two lowest states
of $V_1$ and then we re-add the two states at $E_2$ and $E_1$ by SUSY
transformations.  The most general solutions of the Schr\"{o}dinger
equation for the potential $V_3$ are given by $\Phi_2(\lambda_2) = ({\cal
I}_2 + \lambda_2)/\psi_2$ at energy $E_2$, and $A_2\Phi_1(\lambda_1)$ at
energy $E_1$ (see Fig. 7.3). The quantities ${\cal I}_i$ are defined by

\be
{\cal I}_i(x)~\equiv~\int_{-\infty}^{x}~\psi_i^2(x^\prime) dx^\prime~~.
\ee
Here the SUSY operator $A_i$ relates solutions of the potentials
$V_i$ and $V_{i+1}$,

\be
A_i = \frac{d}{dx} - (\ln\psi_i)^\prime~.
\ee
Then, as before, we
find an isospectral one-parameter family $\hat{V}_2(\lambda_2)$,

\be
\hat{V}_2(\lambda_2) = V_2 - 2\frac{d^2}{dx^2}\ln({\cal I}_2 + \lambda_2)~.
\ee
The solutions of the Schr\"{o}dinger equation for potentials $V_3$ and
$\hat{V}_2(\lambda_2)$ are related by a new SUSY operator

\be
\hat{A}_2^{\dag} (\lambda_2) = -\frac{d}{dx} + (\ln\Phi_2(\lambda_2))^\prime~.
\ee
Therefore, the solution $\Phi_1(\lambda_1, \lambda_2)$ at $E_1$ for
$\hat{V}_2(\lambda_2)$ is

\be
\Phi_1(\lambda_1,\lambda_2) = \hat{A}_2^{\dag}
                              (\lambda_2)A_2\Phi_1(\lambda_1)~.
\ee
The normalizable function $1/\Phi_1(\lambda_1, \lambda_2)$ is the ground
state at $E_1$ of a new potential, which results in a two-parameter family
of isospectral systems $\hat{V}_1(\lambda_1, \lambda_2)$,

\begin{eqnarray}
\hat{V}_1(\lambda_1, \lambda_2)
& = & V_1 - 2 \frac{d^2}{dx^2} \ln(\psi_1 \psi_2 \Phi_2 (\lambda_2)
      \Phi_1(\lambda_1, \lambda_2)) \nonumber \\
& = & V_1 - 2 \frac{d^2}{dx^2} \ln(\psi_1({\cal I}_2 + \lambda_2)\Phi_1(
      \lambda_1, \lambda_2))~,
\end{eqnarray}
for $\lambda_i > 0$ or $\lambda_i < -1$.  A useful alternative expression is

\be
\hat{V}_1(\lambda_1, \lambda_2) = -\hat{V}_2(\lambda_2) + 2(\Phi_1^\prime
(\lambda_1, \lambda_2)/\Phi_1(\lambda_1, \lambda_2))^2 + 2E_1~.
\ee
The above procedure is best illustrated by the pyramid structure in Fig. 7.3.
It can be generalized to an $n$-parameter family of isospectral
potentials for an initial system with $n$  bound states.  The formulas for
an $n$-parameter family are

\be
\Phi_i(\lambda_i) = ({\cal I}_i + \lambda_i)/\psi_i~,~~i = 1, \cdots, n~;
\ee

\be
A_i = \frac{d}{dx} - (\ln\psi_i)^\prime~;
\ee

\be
\hat{A}_i^{\dag} (\lambda_i, \cdots, \lambda_{n}) = - \frac{d}{dx} + (\ln
\Phi_i(\lambda_i, \cdots, \lambda_{n}))^\prime~;
\ee

\begin{eqnarray}
\small
& &
\Phi_i(\lambda_i, \lambda_{i+1}, \cdots, \lambda_{n}) \nonumber\\
& = & \hat{A}_{i+1}^{\dag} (\lambda_{i+1}, \lambda_{i+2}, \cdots,
      \lambda_{n}) \hat{A}_{i+2}^{\dag} (\lambda_{i+2}, \lambda_{i+3},
   \cdots, \lambda_{n}) \cdots\hat{A}_{n}^{\dag} (\lambda_{n})\nonumber\\
& \times & A_{n}A_{n-1} \cdots A_{i+1}\Phi_i(\lambda_i)~;
\normalsize
\end{eqnarray}

\be
\hat{V}_1(\lambda_1, \cdots, \lambda_{n}) = V_1
- 2 \frac{d^2}{dx^2} \ln(
\psi_1\psi_2 \cdots \psi_{n}\Phi_{n}
(\lambda_{n}) \cdots \Phi_1
(\lambda_1, \cdots, \lambda_{n}))~.
\ee
The above equations \cite{Keung89} summarize the main results of this section.

\subsection{$n$-Soliton Solutions of the KdV Equation}
\label{sec7.3}

As an application of isospectral potential families, we consider reflectionless
potentials of the form

\be
V_1 = -n(n + 1)\sech^2x~,
\ee
where $n$ is an integer, since these potentials are of special physical
interest.  $V_1$ holds $n$ bound states, and we may form a
$n$-parameter family of isospectral potentials.  We start with the simplest
case $n = 1$.  We have $V_1 = -2\sech^2x, E_1 = -1$ and $\psi_1 =
2^{-\frac{1}{2}}\sech x$.  The corresponding 1-parameter family is

\be
\hat{V}_1(\lambda_1) = -2\sech^2 (x + \frac{1}{2}\ln(1 + \frac{1}{\lambda_1}))
{}~.
\ee
Clearly, varying the parameter $\lambda_1$ corresponds to translations of
$V_1(x)$.  As $\lambda_1$ approaches the limits $0^{+}$ (Pursey limit) and
$-1^-$ (Abraham-Moses limit), the minimum of the potential moves to
$-\infty$ and $+\infty$ respectively.

For the case $n = 2$, $V_1 = -6 \sech^2x$ and there are two bound states at
$E_1 = -4$ and $E_2 = -1$.  The SUSY partner potential is $V_2 =
-2\sech^2x$.  The ground state wave functions of $V_1$ and $V_2$ are
$\psi_1 = \frac{\sqrt 3}{2} \sech^2x$ and $\psi_2 = \frac{1}{\sqrt 2}
\sech x$.  Also, ${\cal I}_1 = \frac{1}{4}(3\tanh x - \tanh^3x + 2)$ and
${\cal I}_2 = \frac{1}{2}(\tanh x+1)$.  After some algebraic work, we
obtain the 2-parameter family

\begin{eqnarray*}
\hat{V}_1(\lambda_1, \lambda_2)
{}~ = ~ -12 \frac{[3+4\cosh(2x-2\delta_2)+
      \cosh(4x-2\delta_1)]} {[\cosh(3x - \delta_2 - \delta_1)
      +3\cosh(x+\delta_2 - \delta_1)]^2}~, \\
\delta_i
\equiv - \frac{1}{2} \ln (1 + \frac{1}{\lambda_i})~,~~i = 1,2~~.
\end{eqnarray*}
As we let $\lambda_1 \rightarrow -1$, a well with one bound state at $E_1$
will move in the $+x$ direction leaving behind a shallow well with one
bound state at $E_2$.  The movement of the shallow well is essentially
controlled by the parameter $\lambda_2$.  Thus, we have the freedom to move
either one of the wells.

It is tedious but straightforward to obtain the result for arbitrary $n$
and get $\hat{V}_1 (\lambda_1, \lambda_2, \cdots, \lambda_{n}, x)$.  It
is well known that one-parameter $(t)$ families of isospectral potentials
can also be obtained as solutions of a certain class of nonlinear evolution
equations \cite{Gardner74,Eckhaus81,Lamb80,Calogero82}.
These equations have the form $(q = 0,1,2,\cdots)$

\be\label{e1}
- u_t = (L_u)^q ~u_x
\ee
where the operator $L_u$ is defined by

\be
L_u f(x) = f_{xx} - 4uf + 2u_x \int_x^\infty dy f(y)
\ee
and $u$ is chosen to vanish at infinity.  [For $q=0$ we simply get $-u_t =
u_x$, while for $q=1$ we obtain the well studied Korteweg-de Vries (KdV)
equation].  These equations are also known to possess pure (i.e.,
reflectionless) multisoliton solutions.  It is possible to show that by
suitably choosing the parameters $\lambda_i$ as functions of $t$ in the
$n$-parameter SUSY isospectral family of a symmetric reflectionless
potential holding $n$ bound states, we can obtain an explicit analytic
formula for the pure $n$-soliton solution of each of the above evolution
equations \cite{Wang90}.
These expressions for the multisoliton solutions of eq. (\ref{e1})
are much simpler than any previously obtained using other procedures.
Nevertheless, rather than displaying the explicit algebraic expressions
here, we shall simply illustrate the 3-soliton solution
of the KdV equation.  The potentials
shown in Fig. 7.4 are all isospectral and reflectionless
holding bound states at $E_1$ = -25/16, $E_2 = -1, E_3$ = -16/25 .  As $t$
increases, note the clear emergence of three independent solitons.

In this section, we have found $n$-parameter isospectral families by
repeatedly using the supersymmetric Darboux procedure for removing and
inserting bound states. However, as briefly mentioned in Sec. 7.1,
there are two other closely related,
well established procedures for deleting and adding bound states. These are
the Abraham-Moses procedure \cite{Abraham80} and the Pursey procedure
\cite{Pursey86}.
If these alternative procedures are used, one gets new potential
families all having the same bound state energies but different reflection
and transmission coefficients. Details can be found in reference
\cite{Khare89b}.

\section {Path Integrals and Supersymmetry}
\label{sec8}

In this section, we will describe the Lagrangian formulation of SUSY QM and
discuss three related path integrals: one for the generating functional of
correlation functions, one for the Witten index - a topological
quantity which determines whether SUSY is broken, and one for a related
``classical"  stochastic differential equation, namely the Langevin equation.
We will also briefly discuss the superspace formalism for SUSY QM.

Starting from the matrix SUSY Hamiltonian which is  $1/2$ of our previous
$H $ [eq. (\ref{susyh2})] for convenience:
\[
H= { 1 \over 2} p^2+ {1 \over 2} W^2(x)  I  - { 1 \over 2}
[\psi,\psi^{\dag}]  W^{
\prime}(x),
\]
we obtain the Lagrangian
\be
L= {1 \over 2} \dot{x}^2 + i \psi^{\dag}\partial_t \psi -
{1 \over 2} W^2(x) +{ 1 \over 2} [\psi,\psi^{\dag}]  W^{\prime}(x).
\ee
It is most useful to consider the generating functional of correlation
functions in Euclidean space.  We rotate $t \rightarrow i \tau$
and obtain for the Euclidean path integral:
\be
Z[j,\eta,\eta^*] = \int [dx] [d\psi] [d \psi^*]\exp[- S_E +\int jx+\eta
\psi^*+\eta^* \psi],
\ee
where
\[
S_E = \int_0 ^{\tau} d\tau \left({1 \over 2}  x_{\tau} ^2+{1 \over 2} W^2(x)
- \psi^{\ast}[\partial_{\tau}  -
W^{\prime}(x)]\psi \right) \]
and $\psi$ and $\psi^{\ast}$ are now elements of a Grassman algebra:
\be
\{ \psi^{\ast},\psi \} = \{ \psi,\psi \} =\{ \psi^{\ast},\psi^{\ast} \} =0,
\ee
and
\[  x_{\tau} = { dx \over d \tau} \].
The Euclidean action is invariant under the following SUSY
transformations \cite{Sourlas85,Parisi82}  which mix bosonic and fermionic
degrees of freedom:

\[ \delta x = \epsilon^{\ast}  \psi + \psi^{\ast} \epsilon, \]
\[ \delta \psi^{\ast} = - \epsilon^{\ast} \left( \partial_{\tau} x
+ W(x) \right), \]
\be
 \delta \psi = - \epsilon \left( -\partial_{\tau} x + W(x) \right),
\ee
where $\epsilon$ and $\epsilon^{\ast}$ are two infinitesimal anticommuting
parameters. These transformations correspond to $N=2$ supersymmetry.

The path integral over the fermions can now be  explicitly performed
using a cutoff lattice which is periodic in the the coordinate $x$ but
antiperiodic in the fermionic degrees of freedom at $ \tau = 0$ and
$\tau =T$.  Namely we evaluate the fermionic path integral:
\be
\int [d\psi] [d \psi^*] e^{ \int_0 ^T  \, d \tau \psi^{\ast}[\partial_{\tau}  -
W^{\prime}(x)]\psi}
 \ee
by calculating the determinant of the operator
$[\partial_{\tau}  - W^{\prime}(x)]$  using eigenvectors which are
antiperiodic.

We have, following Gildener and Patrascioiu \cite{Gildener77},
that
\[
{\rm{det}} [\partial_{\tau}  - W^{\prime}(x)] = \prod_m \lambda_m~,
\]
where
\[
[\partial_{\tau}  - W^{\prime}(x)] \psi_m = \lambda_m \psi_m~,
\]
so that
\be
\psi_m (\tau) = C_m {\rm{exp}} [ \int_0 ^{\tau} d \tau^{\prime}[\lambda_m +
W'].
\ee
Imposing the antiperiodic boundary conditions:
\[
\psi_m(T) = - \psi_m (0)
 \]
yields:
\be
\lambda_m = { i (2m+1) \pi \over T} - {1 \over T} \int_0 ^T d \tau W'(x).
\ee
Regulating the determinant by dividing by the determinant for the case where
the potential is zero we obtain:
\be
{\rm{det}}[{\partial_{\tau}  - W^{\prime}(x) \over \partial_{\tau} }] =
{\rm{cosh}} \int_0 ^ T d \tau {W'(x) \over 2}.
\ee
Rewriting the $cosh$ as a sum of two exponentials we find, as expected that
$Z$ is the sum of the partition functions for the two pieces of the
supersymmetric Hamiltonian.  Namely when the external sources
are zero:
\be {\rm{Tr}} e^{-H_1T}+{\rm{Tr}} e^{-H_2 T} \equiv Z_{-} + Z_{+}.
\ee

For the case when SUSY is unbroken, only the ground state
of $H_1$ contributes as $ T \rightarrow \infty $
We also have:
\be
Z_{\pm} = \int [dx] {\rm{exp}} [-S_E^{\pm}]
\ee
where
\[
S_E^{\pm} = \int_0 ^T d\tau\left( {\dot{x}^2 \over 2 } + {W^2(x) \over 2} \pm
{W'(x) \over 2} \right).
\]

A related path integral is obtained for the noise averaged correlation
functions coming from a classical stochastic equation, the Langevin equation.
If we have the stochastic differential equation
\be
\dot{x} = W(x(\tau)) + \eta (\tau)
\ee
where $ \eta(\tau)$ is a random stirring forcing obeying
Gaussian statistics, then the correlation functions
of $x$ are exactly the same as the correlation functions
obtained from the Euclidean quantum mechanics related to the
Hamiltonian $H_1$.  To see this we realize that Gaussian noise is described
by a probability functional:
\be
P[\eta] = N \rm{exp}  [ - {1 \over 2 } \int_0 ^ T d \tau  {\eta^2(\tau)
\over F_0 }]
\ee
normalized so that:
\[
\int D \eta P[\eta] = 1, \]
\[
\int D \eta P[\eta] f(\tau)  = 0, \]
\[
\int D \eta P[\eta] f(\tau) f(\tau^{\prime})  =
F_0 \delta(\tau- \tau^{\prime}). \]

The correlation functions averaged over the noise are:

\be
< x(\tau_1) x(\tau_2)... > =  \int D \eta P[\eta] x (\tau_1) x(\tau_2)...
\ee
where we have in mind first solving the Langevin equation
explicitly for $x(\eta(\tau))$ and then averaging over the noise as discussed
in Sec. 2.2.    Another way to calculate the correlation function is
to change variables in the functional integral from $\eta$ to
$x$.
\be
< x(\tau_1) x(\tau_2)... > =  \int D[x]~ P[\eta]~ Det \mid { d \eta(\tau)
 \over dx (\tau^{\prime}) } \mid~ x (\tau_1) x(\tau_2)...
\ee
This involves calculating the functional determinant,
\be
Det |{ d \eta(\tau)  \over dx (\tau^{\prime}) } |
\ee
subject to the boundary condition that the Green's function
obey causality, so one has retarded boundary conditions.
One has
\be
Det |{ d \eta  \over dx }| =  {\rm exp} \int dt {\rm ~Tr ~ln} \left
([{d \over d \tau} - W^{\prime}(x(\tau))
] \delta(\tau-\tau^{\prime}) \right) d\tau. \ee
When there are no interactions the retarded boundary conditions yield
\be
 G_0 (\tau - \tau^{\prime}) = \theta(\tau - \tau^{\prime}).
\ee
Expanding $ {\rm{ln}} (1- G_ 0 W')$ one finds  because of the retarded
boundary conditions that only the first term in the expansion contributes so
that
\be Det |{ d \eta(\tau)  \over dx (\tau^{\prime}) } | = {\rm exp}
[ {1 \over 2} \int_0 ^ T d\tau
W'(x)]. \ee
Choosing $F_0 = \hbar$ so that
\bea
P[\eta] &=& N \exp [ - {1 \over 2 } \int_0 ^ T d \tau
{\eta^2(\tau) \over F_0}] \nonumber \\
 &=& N \exp  [ - {1 \over 2 \hbar} \int_0 ^ T d \tau  (\dot{x}
-W(x))^2 ]\nonumber \\
 &=& N \exp  [ - {1 \over 2 \hbar} \int_0 ^ T d \tau  (\dot{x}^2 +W^2(x))],
\eea
we find that the generating functional for the correlation functions is
exactly the generating functional for the correlation functions for  Euclidean
quantum mechanics corresponding to the Hamiltonian $H_1$:
\be
Z[j] =  N \int D[x] ~{\rm exp} [ - {1 \over 2 } \int_0 ^ T d \tau
\left(\dot{x}^2 +W^2(x)- W^{\prime}(x) -2 j(\tau) x (\tau)\right)]
\ee
Thus we see that we can determine the correlation functions of $x$ for the
Hamiltonian $H_1$ by either evaluating the path integral or solving
the Langevin equation and averaging over Gaussian noise.

An equation related to the Langevin equation is the Fokker-Planck
equation,  which defines the classical probability function $P_c$ for the equal
time  correlation functions of $H_1$. Defining the noise average:
\be
P_c(z) = < \delta(z-x(t) >_{\eta} = \int D \eta \delta (z-x(t)) P[\eta]
\ee
one obviously has:
\[
\int dz z^n P_c(z,t) = \int D \eta [x(t)]^n P[\eta] = < x^n>.
\]
One can  show \cite{Ma75} that $P_c$ obeys the Fokker-Planck equation:
\be
{\partial P \over \partial t} = { 1 \over 2} F_0 {\partial^2 P \over
\partial z^2} +
{\partial \over \partial z} [W(z) P(z,t))]
\ee
For an equilibrium distribution to exist at long times $t$ one requires that
\[
P(z,t) \rightarrow \hat{P}(z) \]
and
\[ \int \hat{P} (z) dz =1. \]

Setting $ {\partial P \over \partial t} =0$ in the Fokker-Planck equation,
we obtain
\be
\hat{P}(z) = N e^{-2 \int_0 ^z W(y) dy} = \psi_0(z)^2 ~.
\ee
Thus at long times only the ground state wave function contributes (we are
in Euclidean space) and the probability function is just the usual ground
state wave function squared. We see from this that when SUSY is broken, one
cannot define an equilibrium distribution for the classical stochastic system.

A third path integral for SUSY QM is related to the
Witten index. As we discussed before, one can introduce a ``fermion'' number
operator via
\be
   n_F =  {1-\sigma_3  \over 2} = { 1- [\psi,\psi^{\dag}] \over 2}.
\ee
Thus
\be
(-1)^F = [\psi,\psi^{\dag}] =\sigma_3.
\ee

The Witten index is given by $\triangle = \rm{Tr} (-1)^F$. As we discussed
earlier, the Witten index needs to be regulated and the regulated index is
defined as: \be
\triangle(\beta)  = {\rm{Tr}} (-1)^F e^{-\beta H} = {\rm{Tr}}(e^{-\beta H_1 }-
e^{-\beta H_2})~.
\ee
In Sec 2.2 we showed how to determine $\triangle(\beta)$ using heat
kernel methods and how it was useful in discussing non-perturbative breaking of
SUSY. Here we will show that the Witten index can also be obtained using
the path integral representation of the generating functional of
SUSY QM where the fermion determinant is now evaluated
using periodic boundary conditions to incorporate the factor $(-1)^F$. It is
easy to verify a posteriori that this is the case. Consider the path integral:
\be
\triangle(\beta) = \int [dx] [d\Psi] [d \Psi^*] e^{\int_0 ^{\beta}
L_E(x,\Psi,\Psi^*) d \tau},
\ee
where
\[
L_E =  {1 \over 2}  x_{\tau} ^2 + {1 \over 2} W^2 -
\Psi^{\ast}[\partial_{\tau}  -
W^{\prime}(x)]\Psi. \]
To incorporate the $(-1)^F$ in the trace, one changes the boundary conditions
for evaluating the fermion determinant
 at $ \tau = 0, \beta$ to periodic ones:
\[ x(0)= x(\beta) \hspace{.5in} \Psi(0) = \Psi(\beta) \]
The path integral over the fermions can again be  explicitly performed
using a cutoff lattice which is periodic in the fermionic degrees of freedom at
$ \tau = 0$ and $\tau =\beta$.
We now impose these boundary conditions on
the determinant of the operator $[\partial_{\tau}  - W^{\prime}(x)]$
using eigenvectors which are periodic.

We again have
\[
{\rm{det}} [\partial_{\tau}  - W^{\prime}(x)] = \prod_m \lambda_m.
\]
Imposing periodic boundary conditions:
\be
\lambda_m = { i (2m) \pi \over \beta} - {1 \over \beta} \int_0 ^{\beta}
d \tau W'(x).
\ee
Regulating the determinant by dividing by the determinant for the case where
the potential is zero we obtain:
\be
det\left[{\partial_{\tau}  - W^{\prime}(x) \over \partial_{\tau} }\right] =
\rm{sinh} \int_0 ^ {\beta} d \tau {W'(x) \over 2}~.
\ee
Again rewriting the $sinh$ as a sum of two exponentials we find, as expected
that we obtain the regulated Witten index:
\be
\triangle(\beta) = Z_{-} - Z_{+} =\rm{Tr} e^{- \beta H_1}- \rm{Tr} e^{- \beta
H_2 }.
\ee

\subsection{Superspace Formulation of Supersymmetric Quantum Mechanics}
\label{sec8.1}

 One can think of SUSY QM as a degenerate case of supersymmetric
field theory in $d=1$ in the superspace formalism of Salam and Strathdee
\cite{Salam75} (This idea
is originally found in  unpublished lecture notes of S. Raby \cite{Raby81}).
superfields are defined on the space $(x_n; \theta_a)$ where $x$ is the space
coordinate  and $\theta_a$ are anticommuting spinors.  In the degenerate case
$d=1$ the  field is replaced by $x(t)$ so that the only coordinate is time. The
anticommuting variables are $ \theta$ and $ \theta^{\ast}$ where
\[
\{ \theta, \theta^{\ast} \} = \{ \theta,\theta \} = [\theta,t] = 0.
\]

Consider the following SUSY transformation:
\be
t' = t - i ( \theta^{\ast} \epsilon - \epsilon^{\ast} \theta),~~
\theta^{\prime} = \theta + \epsilon,~~ \theta^{\ast \prime}
= \theta^{\ast} +
\epsilon^{\ast}.
\ee
If we assume that finite SUSY transformations can be parametrized by
\[ L = e^{i(\epsilon^{\ast} Q^{\ast} + Q \epsilon)}~, \]
then from
\be
\delta A = i [\epsilon^{\ast} Q^{\ast} + Q \epsilon, A]
\ee
we infer that the operators $Q$ and $Q^{\ast}$ are given by:
\be
 Q= i \partial_{\theta} - \theta^{\ast} \partial_t ,~~
 Q^{\ast} = -i \partial_{\theta^{\ast}} - \theta \partial_t~.
\ee
Now these charges obey the familiar SUSY QM algebra:
\be
\{ Q, Q^{\ast} \} = 2 i \partial_t = 2H , \hspace{.2in} [Q,H]=0.
\ee
The Lagrangian in superspace is determined as follows.  A superfield made
up of $x$ and $\theta$
and $\theta^{\ast}$ can at most be a bilinear in the Grassman variables:
\be
\phi(x, \theta, \theta^{\ast} ) = x(t) + i \theta \psi(t) - i\psi^{\ast}
\theta^{\ast} +
\theta \theta^{\ast} D(t).
\ee
Under a SUSY transformation, the following derivatives are invariant:
\[D_{\theta} = \partial_{\theta} - i \theta^{\ast} \partial_t \]
or in component form:
\be
D_{\theta} \phi = i \psi - \theta^{\ast} D - i \theta^{\ast} \dot{x}
+\theta^{\ast} \theta
\dot{\psi},
\ee
and
\[D_{\theta^{\ast}} = \partial_{\theta^{\ast}} - i \theta \partial_t \]
or in component form:
\be
[D_{\theta} \phi]^{\ast} = -i \psi^{\ast} - \theta D + i \theta \dot{x}
+\theta^{\ast}
\theta \dot{\psi}^{\ast}.
\ee
The most general invariant action is:
\be
S = \int dt d \theta^{\ast} d \theta \left({1 \over 2} | D_{\theta} \phi |^2
- f(\phi)\right)
\ee
Again the expansion in terms of the Grassman variables causes a Taylor
expansion of $f$ to truncate at the second derivative level. Integrating over
the Grassman degrees of
freedom using the usual path integral rules for Grassman variables:
\[
\int \theta d \theta = \int \theta^{\ast} d \theta^{\ast} = 1,~~~
\int d \theta = \int d \theta ^{\ast} = 0,
\]
one obtains
\be
S= \int dt( {1 \over 2} \dot{x} ^2+ \psi^{\ast}[\partial_{t}  -
f^{\prime \prime} (x)]\psi + {1 \over 2} D^2 + D f^{\prime} (x) ).
\ee
Eliminating the constraint variable $ D= - f'(x) = W(x)$ we obtain our previous
result for the  action (now in Minkowski space):
\be
S= \int dt \left( {1 \over 2} \dot{x} ^2+ \psi^{\ast}[\partial_{t}  -
W'(x)]\psi - {1 \over 2} W^2 \right)
\ee
A more complete discussion of this can be found in ref. \cite{Cooper83}.

\section{Perturbative Methods for Calculating Energy Spectra and Wave
Functions}
\label{sec9}

The framework of supersymmetric quantum mechanics has been very useful in
generating several new perturbative methods for calculating the energy spectra
and wave functions for one dimensional potentials. Four such methods are
described in this section.

In Secs. 9.1 and 9.2, we discuss two approximation methods (the variational
method and the $\delta-$ expansion) for determining the wave functions and
energ
 y
eigenvalues of the anharmonic oscillator making use of SUSY QM. Sec. 9.3
contains a description of a SUSY QM calculation of the energy splitting and
rate of tunneling in a double well potential. The result is a rapidly
converging series which is substantially better than the usual WKB tunneling
formula. Finally, in Sec. 9.4, we describe how the large $N$ expansion ($N$ =
number of spatial dimensions) used in quantum mechanics can be further
improved by incorporating SUSY.

\subsection {Variational Approach}
\label{sec9.1}

The anharmonic oscillator potential $V(x) = g  x^4$ is not exactly
solvable. To determine the superpotential one has to first subtract
the ground state energy $E_0$ and solve the Riccati equation for W(x):
\be
V_{1}(x) =g  x^4 - E_0
\equiv W ^2- W^{\prime},
\ee
Once the ground state energy and the superpotential is known to some
order of accuracy, one can then determine the partner potential and its ground
state wave function approximately. Then, using the SUSY operator
\[  {d \over dx}- W(x) \]
one can construct the first excited state of the anharmonic oscillator in the
usual manner. Using the hierarchy of Hamiltonians discussed in Sec. 3, one can
construct from the approximate ground state wave functions of the hierarchy and
the approximate superpotentials $W_n$ all the excited states of the anharmonic
oscillator approximately.

First let us see how this works using a simple variational approach.
For the original potential, we can determine the optimal
Gaussian wave function quite easily.  Assuming a trial wave function of the
form
\be
\psi_0 =({2 \beta \over \pi})^{1/4}   e^{- \beta x^2}, \label{eqpsi0}
\ee
we obtain
\be
< H > = < {p^2 \over 2} + g x^4 > = {\beta \over 2} + {3g \over 16 \beta^2}.
\ee
(In this subsection, we are taking $m=1$ in order to make contact with
published numerical results). Minimizing the expectation value of the
Hamiltonia
 n
with respect to the parameter $ \beta$ yields
\[
 E_0 = ({3 \over 4})^{4/3} g^{1/3} ~,  \hspace{.5in}   \beta = ({3 \over
4})^{1/3} g^{1/3} .  \]
This is rather good for this crude approximation since the exact ground
state energy of the anharmonic oscillator determined numerically is
$ E_0 = .668 g^{1/3} $ whereas  $({3 \over 4})^{4/3} =.681~$.
The approximate potential $W$ resulting from this variation calculation is
\be
W(x) = -  { d log \psi_0  \over dx}= 2 \beta x,
\ee
which leads to a Gaussian approximation to the potential

\be
V_{1G} = 4  \beta x^2  - 2 \beta.
\ee
The  approximate supersymmetric partner potential is now
\be
V_{2G} = 4  \beta x^2  + 2 \beta.
\ee
Since $V_{2G}$ differs from $V_{1G}$ by a constant, the approximate ground
state wave function for $V_2$ is given by eq. (\ref{eqpsi0}).
The approximate ground state energy of the second potential is now
\be
<H_2> = _v <0| {p^2 \over 2} + V_{2G} |0>_v = H_1 + 4\beta.
\ee
Thus we have approximately that the energy difference between the ground
state and first excited state of the anharmonic oscillator is
\[
E_1 - E_0 = 4 \beta = 4 ({3 \over
4})^{1/3} g^{1/3}.
\]
The approximate (unnormalized) first excited state wave function is
\be
\psi_1 ^{(1)} = [{d \over dx} - 2 \beta x ]\psi_{0v}
 \propto - 4 \beta x  e^{ -\beta  x^2}.
\ee

This method can be used to
find all the excited state wave functions and energy levels of the anharmonic
oscillator  by using the methods discussed in ref. \cite{Gozzi93}.

Let us now look at a more general class of trial wave functions.
If we choose for the trial ground
state wave functions of the hierarchy of Hamiltonians,
\begin{equation}
  \begin{array}{cc}
     \psi_{0}^{(k)} = N_k \, \exp \Bigl[
        - \frac{1}{2} \Bigl( \frac{x^2}{\rho_k} \Bigr)^{n_k} \Bigr]
     \>, \qquad &
     N_k = \Bigl[ 2 \sqrt{\rho_k} \Gamma( 1 + \frac{1}{2 n_k} ) \Bigr]^{-1/2}
     \>.
  \end{array}
\label{eq:trialwf}
\end{equation}
we obtain much better agreement for the low lying eigenvalues and
eigenfunctions. It is convenient in this case to  first scale the Hamiltonian
for the anharmonic oscillator,
\begin{equation}
  H = - \frac{1}{2} \frac{d^2}{dx^2} + g x^4  \>,
\end{equation}
by letting $x \rightarrow x/g^{1/6}$ and $H \rightarrow g^{1/3} H$.
Then we find the ground state energy of the anharmonic oscillator and the
variational parameters $\rho_1$ and $n_1$ by forming the functional
\begin{equation}
     E_0(\rho_1,n_1) =
        \langle \, \psi_{0} | - \frac{1}{2} \frac{d^2}{dx^2} +   x^4 |
        \psi_{0} \, \rangle.
\end{equation}
Thus we first determine $\rho_1$ and $n_1$ by requiring
\begin{equation}
  \begin{array}{cc}
     {\partial E_0 \over \partial \rho_1} = 0  \>, \quad &
   {\partial E_0 \over \partial n_1} = 0  \>.
  \end{array}
\end{equation}
The equation for the energy functional for the anharmonic oscillator is
\begin{equation}
   E_0(\rho_1,n_1) =
     { n_1^2 \over \ 2 \rho_1 }
     { \Gamma( 2-{1 \over 2n_1}  ) \over \Gamma( { 1 \over 2n_1 } ) } +
     \rho_1^2
     { \Gamma( { 5 \over 2n_1 } ) \over \Gamma( { 1 \over 2n_1 } ) }.
\end{equation}
Minimizing this expression, we obtain the following variational result:
\begin{equation}
  E_0 = 0.66933, \hspace{.2in} n_1=1.18346, \hspace{.2in} \rho_1 = 0.666721~.
\end{equation}
This ground state energy is to be compared with a numerical evaluation which
yie
 lds
$0.667986~$.
Since the trial wave function for all ground states is given by eq.
(\ref{eq:trialwf}), the variational superpotential for all $k$ is
\begin{equation}
   W_{kv} = n_k |x|^{2n_k -1} (\rho_k)^{-n_k}  \>.
\end{equation}
Since we are interested in the energy differences $ E_n-E_{n-1}$  of the
anharmonic oscillator, we consider the variational Hamiltonian
\be
   \bar{H}_{vk+1} =  \frac{1}{2} A_{kv} A_{kv}^{\dagger}
\ee
which approximately determines these energy differences.
We obtain the approximate energy splittings by minimizing the energy functional
\begin{equation}
  \delta E_k (\rho_k,n_k) = \frac{1}{2}
     \langle \, \psi_{0}^{(vk+1)} | -\frac{d^2}{dx^2} +
            W^2_{vk} + W'_{vk} | \psi_{0}^{(vk+1)} \, \rangle   \>.
\end{equation}
Performing the integrals one obtains the simple recursion
relation:
\[
 \delta E_k (\rho_k,n_k) =
   \frac{ n_k^2 }{ 2 \rho_k }
   \frac{ \Gamma\left( 2-{ 1 \over 2n_k } \right) }
        { \Gamma\left( { 1 \over 2n_k } \right) } +
   \frac{ n_{k-1}^2 }{ 2 \rho_k }
   \left( \frac{ \rho_k }{ \rho_{k-1} } \right)^{ 2n_{k-1} }
   \frac{ \Gamma\left( { 4n_{k-1}-1 \over 2n_k } \right) }
        { \Gamma\left( { 1 \over 2n_k } \right) }
\]
\begin{equation}
   + \frac{ n_{k-1} }{ 2 \rho_{k} }  ( 2n_{k-1} - 1 )
     \left( \frac{ \rho_k }{ \rho_{k-1} } \right)^{ n_{k-1} }
     \frac{ \Gamma\left( \frac{ 2n_{k-1}-1 }{ 2n_k } \right) }
          { \Gamma\left( \frac{ 1 }{ 2n_k } \right) }.
\end{equation}
One can perform the minimization in $\rho$ analytically leaving one
minimization
to perform numerically.

The results for the variational parameters and for the energy
differences are presented
in Table 9.1 for the first three energy eigenvalues and compared with a
numerical  calculation, based on a shooting method.\\ \\

\subsection{$\delta$ Expansion Method}
\label{sec9.2}

In this method \cite{Cooper90} we consider the anharmonic oscillator
as an analytic continuation from the harmonic oscillator in the paramater
controlling the anharmonicity.  That is we consider simultaneously potentials
of the form
\be
V_{1}(x) = M^{2+ \delta} x^{2+2 \delta} - C(\delta) \equiv W^2(x,\delta) -
 W^{\prime} \label{eqric1},
\ee
where $M$  is a scale parameter, $\delta$ measures the anharmonicity,
and $C$ is the ground state energy of the anharmonic oscilator.
$C$ is subtracted as usual from the potential so it can be factorized.
The standard anharmonic oscillator corresponds to $\delta=1$ and
$M=(2g)^{1/3}$.
To approximately determine $W(x)$ from $V_1(x)$ we assume that
both $W(x)$ and  $V_1(x)$ have a Taylor series  expansion in $\delta$.
Thus we write:
\be
V_1(x) = M^2 x^2 \sum_{n=0} ^{\infty} { \delta^n [\rm{ln} (Mx^2)]^n \over n!}
-\sum_{n=0} ^{\infty} 2 E_n \delta^n,
\ee
where  $E_n$ corresponds to the order Taylor expansion of the dependence of
the ground state energy on the parameter $\delta$.
We assume
\be
W(x) = \sum_{n=0} ^{\infty} \delta^n W_{(n)}(x),
\ee
and insert these
expressions in eq. (\ref{eqric1}) and match terms order by order.
At lowest order in $\delta$ the problem reduces to the supersymmetric
harmonic oscillator. We have:
\be
W_0^2 - W_0 ' =  M^2 x^2-2 E_0,
\ee
whose solution is
\be
W_0(x)= Mx, \hspace{.5in} E_0 = {1 \over 2} M
\ee
To next order we have the differential equation:
\be
{dW_1 \over dx} - 2 W_1 W_0 = - M^2 x^2 {\rm ln} (Mx^2) + 2E_1
\ee
which is to be solved with the boundary condition$W_n(0)=0$. The order $\delta$
contribution to the
energy eigenvalue $E_1$ is determined by requiring that the ground state wave
function be square integrable. Solving for $W_1$ we obtain
\be
W_1(x) = - e^{Mx^2} \int_0 ^x dy e^{-My^2} [M^2y^2 {\rm ln} (My^2)- 2E_1].
\ee
To first order in $\delta$ the ground state wave function is now :
\[
\psi_0(x) = e^{-Mx^2} (1- \delta \int_0^x dy W_1(y))
\]
Imposing the condition that $\psi_0$ vanishes at infinity, we obtain:

\be
E_1 = {1 \over 4} M \psi(3/2)~, \hspace{.5in} \psi(x) =
\Gamma^{\prime}(x)/\Gamma(x). \ee
Writing $M=(2g)^{1/3}$, we find that the first two terms in the $\delta$
expansion for the ground state energy are
\be
E= {1\over 2} (2g)^{1/3}[1+ {1 \over 2} \psi(3/2) \delta].
\ee
At $\delta =1$, we get
\be
E=.6415 g^{1/3}.
\ee
A more accurate determination of the ground state energy can be obtained
by calculating up to order $\delta^2$ and then analytically continuing in
$\delta$ using Pad\'e approximants. This is discussed in ref. \cite{Bender88}.

\subsection{Supersymmetry and Double Well Potentials}
\label{sec9.3}

Supersymmetric quantum mechanics has been profitably used to obtain a novel
perturbation expansion for the probabililty of tunneling in a double well
potential \cite{Keung88}. Since double
wells are widely used in many areas of physics and
chemistry, this expansion has found many applications ranging from
condensed matter physics to the computation of chemical reaction rates
\cite{Caldeira81,Chakravarty82,Bray82,Chakravarty84,%
Schon90,Glasstone41,Miller74}
,\cite{Kampen77,Affleck81,Bernstein84,Kumar86,Marchesoni88,%
Boyanovsky92,Schonhammer90}
{}.
In what follows, we shall restrict our attention to symmetric double
wells, although an extension to asymmetric double wells is relatively
straightforward \cite{Gangopadhyaya93}.

Usually, in most applications the quantity of interest is the energy
difference $t \equiv E_1 - E_0$ between the lowest two eigenstates, and
corresponds to the tunneling rate through the double-well barrier.  The
quantity $t$ is often small and difficult to calculate numerically,
especially when the potential barrier between
the two wells is large.  Here, we
show how SUSY facilitates the evaluation of $t$.  Indeed, using
the supersymmetric partner potential $V_2(x)$, we obtain a systematic,
highly convergent perturbation expansion for the energy difference $t$.
The leading term is more accurate than the standard WKB tunneling formula,
and the magnitude of the nonleading terms gives a reliable handle on the
accuracy of the result.

First, we briefly review the standard approach for determining $t$ in the
case of a symmetric, one-dimensional double-well potential, $V_1(x)$, whose
minima are located $x = \pm x_0$.  We define the depth, $D$, of $V_1(x)$
by $D \equiv V_1(0) - V_1(x_0)$.  An example of such a potential is shown
in Fig. 9.1.  For sufficiently deep wells, the double-well structure produces
closely spaced pairs of energy levels lying below $V_1(0)$.  The number of
such pairs, $n$, can be crudely estimated from the standard WKB bound-state
formula applied to $V_1(x)$ for $x > 0$:

\be
n \pi = \int_0^{x_c} [V_1(0) - V_1(x)]^{1/2} dx ,
\ee
where $x_c$ is the classical turning point corresponding to energy $V_1(0)$
and we have chosen units where $\hbar = 2m = 1$.  We
shall call a double-well potential ``shallow" if it can hold at most one
pair of bound states, i.e., $n \leq 1$.  In contrast, a ``deep" potential
refers to $n \geq 2$.

The energy splitting $t$ of the lowest-lying pair of states can be obtained
by a standard argument \cite{Landau77}.
Let $\chi(x)$ be the normalized eigenfunction for
a particle moving in a single well whose structure is the same as the
right-hand well of $V_1(x)$ (i.e., $x > 0$).  If the probability of barrier
penetration is small, the lowest two eigenfunctions of the double-well
potential $V_1(x)$ are well approximated by
\be
\psi_{0,1}^{(1)}(x) = [\chi(x) \pm \chi(-x)]/\sqrt2 ~~.
\ee
By integration of Schr\"{o}dinger's equation for the above eigenfunctions,
it can be shown that \cite{Landau77}
\be \label{u3}
t \equiv E_1 - E_0 = 4\chi(0)\chi^\prime(0)~,
\ee
where the prime denotes differentiation with respect to $x$.  This result
is accurate for ``deep" potentials, but becomes progressively worse as the
depth decreases.  Use of WKB wave functions in eq. (\ref{u3})
yields the standard result:
\be
t_{\rm WKB} = \{[2V^{\prime\prime}_1 (x_0)]^{1/2}/\pi \} \exp\left(- 2
\int_0^{x_0} [V_1(x) - V_1(x_0)]^{1/2} dx \right )~.
\ee
The same result can also be obtained via instanton
techniques \cite{Coleman85}.

Using the supersymmetric formulation of quantum mechanics for a given
Hamiltonian, $H_1 = -d^2/dx^2 + V_1(x)$, and its zero-energy ground state
wave function $\psi_0(x)$, we know that the supersymmetric
partner potential $V_2(x)$ is given by
\begin{eqnarray} \label{u5}
V_2(x) &=& V_1(x) -2(d/dx)(\psi_0^\prime/\psi_0) \nonumber\\
       &=& - V_1(x) + 2(\psi_0^\prime/\psi_0)^2~.
\end{eqnarray}

Alternatively, in terms of the superpotential $W(x)$ given by
$W(x) = -\psi_0^\prime/\psi_0~$
we can write
\be \label{u7}
V_{2,1}(x) = W^2(x) \pm dW/dx~.
\ee

{}From the discussion of unbroken SUSY in previous sections, we
know that the energy spectra of the potentials $V_2$ and $V_1$ are
identical, except for the ground state of $V_1$ which is missing from the
spectrum of $V_2$ \cite{Witten81}.
Hence, for the double-well problem, we see that if
$V_1(x)$ is ``shallow" (i.e., only the lowest two states are paired), then
the spectrum of $V_2$ is well separated.  In this case, $V_2$ is relatively
structureless and simpler than $V_1$.
Previous papers \cite{Bernstein84,Kumar86,Marchesoni88} have implicitly
treated just the case of shallow potentials, and, not surprisingly, have
found that the use of SUSY simplifies the evaluation of the energy
difference $t$.  In contrast, let us now consider the case of a deep double
well as shown in Fig. 9.1 .  Here, the spectrum of $V_2$ has a single unpaired
ground state followed by paired excited states.  In order to produce this
spectrum, $V_2$ has a double-well structure together with a sharp ``$\delta-$
function like" dip at $x = 0$.  This central dip produces the unpaired ground
state, and becomes sharper as the potential $V_1(x)$ becomes deeper.

As a concrete example, we consider the class of potentials whose ground
state wave function is the sum of two Gaussians, centered around $\pm x_0$,

\be \label{u8}
\psi_0(x)~\sim~e^{-(x-x_0)^2} + e^{-(x+x_0)^2}~.
\ee
The variables $x$ and $x_0$ have been chosen to be dimensionless.  The
corresponding superpotential $W(x)$, and the two  supersymmetric partner
potentials $V_1 (x)$ and $V_2 (x)$, are given respectively by
\be
W(x) = 2[x - x_0\tanh(2xx_0)],
\ee

\be \label{u10}
V_{2,1}(x) = 4{[x - x_0\tanh(2xx_0)]}^2 \pm 2[1-2x_0^2\sech^2(2xx_0)]~.
\ee

The minima of $V_1(x)$ are located near $\pm x_0$ and the well depth (in
the limit of large $x_0$) is $D \simeq  4x_0^2$.  We illustrate the potentials
$V_1(x)$  and $V_2(x)$ in Fig. 9.2 for the two choices
$x_0 =  1.0$ and $x_0 = 2.5$.  We see that in the limit of large $x_0$, for
both
$V_1(x)$ and $V_2(x)$, the wells become widely separated and deep and that
$V_2(x)$ develops a strong central dip.

The asymptotic behavior of the energy splitting, $t$, in the limit $x_0
\rightarrow \infty$ can be calculated from eq. (\ref{u3}),
with $\chi (x)$ given by one of the (normalized) Gaussians
in eq. (\ref{u8}).  We find that

\be \label{u11}
t \rightarrow 8x_0(2/\pi)^{1/2} e^{-2x_0^2}~.
\ee
The same result can be obtained by observing that $V_1(x) \rightarrow
4(\mid x \mid - x_0)^2$ as $x_0 \rightarrow \infty$.  This potential has a
well known \cite{Merzbacher70} analytic
solution, which involves solving the parabolic cylindrical differential
equation. After carefully handling the boundary conditions, one obtains
the separation of the lowest two energy levels
to be $8x_0(2/\pi)^{1/2}\exp(-2x_0^2)$, in agreement with eq. (\ref{u11}).

We now turn to the evaluation of $t$ via the ground state energy of the
supersymmetric partner potential $V_2(x)$.  In general, since $V_2(x)$ is
not analytically solvable, we must solve an approximate problem and
calculate the corrections perturbatively.  The use of SUSY,
coupled with the observation that the magnitude of $t$ is in general small,
allows us to construct a suitable unperturbed problem.  Consider the
Schr\"{o}dinger equation for $V_2(x)$ and $E = 0$. From
supersymmetry [eq. (\ref{u5})]
we see immediately that $1/\psi_0$ is a solution.  Since $t$ is small, we
expect this solution to be an excellent approximation to the correct
eigenfunction for small values of $x$.  However, $1/\psi_0$ is not
normalizable and hence is not acceptable as a starting point for
perturbation theory.  One possibility is to regularize the behavior
artificially at large $\mid x \mid$ \cite{Bernstein84}.  This
procedure is cumbersome and
results in perturbation corrections to the leading term which are
substantial.  Instead, we choose for our unperturbed problem the second
linearly independent solution of the Schr\"{o}dinger equation
given by \cite{Hildebrand76}
\be
\phi(x) = \frac{1}{\psi_0}~~\int_x^\infty~\psi_0^2(x^\prime)dx^\prime,~~x >
0~,
\ee
and $\phi(x) = \phi(-x)$ for $x < 0$.  Clearly, $\phi(x)$ is well behaved
at $x = \pm \infty$ and closely approximates $1/\psi_0$ at small $x$;
thus we expect it to be an excellent approximation of the exact ground
state wave function of $V_2(x)$ for all values of $x$.  The derivative of $
\phi(x)$ is continuous except at the origin, where, unlike the exact
solution, it has a discontinuity $\phi \mid_{0+} - \phi \mid_{0-} = -2
\psi_0(0)$.  Hence $\phi (x)$ is actually a zero-energy solution of the
Schr\"{o}dinger equation for a potential $V_0(x)$ given by

\be
V_0(x) = V_2(x) - 4{\psi_0^2}(0)\delta(x)~,
\ee
where we have assumed that $\psi_0(x)$ is normalized.  We calculate the
perturbative corrections to the ground state energy using $\Delta V = +
4{\psi_0^2}(0)\delta(x)$ as the perturbation.  Note that the coefficient
multiplying the $\delta$-function is quite small so that we expect our
perturbation series to converge rapidly.

For the case of a symmetric potential such as $V_2(x)$, the perturbative
corrections to the energy arising from $\Delta V$ can be most simply
calcualted by use of the logarithmic perturbation-theory \cite{Imbo84}
formulation of
the usual Rayleigh-Schr\"{o}dinger series.  The first and second order
corrections to the unperturbed energy $E = 0$ are

\be \label{u14a}
E^{(1)} = \frac{1}{2\xi(0)}~,~ \xi(x) \equiv~~\int_x^\infty~~\phi^2(x^\prime
)dx,~~
E^{(2)} = -2 \int_0^\infty~~\left[ \frac{E^{(1)}\xi(x^\prime)}{\phi
(x^\prime)} \right]^2~dx^\prime~~.
\ee

For our example, we  numerically evaluate these corrections in order to
obtain an estimate of $t$.
The results are shown in Fig. 9.3 for values of $x_0 \leq 2$.  Estimates of
$t$ correct to first, second, and third order calculated from logarithmic
perturbation theory are compared with the exact result for $V_2$, obtained
by the Runge-Kutta method.  The asymptotic behavior of $t$
given by eq. (\ref{u11}) is also shown.
This asymptotic form can also be recovered from eq. (\ref{u14a})
by a suitable approximation of the integrand in the large-$x_0$ limit.
Even for values of $x_0 \leq 1/\sqrt{2}$, in which case $V_1(x)$ does not
exhibit a double-well structure, the approximation technique is
surprisingly good.  The third-order perturbative result and the exact
result are indistinguishable for all values of $x_0$.

In conclusion, we have demonstrated how SUSY can be used to calculate
$t$, the energy splitting for a double-well potential.  Rather
than calculating this splitting as a difference between the lowest-
lying two states of $V_1(x)$, one can instead develop a perturbation series
for the ground state energy $t$ of the partner potential $V_2(x)$.  By
choosing as an unperturbed problem the potential whose solution is the
normalizable zero-energy solution of $V_2(x)$, we obtain a very simple
$\delta-$function perturbation which produces a rapidly convergent series
for $t$ \cite{Keung88}. The procedure is quite general and is
applicable to any arbitrary
double-well potential, including asymmetric ones \cite{Gangopadhyaya93}.
The numerical results are very accurate for both
deep and shallow potentials.

\subsection{Supersymmetry and Large-$N$ Expansions}
\label{sec9.4}

The large-$N$ method, where $N$ is the number of spatial dimensions, is a
powerful technique for analytically determining the eigenstates of the
Schr\"{o}dinger equation, even for potentials which have no
small coupling constant
and hence not amenable to treatment by standard perturbation theory \cite
{Mlodinow80,Witten80,Bender82,Ader83,Gangopadhyay84,%
Yaffe82,Sinha-Roy84,Miramontes84,Hikami79,Dutt85}.
A slightly modified, physically motivated approach, called the
``shifted large-$N$ method" \cite{Sukhatme83,Imbo84a,Imbo84b,Imbo85a}
incorporates exactly known analytic results
into $1/N$ expansions, greatly enhancing their accuracy, simplicity and
range of applicability. In this subsection, we will descibe how the rate
of convergence of shifted $1/N$ expansions can be still further improved
by using the ideas of
SUSY QM \cite{Imbo85b}.

The basic idea in obtaining a $1/N$ expansion
in quantum mechanics consists of solving the Schr\"{o}dinger
equation in $N$ spatial dimensions, assuming $N$ to be large, and taking
$1/N$ as an ``artificially created" expansion parameter for doing
standard perturbation theory. At the end of the calculation, one
sets $N = 3$ to get results for problems of physical interest in three
dimensions.

For an arbitrary spherically symmetric potential $V(r)$ in $N$ dimensions,
the radial Schr\"{o}dinger equation contains the effective potential

\be \label{veff}
V_{\rm eff}(r)=V(r)+\frac {(k-1)(k-3)\hbar^2}{8mr^2}~,~ k=N+2l ~.
\ee

It is important to note that $N$ and $l$
always appear together in the combination
$k=N+2l$. This means that the eigenstates, which could in principle
have depended on the three quantities $N,l,n,$ in fact only depend on
$k$ and $n$, where $n$ is the radial quantum number which can take
values 0,1,2,...
One now makes a systematic expansion
of eigenstates in the parameter $1/\overline{k}$, where $\overline{k} = k-a$.
Of course, for very large values of $N$,
the two choices $\overline{k}$ and $k$ are equivalent.
However, for $N=3$ dimensions, a properly chosen shift $a$ produces great
improvements in accuracy and simplicity. At small values of $r$, the $n=0$
wave function $\psi_0(r)$ has the behavior $r^{(k-1)/2}$. If one sets
\be \label{rr}
\psi_0(r)=r^{(k-1)/2}\Phi_0(r)
\ee
where $\Phi_0(r)$ is finite at the origin, then eq. (\ref{rr}) readily
gives the supersymmetric partner potential of $V_{\rm eff}(r)$ to be
\be \label{part}
V_2(r)=V(r)+\frac {(k+1)(k-1)\hbar^2}{8mr^2}
-\frac{\hbar^2}{m}\frac {d^2}{dr^2}ln \Phi_0(r)
\ee

$V_2(r)$ and $V_{\rm eff}(r)$ have the same energy values
[except for the ground state].
However, large-$N$ expansions with the partner potential $V_2(r)$ are
considerably better
since the angular momentum barrier in eq. (\ref{part}) is given by
$(k'-1)(k'-3)\hbar^2/{8mr^2}$,
where $k'=k+2$. So, effectively, one is working in two extra spatial
dimensions! Thus, for example, in order to calculate the energy of the state
with quantum numbers $k,n$ of $V_{\rm eff}(r)$ one can equally well use
$k'=k+2, n-1$ with $V_2(r)$. To demonstrate this procedure, let us give an
explicit example. Using the usual choice of units $\hbar=2m=1$,
the s-wave Hulthen effective potential
in three dimensions and its ground state wave function are:
\be
V_{\rm eff}^H(r)=-\frac{2 \delta e^{-\delta r}}{1-e^{-\delta r}}
         +\frac{(2-\delta)^2}{4}~,~\psi_0(r) \sim (1-e^{-\delta r})~
         e^{-\frac {(2-\delta)r}{2}}~,
\ee
where the parameter $\delta$ is restricted to be less than 2.
The supersymmetric partner potential is
\be
V_2^H=V_1^H+\frac{2 \delta ^2e^{-\delta r}}{(1-e^{-\delta r)^2}}~.
\ee
As $r$ tends to zero, $V_2^H$ goes like $2r^{-2}$, which as mentioned above,
corresponds to the angular momentum barrier $(k'-1)(k'-3)\hbar^2/8mr^2$ for
$k'=5$ ($N=5,l=0$). Let us compute the energy of the first excited state
of $V_{\rm eff}^H(r)$. For the choice $\delta=0.05$,
the exact answer is known to be 0.748125 \cite{Lam71}.
The results upto leading, second and third order using a shifted
$1/N$ expansion for $V_{\rm eff}^H$ are 0.747713, 0.748127 and
0.748125 . The corresponding values using the supersymmetric partner
potential $V_2^H$ are all 0.748125 !
It is clear that although excellent results are obtained with the
use of the shifted $1/N$ expansion for the original
potential $V_{\rm eff}^H(r)$ in
three dimensions, even faster convergence is obtained by using the
supersymmetric partner potential, since we are now effectively
working in five dimensions instead of three. Thus,
SUSY has played an important role in making a
very good expansion even better \cite{Imbo85b}. In fact, for many applications,
considerable analytic simplification occurs since it is
sufficient to just use the leading term in the shifted $1/N$ expansion
for $V_2(r)$. Other examples can be found in ref. \cite{Imbo85b}.

\section{Pauli Equation and Supersymmetry }
\label{sec10}

So far, we have discussed the concept of SUSY for Schr\"{o}dinger Hamiltonians
in one-dimension and for
central potentials in higher dimensions which are essentially again
one-dimensional problems. In this section we shall show that the
concept of SUSY can also be applied to some geniune two-dimensional
problems. In particular we show that the Pauli Hamiltonian which
deals with the problem of a charged particle in external magnetic
field can  always be put in SUSY form provided the gyromagnetic ratio
is equal to two. It is worth emphasizing here that not only the
uniform magnetic field problem (i.e. the famous Landau level problem)
but the nonuniform magnetic field problems can also be put in SUSY
form. Using the concepts of SUSY and shape invariance we show that
some of the nonuniform magnetic field problems can be solved
analytically.

The Pauli Hamiltonian for the motion of a charged particle in
external magnetic field in two dimensions is given by $(\hbar = 2m=e=1)$
\be \label{eq10.1}
H = (p_x+A_x)^2+(p_y+A_y)^2+{g\over 2}{(\vec\nabla \times \vec A)}_z\sigma_z
\ee
It is easily seen that this $H$ along with the supercharges $Q^1$ and $Q^2$
defined by \cite{Crumbrugghe83,Khare84}
\bea \label{eq10.2}
Q^1 &=& {1\over \sqrt 2}[-(p_y+A_y)\sigma_x+(p_x+A_x)\sigma_y]\nonumber \\
Q^2 &=& {1\over\sqrt 2}[(p_x+A_x)\sigma_x+(p_y+A_y)\sigma_y]
\eea
satisfy the $N=1$ supersymmetry algebra provided the gyromagnetic ratio
$g$ is two
\be \label{eq10.3}
\{ Q^{\alpha},Q^{\beta}\} = H \delta^{\alpha\beta},[H,Q^{\alpha}]=0;
\alpha,\beta=1,2
\ee
It is interesting to note that SUSY fixes the value of $g$. The above
Hamiltonian has an additional $0(2)\otimes 0(2)$ symmetry coming from
$\sigma_z$ and an 0(2) rotation in the $A^1,A^2$ plane
$(A^1\equiv p_x+A_x; A^2\equiv p_y+A_y)$.
Let us first consider the problem in an asymmetric gauge i.e.
\be
A_y(x,y) = 0, A_x(x,y) =W(y)
\ee
where $W(y)$ is an arbitrary function of $y$. In this case the Pauli
Hamiltonian takes the form
\be
H = (p_x+W(y))^2+p^2_y-W'(y)\sigma_z
\ee
Since this $H$ does not depend on $x$, hence the eigenfunction $\tilde\psi$
can be factorized as
\be
\tilde\psi(x,y) = e^{ikx}\psi(y)
\ee
where $k$ is the eigenvalue of the operator $p_x (-\infty\leq
k\leq\infty)$. The Schr\"{o}dinger equation for $\psi(y)$ then takes the form
\be
[-{d^2\over dy^2}+(W(y)+k)^2-W'(y)\sigma]\psi(y)=E\psi(y)
\ee
where $\sigma(=\pm 1)$ is the eigenvalue of the operator $\sigma_z$.
Thus we have reduced the problem to that of SUSY in one dimension
with superpotential $W(y)+k$ where $W(y)$ must be independent of $k$. This
constraint on $W(y)$ strongly restricts the allowed forms of shape
invariant $W(y)$ for which the spectrum can be written down
algebraically. In particular from Table 4.1 we find that the only
allowed forms are (i) $W(y) = {\omega}_c y +c_1$ (ii) $W(y) = a \tanh y +c_1$
(iii)
 $W(y) = a \tan y +c_1$ (iv) $W(y) = c_1 -c_2 \exp{(-y)}$ for which $W(y)$ can
b
 e
written in terms of simple functions and for which the spectrum can
be written down algebraically \cite{Stanciu67,Cooper88}. In
particular when
\be
W(y) = {\omega}_c y +c_1
\ee
which corresponds to the uniform magnetic field, then the energy
eigenvalues known as Landau levels are given by
\be
E_n=(2n+1+\sigma){\omega}_c;~~~ n=0,1,2...
\ee
Note that the ground state and all excited states are infinite-fold
degenerate since $E_n$ does not depend on $k$ which assumes a
continuous sequence of value $(-\infty \leq k\leq\infty)$.

The magnetic field corresponding to the other choices of $W$ are
(ii) $B = - a \sech^2 y$ (iii)$B = -a \sec^2 y ({-{\pi}\over 2}\leq y\leq
{\pi\over 2})$ (iv) $B = +c_2 \exp{(-y)}$ and as mentioned above, all
these problems can be solved algebrically by using the results of
Sec. 4.

Let us now consider the same problem in the symmetric gauge. We choose
\be
A_x = {\omega}_c y f(\rho),~~ A_y =-{\omega}_c xf(\rho)
\ee
where $\rho^2 = x^2+y^2$ and ${\omega}_c$ is a constant. The corresponding
magnetic field $B_z$ is then given by
\be
B_z(x,y)\equiv\partial_xA_y- \partial_yA_x =-2{\omega}_cf(\rho)-{\omega}_c\rho
f
 '(\rho)
\ee
In this case the Hamiltonian (\ref {eq10.1}) can be shown to take the form
\be
H = -({d^2\over dx^2}+{d^2\over dy^2})+{\omega}^2_c\rho^2f^2-2{\omega}_cf
L_z-(2\omega_cf+{\omega}_c\rho f')\sigma_z
\ee
where $L_z$ is the $z$-component of the orbital angular momentum
operator. Clearly the corresponding Schr\"{o}dinger problem can be solved
in the cylindrical coordinates $\rho,\phi$. In this case, the eigenfunction
$\psi(\rho,\phi)$ can be factorized as
\be
\psi(\rho,\phi) =R(\rho)e^{im\phi}/\sqrt\rho
\ee
where $m=0,\pm 1,\pm 2,..$ is the eigenvalue of $L_z$. In this case
the Schr\"odinger equation for $R(\rho)$ takes the form
\be
[-{d^2\over d\rho^2}+{\omega}_c^2\rho^2f^2-2{\omega}_cfm-(2{\omega}_c f
+{\omega}_c\rho
f'(\rho))\sigma +{m^2-1/4\over \rho^2}]R(\rho) = E R(\rho)
\ee
where $\sigma(=\pm 1)$ is the eigenvalue of the operator $\sigma_z$.
On comparing with Table 4.1, it is easily checked that there is only one
shape invariant potential $(f(\rho)=1)$ for which the spectrum can be
written down algebraically. This case again corresponds to the famous
Landau level problem i.e. it corresponds to the motion of a charged
particle in the $x-y$ plane and subjected to a uniform magnetic field (in
the symmetric gauge) in the $z$-direction. The energy eigenvalues are
\be
E_n=2(n+m+\mid m\mid){\omega}_c; n=0,1,2...
\ee
so that all the states are again infinite-fold degenerate.
It is worth noting that the nonuniform magnetic fields can also give this
equi-spaced spectrum \cite {Khare93e}. However, they do so only for one
particular value of $m$ while for other values of $m$, the spectrum is
in general not equi-spaced.

The fact that in this example there are infinite number of degenerate
ground states with zero energy can be understood from the
Aharonov-Casher theorem \cite {Aharonov78} which states that if the total
flux defined by $\Phi =\int B_z dxdy = n+ \epsilon (0\leq \epsilon < 1)$ then
there are precisely $n-1$ zero energy states. Note that in our case
$\Phi$ is infinite.

\section {Supersymmetry and the Dirac Equation}
\label{sec11}

There have been many applications of SUSY QM in the context of the Dirac
equation. In view of the limitations of space, we shall concentrate
on only a few of these applications \cite{Thaller92}. In particular,
we discuss the
supersymmetric structure of the Dirac Hamiltonian $(H_D)$ and show
how the methods used to obtain analytical solutions of the
Schr\"{o}dinger equation can be extended to the Dirac case. First of all,
we consider the Dirac equation in 1+1 dimensions with Lorentz scalar
potential $\phi(x)$. We show that whenever the one-dimensional
Schr\"{o}dinger equation is analytically solvable for a potential $V(x)$,
then there always exists a corresponding Dirac scalar potential
problem which is also analytically solvable \cite{Cooper88}. It turns
out that, on the one hand, $\phi (x)$ is essentially the
superpotential of the Schr\"{o}dinger problem and on the other hand, it
can be looked upon as the kink solution of a scalar field theory in
1+1 dimensions. Next we discuss the celebrated problem of the
Dirac particle in a Coulomb field \cite{Sukumar85b} and show that its
eigenvalues and eigenfunctions can be  simply obtained by using
the concept of SUSY and shape invariance as developed in Secs. 2 and
4. We also discuss the problem of the Dirac equation in an external
magnetic field in two dimensions and show that there is always a
supersymmetry in the problem in the massless case. We also classify a
number of magnetic field problems whose solutions can be algebraically
obtained by
using the concepts of SUSY and shape invariance \cite{Cooper88}.
In addition, we show that the Euclidean Dirac operator in four
dimensions, in the background of gauge fields, can always be cast in the
language of SUSY QM. Finally, we discuss the path integral formulation of the
fermion propagator in an external field and show how the previously known
results for the constant external field can be very easily obtained by using
the ideas of SUSY \cite{Cooper88}.

\subsection{Dirac Equation With Lorentz Scalar Potential}
\label{sec11.1}

The Dirac Lagrangian in 1+1 dimensions with a Lorentz scalar potential
$\phi (x)$ is given by
\be \label{eq11.1}
{\cal L} = i \overline\psi \gamma^{\mu}\partial_{\mu}\psi -
\overline\psi \psi\phi.
\ee
The scalar potential $\phi (x)$ can be looked upon as the static,
finite energy, kink solution corresponding to the scalar field Lagrangian
\be \label{eq11.2}
{\cal L}_{\phi} = {1\over 2} \partial_{\mu} \phi (x)\partial^{\mu}\phi (x)
-V(\phi).
\ee

Such models have proved quite useful in the context of the
phenomenon of fermion number fractionalization
\cite{Jackiw76,Jackiw81,Campbell82} which has been seen in certain
polymers like polyacetylene. Further, a variant of this model
is also relevant in the context of supersymmetric field theories in
1+1 dimensions \cite{Witten78,dAdda78}. Note that the coupling constant in
eqs. (\ref{eq11.1}) and (\ref{eq11.2}) has been absorbed in $\phi$ and
$V(\phi)$ respectively.

The Dirac equation following from eq. (\ref{eq11.1}) is

\be \label{eq11.3}
i\gamma^{\mu}\partial_{\mu} \psi(x,t) - \phi (x)\psi (x,t) = 0.
\ee
Let
\be \label{eq.11.4}
\psi(x,t) = \exp(-i{\omega}t) \psi (x)
\ee
so that the Dirac equation reduces to
\be \label{eq11.5}
\gamma^0 {\omega}\psi (x) + i\gamma^1 {d\psi (x)\over dx}- \phi (x) \psi (x)
=0.
\ee
We choose
\be \label{eq11.6}
\gamma^0 = \sigma^1 = \pmatrix{0 & 1\cr 1& 0\cr}, \gamma^1 =
i\sigma^3 = \pmatrix{i & 0\cr 0 & -i\cr}, \psi (x) =
\pmatrix{\psi_1(x) \cr \psi_2(x)\cr}
\ee
so that we have the coupled equations
\bea \label{eq11.7}
 A \psi_1 (x)  & = & {\omega} \psi_2 (x) \nonumber \\
A^{\dag} \psi_2 (x) & = & {\omega} \psi_1 (x),
\eea
where
\be \label{eq11.8}
A = {d\over dx} + \phi (x) , A^{\dag} = -{d\over dx} + \phi (x).
\ee
We can now easily decouple these equations. We get
\be \label{eq11.9}
A{^{\dag}} A \psi_1 ={\omega}^2 \psi_1, AA^{\dag} \psi_2 = {\omega}^2 \psi_2.
\ee
On comparing with the formalism of Sec. 2, we see that there is a
supersymmetry in the problem and $\phi (x)$ is just the
superpotential of the Schr\"{o}dinger formalism. Further $\psi_1$ and
$\psi_2$ are the eigenfunctions of the Hamiltonians $H_-\equiv A^{+} A$
and $H_+\equiv AA^{\dag}$ respectively with the corresponding potentials
being $V_{\mp} (x) \equiv \phi^2 (x) \mp \phi' (x)$. The spectrum of
the two Hamiltonians is thus degenerate except that $H_-(H_+)$ has an
extra state at zero energy so long as $\phi (x\rightarrow \pm\infty)$
have opposite signs and $\phi (x\rightarrow + \infty) > 0 (< 0)$.
Using the results of the Secs. 2 and 4  we then conclude that for
every SIP given in Table 4.1 there
exists an analytically solvable Dirac problem with the corresponding
scalar potential $\phi (x)$ being the superpotential of the
Schr\"{o}dinger problem. In particular, using the reflectionless
superpotential given by
\be \label{eq11.10}
W ( x) = n\tanh x
\ee
one can immediately construct perfectly transparent Dirac potentials
with $n$ bound states \cite{Toyama93}. Further, using the results for
the SIP with scaling ansatz $(a_2= q a_1)$ \cite{Khare93a,Barclay93},
one can also construct perfectly transparent Dirac potentials with an
infinite number of bound states.

\subsection{Supersymmetry and the Dirac Particle in a Coulomb Field}
\label{sec11.2}

The Dirac equation for a charged particle in an electromagnetic field
is given by $(e = \hbar = c = 1)$
\be \label{eq11.11}
[i\gamma^{\mu}(\partial_{\mu}+iA_{\mu}) - m ] \psi = 0
\ee
For a central field i.e. $\vec A = 0$ and $A_0(\vec x,t)= V(r)$, this
equation can be written as \cite{Bjorken64}
\be \label{eq11.12}
i{\partial \psi\over \partial t} = H\psi = ({\vec\alpha{\cdot}\vec p}+\beta
m +V ) \psi
\ee
where
\be \label{eq11.13}
\alpha^i = \pmatrix{0 &\sigma^i\cr \sigma^i & 0\cr}, \beta =
\pmatrix{1& 0\cr 0& -1\cr}
\ee
with $\sigma^i$ being the Pauli matrices. For central fields, this
Dirac equation can be separated in spherical coordinates and finally
for such purposes as the computation of the energy levels one only needs
to concentrate on the
radial equations which  are given by \cite{Bjorken64}
\bea \label{eq11.14}
G'(r) + {kG\over r} - (\alpha_1-V) F & = & 0 \nonumber \\
F'(r) - {kF\over r} - (\alpha_2+V) G & = & 0
\eea
where
\be \label{eq11.15}
\alpha_1 = m + E, \alpha_2 = m - E
\ee
and $G_k$ is the ``large" component in the non-relativistic limit.
Of course the radial functions $G_k$ and $F_k$ must be multiplied by
the appropriate two component angular eigenfunctions to make up the full
four-component solutions of the Dirac equation \cite{Bjorken64}.
These coupled equations are in general not analytically solvable; one of
the few exceptions being the case of the Dirac particle in a
Coulomb field for which
\be \label{eq11.16}
V(r) = - {\gamma \over r}, \gamma = Z e^2
\ee

We now show that the Coulomb problem can be solved algebraically by
using the ideas of SUSY and shape invariance. To that end, we first
note that in the case of the Coulomb potential, the coupled equations
(\ref{eq11.14}) can be written in a matrix form as
\be \label{eq11.17}
\pmatrix{G'_{k}(r)\cr F'_{k}(r)\cr} + {1\over r} \pmatrix{k
& -\gamma\cr \gamma & -k\cr}\pmatrix{G_k \cr F_k\cr}=
\pmatrix{0 & \alpha_1\cr \alpha_2 & 0\cr}\pmatrix{G_k\cr F_k\cr}
\ee
where $k$ is an eigenvalue of the operator $-(\vec\sigma{\cdot}\vec L+1)$
with the allowed values $k = \pm 1, \pm 2, \pm 3$..., and satisfies $\mid
k \mid = J +  {1\over 2}$.
Following Sukumar \cite{Sukumar85b}, we now notice that the matrix
multiplying $1/r$ can be diagonalized by multiplying it by a matrix $D$
from the left and $D^{-1}$ from the right where
\be \label{eq11.18}
D = \pmatrix{k+s & -\gamma\cr -\gamma & k+s\cr} ,~~ s = \sqrt{k^2-\gamma^2}~.
\ee
On multiplying eq. (\ref{eq11.17}) from the left by the matrix $D$ and
introducing the new variable $\rho = E r$ leads to the pair of equations
\bea \label{eq.19}
  A\tilde F & = & ({m\over E}  - {k\over s}) \tilde G~, \nonumber \\
A^{\dag}\tilde G & = & - ({m\over E}+ {k\over S}) \tilde F
\eea
where
\be \label{eq11.20}
\pmatrix{\tilde G \cr\tilde F\cr} = D\pmatrix{G \cr F\cr}
\ee
and
\be \label{eq11.21}
A = {d\over d\rho} -{s\over \rho}  + {\gamma \over s}, A^{\dag} = -
{d\over d\rho} - {s\over \rho} + {\gamma \over s}.
\ee
Thus we can easily decouple the equations for $\tilde F$ and $\tilde
G$ thereby obtaining
\bea \label{eq11.22}
H_- \tilde F  \equiv  A^{+}A\tilde F & = & ({k^2\over s^2}-{m^2\over E^2})
\tilde F ,\nonumber \\
H_+ \tilde G  \equiv  AA^{+}\tilde G & = & ({k^2\over s^2}-{m^2\over E^2})
\tilde G.
\eea
We thus see that there is a supersymmetry in the problem and
$H_{\pm}$ are shape invariant supersymmetric partner potentials since
\be \label{eq11.23}
H_+(\rho;s,\gamma) = H_- (\rho;s+1,\gamma) +{\gamma^2\over s^2}
 - {\gamma^2\over (s+1)^2}.
\ee
On comparing with the formalism of Sec. 4 it is then clear that in
this case
\be \label{eq11.24}
a_2 = s+1 , a_1 = s, R(a_2) = {\gamma^2\over a^2_1} - {\gamma^2\over a^2_2}
\ee
so that the energy eigenvalues of $H_-$ are given by
\be \label{eq11.25}
({k\over s})^2 - ({m^2\over E^2_n}) \equiv E^{(-)}_n =
\sum^{n+1}_{k=2} R(a_k) = \gamma^2 ({1\over s^2}-{1\over (s+n)^2}).
\ee
Thus the Coulomb bound state energy eigenvalues $E_n$ are given by
\be \label{eq11.26}
E_n = {m\over [1+{\gamma^2\over (s+n)^2}]^{1/2}}, n = 0,1,2,...
\ee
It should be noted that  every eigenvalue of $H_-$ is also an
eigenvalue of $H_+$ except for the ground state of $H_-$ which satisfies
\be \label{eq11.27}
A\tilde F = 0 \Longrightarrow \tilde F_0(\rho) = \rho^s exp{(-\gamma\rho/s)}
\ee
Using the formalism for the SIP as developed in Sec. 4, one can also
algebraically obtain all the eigenfunctions of $\tilde F$ and $\tilde
G$.

Notice that the spectrum as given by eq. (\ref{eq11.26}) only depends
on $\mid k\mid $ leading to a doublet of states corresponding to $k
=\mid k \mid$ and $ k = - \mid k \mid$ for all positive $n$. However, for
$n = 0$, only the negative value of $k$ is allowed and hence this
is a singlet state.

\subsection{SUSY and the Dirac Particle in a Magnetic Field}
\label{sec11.3}

Let us again consider the Dirac equation in an electromagnetic field
as given by eq. (\ref{eq11.11}) but now consider the other case when
the vector potential is nonzero but the scalar potential is zero
i.e. $A_0 = 0, \vec A \not = 0$. Now as shown by Feynman and Gell-Mann
\cite{Feynman58} and Brown \cite{Brown58}, the
solution of the four component Dirac
equation in the presence of an external electromagnetic field can be
generated from the solution of a two component relativistically
invariant equation. In particular, if $\psi$ obeys the two component equation
\be \label{eq11.28}
[(\vec P + \vec A)^2 + m^2 + \vec\sigma{\cdot} (\vec B + i\vec E)]\psi =
(\overline E + A_0)^2\psi
\ee
then the four component spinors that are solutions of the massive
Dirac equation are generated from the two component $\psi$ via
\be \label{eq11.29}
\psi_D = \pmatrix{(\vec\sigma{\cdot} (\vec P+\vec A)+\overline
E-A_0+m)\psi\cr  (\vec\sigma{\cdot} (\vec P+\vec A)+\overline
E-A_0-m)\psi\cr}.
\ee
Thus, in order to solve the Dirac equation, it is sufficient to solve
the much simpler two-component eq.(\ref{eq11.28}) and then generate
the corresponding Dirac solutions by the use of eq. (\ref{eq11.29}).
In the special case when the scalar potential $A_0$ (and hence $\vec
E$) vanishes, the two-component equation then has the canonical form
of the Pauli equation describing the motion of a charged particle in
an external magnetic field. If further, $m = 0$ and the
motion is confined to two dimensions, then the Pauli
eq. (\ref{eq11.28}) exactly reduces to the eq. (\ref{eq10.1}) of the
last section. Further, since
\be \label{eq11.30}
(H_D)^2 = [\vec\alpha{\cdot} (\vec P+\vec A)]^2 = H_{Pauli}
\ee
hence, there is a supersymmetry in the massless Dirac problem in
external magnetic fields in two dimensions since $H^2_D, Q_1$ and
$Q_2$ (see eq. (\ref{eq10.2}) satisfy the SUSY algebra as given by
eq. (\ref{eq10.3}). Clearly, this supersymmetry will also be there in
two Euclidean dimensions. We can now immediately borrow all the
results of the last section. In particular, it follows that if the
total flux $\Phi(= \int B_z dxdy)$=$ n+\epsilon ( 0 \leq \epsilon < 1)$ then
there are precisely $n-1$ zero modes of the massless Dirac equation in
two dimensions in the background of the external magnetic field $B$
$(B \equiv B_z)$
\cite{Aharonov78,Jackiw84}. Further, in view of
eqs. (\ref{eq11.28}) and (\ref{eq11.29}) we can immediately write down
the exact solution of the massless Dirac equation in an external
magnetic field in  two dimensions in all the four situations discussed
in the last section when the gauge potential depended on only one
coordinate (say $y$). Further using the results of that section, one can
also algebraically
obtain the exact solution of the Dirac equation in an uniform
magnetic field in the symmetric gauge when the gauge potential depends on
both $x$ and $y$.

Even though there is no SUSY, exact solutions of the Pauli and hence
the Dirac equation are also possible in the massive case. On
comparing the equations as given by (\ref{eq11.28}) (with $A_0=0,
\vec E=0$) and (\ref{eq10.1}) it is clear that the exact solutions in the
massive case are simply obtained from the massless case by replacing
$\overline E^2$ by $\overline E^2-m^2$. Summarizing, we conclude that
the exact solutions of the massive (as well as the massless) Dirac
equation in external magnetic field in two dimensions can be obtained
algebraically in case the magnetic field $B (\equiv B_Z$) has any one
of the following four forms (i) $B =$ constant (ii) $B = - a~ \sech^2 y$
(iii) $B = -a \sec^2 y (-\pi/2 \leq y \leq \pi/2)$ (iv) $B = - c_2
\exp(-y)$. Further, in the uniform magnetic field case, the solution
can be obtained either in the asymmetric or in the symmetric gauge
\cite{Stanciu67,Cooper88}.

\subsection {SUSY and the Euclidean Dirac Operator}
\label{sec11.4}

In the last few years there has been a renewed interest in
understanding the zero modes and the complete energy spectrum of the
square of the Dirac operator for a Euclidean massless fermionic
theory interacting with the background gauge fields
\cite{Alvarez83,Friedan85,Musto86,Hughes86}. Let us
first show that there is always a supersymmetry in the problem of the
Euclidean massless Dirac operator in the background of the gauge
fields. On defining
\be \label{eq11.31}
Q_{\pm} = {1\over 2} (1\pm \gamma_5)\gamma_{\mu} D_{\mu}
\ee
where
\be \label{eq11.32}
D_{\mu} = \partial_{\mu}+iA_{\mu},
\ee
one finds that the Dirac operator can be written as
\be \label{eq11.33}
\gamma_{\mu} D_{\mu} = Q_++Q_-
\ee

Let us take the following representation of the Euclidean $\gamma$ matrices
\be \label{eq11.34}
\gamma_0 = \pmatrix{0 & i\cr -i & 0\cr}, \gamma_i = \pmatrix{0 &
\sigma^i\cr \sigma^i & 0\cr}, \gamma_5 = \pmatrix{1 & 0\cr 0 & -1\cr}
\ee
where $\sigma^i$ are the usual Pauli matrices. It is then easily seen
that the operators $Q_+, Q_-$ and $H\equiv - (\gamma_{\mu}D_{\mu})^2$
satisfy the usual N = 1 SUSY algebra
\be \label{eq11.35}
H = - (\gamma_{\mu}D_{\mu})^2 = \{ Q_+,Q_-\}, [H,Q_+]=0=[H,Q_-]
\ee
This supersymmetry is popularly known as chiral supersymmetry. In
view of the above representation of the $\gamma$ matrices, we find
\be \label{eq11.36}
\gamma_{\mu}D_{\mu} = \pmatrix{0 & iD_0+\vec\sigma{\cdot}\vec D \cr
-iD_0+\vec\sigma{\cdot}\vec D & 0\cr} = \pmatrix{0 & L\cr L^{\dag} & 0\cr}
\ee
and hence
\be \label{eq11.37}
-H=  (\gamma_{\mu}D_{\mu})^2 = D_{\mu}D_{\mu}
+\pmatrix{\vec\sigma{\cdot}(\vec B +\vec E) & 0\cr 0 & \vec\sigma{\cdot}(\vec B
 -\vec
E)}= \pmatrix{LL^{\dag} & 0\cr 0 & L^{\dag}L \cr}
\ee
where we have used the convention
\be \label{eq11.38}
F_{\mu\nu}=\partial_{\mu}A_{\nu}-\partial_{\nu}A_{\mu}, B_i = {1\over
2} \epsilon_{ijk}F_{jk},E_i=F_{0i}
\ee
Using the techniques of Secs. 2 and 4, one can algebraically obtain
the eigenvalues and the eigenfunctions of $H$ for few special cases
\cite{Cooper88}. Further, once we have an eigenfunction of $H$ with a
nonzero eigenvalue $E_1$, then one can easily obtain the
eigenfunctions $\phi_{\pm}$ of $\gamma_{\mu}D_{\mu}$ with eigenvalues
$\pm(E_1)^{1/2}$ by the construction
\be \label{eq11.39}
\phi_{\pm} = ( \gamma_{\mu}D_{\mu}\pm(E_1)^{1/2})\psi
\ee

Finally, it is worth pointing out that in case the gauge potential
$A_{\mu}$ has the special form
\be \label{eq11.40}
A_{\mu} = f_{\mu\nu}\partial_{\nu}\chi
\ee
where the constant 4 by 4 matrix $f_{\mu\nu}$ has the properties:
\be \label{eq11.41}
f_{\mu\nu} = - f_{\nu\mu}, f_{\mu\nu}f_{\nu\alpha} = - \delta_{\mu\alpha}
\ee
then there exists another breakup of the Dirac operator
$\gamma_{\mu}D_{\mu}$ which leads to
the so called complex supersymmetry \cite{Cooper88}.

\subsection{Path Integral Formulation of the Fermion Propagator}
\label{sec11.5}

We  have seen that there is always a supersymmetry associated with
the Euclidean Dirac operator in four as well as two dimensions in the
background of the gauge fields. This SUSY was first successfully
exploited in the context of the study of the chiral anomalies
\cite{Fujikawa80}. In a nut shell, the solvability of the Dirac
operator is related to the ability to integrate the path integral.
The quantities one wishes to calculate are the Green's functions
\be \label{eq11.42}
G(x,x;\tau) = < x \mid e^{-H\tau} \mid x' > \mid_{x=x'}
\ee
and
\be \label{eq11.43}
G_5(x,x;\tau) = <x \mid Tr(\gamma_5 e^{-H\tau})\mid x' >\mid_{x=x'}
\ee

By using Schwinger's proper time formalism we can determine $S(x,x;A)$
from $G$. Similarly, the index $Z_5$ which in the limit $\tau = 0$ is
related to the chiral anomaly, is just the spatial integral over $G_5$.

Here we would like to show that by introducing the fermionic degrees
of freedom, the path integral for the Dirac operator, which is
initially a path ordered integral gets reduced to an ordinary path
integral. This trick was first
introduced by Rajeev \cite{Rajeev87} who was interested in
reformulating quantum electrodynamics as a supersymmetric theory of
loops. Once we introduce the fermionic variables then $G(G_5)$ is
determined by choosing the antiperiodic (periodic) boundary conditions
for the fermions. By integrating over the fermionic degrees of
freedom, a purely bosonic path integral is then obtained. For the case of
a constant external field strength $F_{\mu\nu}$, the bosonic path
integral is a Gaussian, which allows one to  trivially obtain $G$ and $G_5$.

Let us consider the square of the Dirac operator i.e.
$(\gamma_{\mu}D_{\mu})^2$. It can also be written as
\be \label{eq11.44}
H = (P+A)^2 - {1\over 2} \sigma^{\mu\nu}F_{\mu\nu}
\ee
where
\be \label{eq11.45}
\sigma^{\mu\nu} = {1\over 2i} [\gamma^{\mu},\gamma^{\nu}].
\ee
If we do not introduce auxiliary fermions then the related matrix
valued Lagrangian would be
\be \label{eq.11.46}
L(x,\dot{x}) = {1\over 4} \dot{x}_{\mu} \dot{x}_{\mu} -
iA_{\mu}\dot{x}_{\mu}+{1
 \over 2}\sigma^{\mu\nu}F_{
\mu\nu}
\ee
and we would obtain for Feynman's path integral representation
\be \label{eq11.47}
< x\mid e^{-H\tau}\mid x > = \int {\cal D}x_{\mu}(\tau)
P \exp[-\int^{\tau}_0d\tau'L(x,\dot{x})]
\ee
where $P$ denotes path ordering. Now, taking analogy from SUSY quantum
mechanics, it has been noticed \cite{Rajeev87,Cooper88} that
one can introduce the Grassman variables $\psi_{\mu}$ via
\be \label{eq11.48}
\psi_{\mu}={1\over\sqrt
2}\gamma_{\mu},\hspace{.2in} \{\psi_{\mu},\psi_{\nu}\}=\delta_{\mu\nu}.
\ee
Then H can be written as
\be \label{eq11.49}
H = (P+A)^2+{i\over 2}\psi_{\mu}F_{\mu\nu}\psi_{\nu}
\ee
and hence the Lagrangian now becomes
\be \label{eq11.50}
L_{ss} = {1\over 4} \dot{x}_{\mu} \dot{x}_{\mu}-iA_{\mu}\dot{x}_{\mu}
-{i\over 2}\psi_{\mu}(\partial_{\tau} \delta_{\mu\nu}+F_{\mu\nu})\psi_{\nu}
\ee
which is invariant under the SUSY transformations
\be \label{eq11.51}
\delta x_{\mu} = - i\epsilon\psi_{\mu}; \delta\psi_{\mu}=\epsilon
\dot{x}_{\mu}.
\ee
We now obtain for the path integral
\be \label{eq11.52}
G(x,x;\tau)=\int {\cal D}x_{\mu}(t){\cal D} \psi_{\mu}
exp[-\int^{\tau}_0d\tau'L_{ss}(x,\dot{x})],
\ee
where we impose antiperiodic boundary conditions on the fermion at 0
and $\tau$. Since the fermionic path integral is quadratic, we can
perform the functional integral over the fermionic degrees of
freedom exactly for arbitrary $F_{\mu\nu}$.

The result of the fermionic path integral is
\be
 {{\rm{Det}} ^{1/2} (i \partial_{\tau} \delta_{\mu \nu} + F_{\mu \nu}) \over
{\rm{Det'}} ^{1/2} (i \partial_{\tau} \delta_{\mu \nu}) },
\ee
where the prime denotes the omission of the zero mode. To evaluate the
determinant one puts $F_{\mu \nu}$ in skew diagonal form:
\be
F_{\mu \nu}
= \left[\matrix{ 0 & x_1 & 0 & 0 \cr
 -x_1 & 0 & 0 & 0 \cr
 0 & 0 & 0 & x_2 \cr
0 &0 & -x_2 &0 \cr}\right].
\label{skew}
\ee
and imposes either periodic or antiperiodic boundary conditions.
One obtains for the fermion determinant using antiperiodic boundary conditions
relevant for $G$ \be
\prod_i \cosh {1 \over 2} \int_0 ^{\tau} x_i d \tau.
\ee
and for the fermion determinant with  periodic boundary conditions relevant
for the determination of the index:
\be
({2 \over \tau})^n \prod_i  [-\sinh {1 \over 2} \int_0 ^{\tau} x_i d \tau'].
\ee

Thus for any arbitrary external field one can explicitly perform the
fermionic path integral and is left with the purely bosonic path integral
which, for example, could by performed numerically by Monte Carlo
techniques.
For the particular case of a constant external field one can further
explicitly perform the resulting bosonic path integral. When
\be
A_\mu = - {1 \over 2} F_{\mu \nu} x_{\nu},
\ee
with $F_{\mu\nu}$ being a constant matrix,
one obtains for the remaining bosonic path integral:
\be
\prod_i {1 \over 2} {x_i \over \sinh {x_i \tau \over 2}}
\ee

$ G$ and
$G_5$ are easy to calculate in two and four Euclidean dimensions for
the constant external field case \cite{Cooper88} using this method.
 In two dimensions, where
only $F_{01}=B$ exists one has that ($x_1 =B$) \cite{Cooper88}
\be \label{eq11.54}
G = B \tanh (B\tau) ;\hspace{.2in}  G_5 = B = {1\over 2}.
\epsilon_{\mu\nu}F_{\mu\nu} \ee

It is worth remarking here that the same result has been obtained
with more difficulty by Akhoury and Comtet \cite{Akhoury84}.
 In four dimensions one has a simpler derivation of the result of
Schwinger \cite{Schwinger51} rewritten in Euclidean space. One finds
that
\be
x_1 = [{\bf F}+({\bf F}^2 -    {\bf G}^2)^{1/2}] ^ {1/2},~
x_2  = [{\bf F}-({\bf F}^2 -{\bf G}^2)^{1/2}] ^ {1/2},
\ee
where
\[
{\bf F} = {1 \over 4} F_{\mu \nu}  F_{\mu \nu}
\]
\[
{\bf G}= {1 \over 4} F_{\mu \nu}  F^{\ast} _{\mu \nu}
\]
so that
\be
\prod_i  x_i =  {1 \over 4} F_{\mu \nu}  F^{\ast} _{\mu \nu}.
\ee
{}From this we obtain that:
\[
G(x,x;\tau) =  F_{\mu \nu}  F^{\ast} _{\mu \nu}
\prod_{i=1}^2 { \cosh {x_i \tau \over 2}  \over \sinh {x_i \tau \over 2}}
\]
\be
G_5(x,x;\tau) = {1 \over 4} F_{\mu \nu}  F^{\ast} _{\mu \nu}.
\ee

Once one has obtained the Green's function it is easy to reconstruct
the effective action which allows one to obtain the rate of pair
production from a strong external electric field.  The effective
action is given by
\be
S_{eff} = \int dx L(x),\hspace{.2in} L(x) = \int_{\epsilon}^{\infty} ds s^{-1}
e^{-m^2 s} G(s)
\ee

% Sections 12,13,14 [as of December 27, 1993]

\section{Singular Superpotentials}
\label{sec12}

So far, we have only considered nonsingular superpotentials $W(x)$ which give
rise to the supersymmetric partner potentials $V_1(x)$ and  $V_2(x)$.
This choice of the superpotential was based on the ground
state wave function $\psi_0(x)$ of  $V_1(x)$ by
the relation $W(x)=-{\psi}_0^{\prime}(x)/{\psi}_0(x)$. In this
section, we describe a general procedure for constructing all possible
superpotentials which yield a given potential $\tilde {V}(x)$ upto an additive
constant \cite{Panigrahi93}.
This general procedure is based on any arbitrary solution
${\phi}(x)$ of the Schr\"odinger equation for  $\tilde {V}(x)$,
rather than just the ground state wave function.
We shall see that singular superpotentials
are given by $W_{\phi}=-{\phi}^{\prime}/{\phi}$ and the singularities
are located at the zeros of excited wave functions. Such singularities
produce interesting properties for the potentials  $V_{1(\phi)}(x)$ and
$V_{2(\phi)}(x)$ obtained
via the Riccati equations  $V_{1(\phi)}(x)=W^2(x)-W^{\prime}(x)$
and  $V_{2(\phi)}(x)=W^2(x)+W^{\prime}(x)$. Singular superpotentials
are responsible for
negative energy eigenstates for  $V_{1(\phi)}(x)$
and often give rise to a breakdown of
the degeneracy of energy levels for  $V_{1(\phi)}(x)$ and  $V_{2(\phi)}(x)$
\cite{Jevicki84,Casahorran91,Roy85,Imbo86,Panigrahi93}.
Another interesting physical phenomenon occurs
if one considers a singular superpotential resulting from a solution
$\phi(x)$ for an energy $E$ in the classical energy continuum.
\cite{Panigrahi93} The isospectral family of  $V_{1(\phi)}(x)$ is then
found to have bound states (normalised
eigenfunctions) at energy $E$. Thus SUSY QM
provides a systematic procedure
for generating potentials possessing the purely
quantum mechanical phenomenon of bound states in the
continuum\cite{von Neumann29,Stillinger75}.

\subsection{General Formalism, Negative Energy States and Breakdown of
the Degeneracy Theorem}
\label{sec12.1}

Given any nonsingular potential $\tilde {V}(x)$ with eigenfunctions
${\psi}_n(x)$ and eigenvalues $E_n~
(n=0,1,2,...)$, let us now enquire how one can
find the most general superpotential $W(x)$ which will give $\tilde {V}(x)$
upto an additive constant \cite{Panigrahi93}. To answer this question
consider the Schr\"odinger
equation for $\tilde {V}(x)$:
\be \label{eq6}
-{\phi}^{\prime\prime}+ \tilde {V}(x){\phi} = {\epsilon\phi}
\ee
where ${\epsilon}$ is a constant energy to be chosen later. For convenience,
and without loss of generality, we will always
choose a solution $\phi(x)$ of eq. (\ref{eq6}) which vanishes
at $x=-\infty$.
Note that whenever ${\epsilon}$ corresponds to one of the eigenvalues $E_n$,
the solution ${\phi}(x)$ is the eigenfunction ${\psi}_n(x)$.
If one defines the quantity  $W_{\phi}=-{\phi}^{\prime}/{\phi}$ and takes
it to be the superpotential, then clearly the partner potentials generated by
$W_{\phi}$ are
\be \label{eq7}
V_{2(\phi)}=W_{\phi}^2+W_\phi^\prime~,~~~
V_{1(\phi)}=W_{\phi}^2-W_\phi^\prime=\frac {\phi^{\prime\prime}}{\phi}
=\tilde {V}(x)-\epsilon~,
\ee
where we have used eq.~(\ref{eq6}) for the last step.
The eigenvalues of
$V_{1(\phi)}$ are therefore given by
\be \label{eq8}
E_{n(\phi)}~=E_n-\epsilon.
\ee
One usually
takes ${\epsilon}$ to be the ground state energy $E_0$ and $\phi$ to be the
ground state wave function $\psi_0(x)$, which makes $E_{0(0)} =0$ and
gives the familiar case of unbroken SUSY. With this choice, the superpotential
$W_0(x)=-\psi_0^\prime/\psi_0$ is nonsingular, since $\psi_0(x)$ is
normalizable
and has no zeros. The partner potential $V_{2(\phi)}$ has no eigenstate at
zero energy since $A_0 \psi_0(x)=[d/dx+W_0(x)]\psi_0(x)=0$; however,
the remaining eigenvalues of $V_{2(\phi)}$ are degenerate with those of
$V_{1(\phi)}$.

Let us now consider what happens for other choices of $\epsilon$, both below
and above the ground state energy $E_0$.
For $\epsilon < E_0$, the solution $\phi(x)$ has no
nodes, and has the same sign for the entire range $-\infty<x<+\infty$.  The
corresponding superpotential $W_\phi(x)$ is nonsingular.  Hence the eigenvalue
spectra of $V_{1(\phi)}$ and $V_{2(\phi)}$ are completely degenerate and the
energy eigenvalues are given by eq.~(\ref{eq8}).  In particular, {$E_{0
(\phi)} = E_0 - \epsilon$ is positive.  Here, $W_\phi$ has the same sign at
$z = \pm \infty$, and we have the well-studied case of broken
SUSY \cite{Witten81}.

If the constant $\epsilon$ is chosen in the range $E_0 < \epsilon <
E_1$, then the solution $\phi(x)$ of eq.~(\ref{eq8})
will have only one node, say
at the point $x = x_s$.  Near the point $x = x_s$, the node of $\phi(x)$
makes the superpotential $W(x)$ singular
\be \label{eq9}
\phi~=~a(x-x_s),~~~~\phi^\prime~=~a,~~~~W(x) = -(x-x_s)^{-1},
\ee
and the partner potentials have the behavior [using eq.~(\ref{eq7})]
\be   \label{eq10}
V_{1(\phi)}~=~0,~~~~V_{2(\phi)}~=~2(x - x_s)^{-2}.
\ee

It is well-known \cite{Frank71,Narnhofer74,Lathouwers75}
that for any singular potential $V(x) = \lambda(x -
x_s)^{-2}$, the behavior of the wave function at $x = x_s$ is governed by
the value of $\lambda$.  For values $\lambda~>\frac{3}{4}$, the potential
$V(x)$ has a ``strong" singularity which forces $\psi(x_s) = 0$ and the
range of $x$ is effectively broken into two disjoint pieces $x < x_s$ and
$x > x_s$ with no communication between them.  For $-\frac{1}{4}<\lambda
<\frac{3}{4}$, the potential has a singularity of ``intermediate" strength.
It is not strong enough to make $\psi(x_s)$ vanish and in fact the two regions
$x < x_s$ and $x > x_s$ do communicate and one has the full range
$-\infty < x < +\infty$.  Here, in principle, one can have singular wave
functions which require self-adjoint extensions \cite {Frank71};
in what follows,
we will deal with regular solutions.  Finally, for $\lambda < - \frac{1}{4}$,
the Hamiltonian is unbounded from below.

{}From eq. (\ref {eq10}), it is clear that $V_{2(\phi)}$ has a ``strong"
singularity
which makes $\psi(x_s)=0$.  Thus, the problem of finding the eigenstates of
$V_{2(\phi)}$ is really two separate problems; one in the range $-\infty<x_s$
and the other in the range $x_s < x < +\infty$.  Clearly, this is very
different from the range of $V_{1(\phi)}$ which is the whole real axis
$-\infty < x < +\infty$.   Hence, in general, the degeneracy theorem obtained
from SUSY for the spectra of $V_{2(\phi)}$ and $V_{1(\phi)}$ is not valid.
The above discussion can be readily extended to all values of the constant
$\epsilon$ - the only difference being the number of poles present.  A summary
of results is given in Table 12.1.

So far we have considered superpotentials (both nonsingular and singular)
which give rise to a nonsingular potential $V_{1(\phi)}$.  However, if the
potential $V(x)$ itself has singularities, then so must all superpotentials
which produce it.  Consider the case of a simple pole singularity at $x = x_0$.
Then, near $x = x_0$, the singular superpotential $W(x) = g(x - x_0)^{-1}$
gives the following behavior to the corresponding partner potentials:
\be \label{eq11}
V_{1(\phi)}~=~\frac {g(g+1)}{(x-x_0)^2}~~~~,~~~~V_{2(\phi)}~=~
\frac{g(g-1)}{(x-x_0)^2}~~~.
\ee

Then clearly, for $g>\frac{3}{2}$ or $g<-\frac{3}{2}$, both $V_{1(\phi)}$ and
$V_{2(\phi)}$ have ``strong" singularities and both have two disjoint
regions $x < x_0$ and $x < x_0$.  Here, one expects degenerate energy
levels (corresponding to broken or unbroken SUSY).  Similarly, for
$- \frac{1}{2} < g < + \frac{1}{2}$, both $V_{1(\phi)}$ and $V_{2(\phi)}$ have
``intermediate" strength singularities and the whole region
$-\infty < x < +\infty$ is valid for both.  Here again, one obtains degeneracy.
However, for the two regions $- \frac{3}{2} < g < - \frac{1}{2}$ and
$\frac{1}{2} < g < \frac{3}{2}$ only one of the partner potentials has a
``strong" singularity, whereas the other has a singularity of ``intermediate"
strength.  Therefore, the two potentials have different Hilbert spaces and,
in general, degeneracy is not there.  The above discussion is borne out by
the work of Jevicki and Rodrigues \cite {Jevicki84}
who consider a superpotential of the
form $W(x) = g/x - x$.

The general discussion of singular superpotentials, negative energy states
and breakdown of the degeneracy theorem is best illustrated with some
specific examples.

Consider the harmonic oscillator potential $\tilde {V}(x) = x^2$.
The energy eigenvalues
are $E_n = 2n$ and the first few eigenstates are
\be \label{eq12}
\psi_0(x)~=~\frac{e^{-x^2/2}}{\pi^{1/4}},~~\psi_1(x)~=~\frac{\sqrt{2}~x~
e^{-x^2/2}}{\pi^{1/4}},~~\psi_2(x)~=~\frac{(2x^2 - 1)e^{-x^2/2}}{\sqrt{2}~
\pi^{1/4}}.
\ee
Using the solution $\phi = \psi_0(x)$, one gets the usual nonsingular
superpotential $W_0 = -\psi_0\prime/\psi_0 = x$, leading to the following
partner potentials and eigenvalues:
\be \label{eq13}
V_{1(0)} = x^2 - 1,~~V_{2(0)} = x^2 + 1,~~E_{n(0)}^{(1)} = 2n,~~
E_{n(0)}^{(2)} = 2n + 2,~~n = 0,1,2,...
\ee
This is standard unbroken SUSY.  The harmonic oscillator potential can also
be obtained if one starts from the solution $\phi = \psi_1(x)$.  One gets
the singular superpotential $W_1 = -\psi_1^\prime/\psi_1 = x - 1/x$.  It has
a pole at $x=0$ and is a special case of the form discussed in Ref.
\cite {Jevicki84}. The
partner potentials are
\be \label{eq14}
V_{1(1)} = x^2 - 3  ,  V_{2(1)} = x^2 + {2 \over x^2} - 1 ,
\ee
and both are exactly solvable \cite {Dutt88,Gendenshtein83}.
The two eigenvalue spectra are
\be \label{eq15}
E_{n(1)}^{(1)} = 2n-2,~~~ E_{n(1)}^{(2)} = 4n+4,~~~n=0,1,2,...
\ee
There occurs a negative energy state in $V_{1(1)}$ at $E_{0(1)}^{(1)} = -2$.
This is expected since we chose $\phi$ to be the first excited state and
$E_{1(1)}^{(1)} =0$, which pushes the ground state to a negative energy.
Proceeding along the same lines, we can get the harmonic oscillator
potential using yet another superpotential.  Taking $\phi = \psi_2 (x)$,
one gets $W_2= -\psi_2^\prime/\psi_2=x -4x/(2x^2 -  1)$, which has poles at
$x=\pm 1/\sqrt{2}$.  The corresponding potentials are
\be
V_{1(2)} = x^2 - 5,~~~V_{2(2)} = x^2 -3 +\frac {8(2x^2+1)} {(2x^2-1)^2}.
\ee
The eigenvalue spectrum of $V_{1(2)}$ is $E_{n(2)}^{(1)} = 2n - 4$ indicating
two negative energy states at $-4$ and $-2$, which was anticipated.  The
potential $V_{2(2)}$ is not analytically solvable.  The ``strong" singularities
at $x = \pm 1/\sqrt{2}$ break the x-axis into three disjoint regions
$-\infty < x < -1/\sqrt{2}, -1/\sqrt{2} < x < 1/\sqrt{2}, 1/\sqrt{2} < x <
+\infty$ and the degeneracy theorem breaks down.

In summary, the potentials $V_{1(0)}$ and $V_{2(0)}$ give a realization of
SUSY QM with a non-degenerate zero energy ground state and pairing of
excited states.  For the case of $V_{1(1)}$ and $V_{2(1)}$ the eigenvalues
are given by eq. (\ref {eq15}).
There is partial degeneracy of the spectrum [see Fig. 12.1]
due to the fact that the potentials are symmetric.  The states which
are missing in $V_{2(1)}$ are the even parity states, since the $2/x^2$
barrier requires the vanishing of wave functions at $x = 0$.  Finally, for
the partner potentials $V_{1(2)}$ and $V_{2(2)}$, degeneracy is completely
absent.

Our second example is the Morse potential $\tilde{V}(x)=A^2+B^2e^{-2\alpha x}
-B(\alpha +2A)e^{-\alpha x}$.  The superpotential based on the
ground state is $W_0(x) = A - Be^{-\alpha x}$.  For concreteness, we shall
take $A = 4, \alpha = 1$ and $B = 1$.  The corresponding energy eigenvalues
are $E_{n(0)}^{(1)} = 16 - (4-n)^2$; there are four bound states with
eigenvalues $0,7,12$ and $15$ and a continuum starting above $16$.  The two
lowest eigenfunctions are \cite {Dutt88}
\be
\psi_0(x) \sim e^{-(4x+e^{-x})},~~\psi_1(x) \sim
e^{-(3x+e^{-x})}~(7-2e^{-x}).
\ee
The supersymmetric partner potentials constructed from the
nonsingular superpotential $W_0(x)=4-e^{-x}$ are
\be
V_{1(0)} = 16-9e^{-x} + e^{-2x},~~~V_{2(0)} = 16-7e^{-x}+ e^{-2x}.
\ee
The spectrum of $V_{2(0)}$ is identical to that of $V_{1(0)}$,
except that
there is no state at zero energy.  An alternative singular superpotential
$W_1(x)$ which also yields the Morse potential is
\be \label{eq19}
W_1(x) = -~\frac{\psi_1^\prime}{\psi_1}~=~
\frac{21 - 15e^{-x} + 2e^{-2x}}{7-2e^{-x}}.
\ee
There is a simple pole at $x_s = -\ln(3.5)$.  The
partner potentials are
\be \label{eq20}
V_{1(1)} = 9-9e^{-x} + e^{-2x}, ~~~V_{2(1)} = \frac{441-567e^{-x} +
281e^{2x} - 56e^{-3x} + 4e^{-4x}}{(7-2e^{-x})^2}
\ee
By construction, the potential $V_{1(1)}$ is the Morse potential and
$V_{2(1)}$ has a singularly at $x_s = -\ln(3.5)$.  Expanding
$V_{2(1)}$ about the singular point gives $V_{2(1)}~\sim~2/(x-x_s)^2$,
which requires the wave function to vanish at $x = x_s$.
This effectively breaks the
potential and the real axis into two parts:~~$V_{2(1)}$ (left) for
$-\infty < x < x_s$ and $V_{2(1)}$ (right) for $x_s < x < +\infty$.
The potentials and their eigenstates are plotted in Figure 12.2.
The potential $V_{1(1)}$ has energy levels located at -7, 0, 5, 8
and a continuum above 9.
As expected from Table 12.1, there is a negative energy state at -7.  We have
calculated the energy levels of $V_{2(1)}$ (right) and $V_{2(1)}$ (left)
numerically.  They are $E_{(1)}^{(2)}$ (right) = 6.08, 8.65  with a
continuum above 9 and $E_{(1)}^{(2)}$ (left) = 22.96, 50.69, 82.16,
116.8,....  The potential $V_{2(1)}$ (left) has an infinite number of
bound states and as is obvious from the spectra, the degeneracy between
$V_{2(1)}$ and $V_{1(1)}$ is completely broken by the ``strong" singularity
in $W_1(x)$.  Note that since the Morse potential is asymmetric, no partial
degeneracy remains.  This is unlike our first example (harmonic oscillator)
which had symmetric
potentials and $V_{2(1)}$ and $V_{1(1)}$ had a partial degeneracy.

Our last example is the reflectionless potential $\tilde {V}(x)=
A^2 - A(A+1)\sech^2x$.
The motivation for considering this example is to make contact with
singular superpotentials considered by
Casahorran and Nam \cite {Casahorran91}.
The ground state wave function is given by $\psi_0 = (\sech x)^A$.  The
nonsingular superpotential $W_0 = A\tanh x$ comes from $\psi_0$.  Using the
property of shape invariance \cite {Dutt88,Gendenshtein83}
one readily obtains $\psi_1(x) =
\tanh x (\sech x)^{A-1}$ as the wave function for the first excited state.
If one takes $\phi = \psi_1$, the resulting superpotential is $W_1
= -\psi_1^\prime/\psi_1 = -2A\cosech~ 2x + (A-1)\coth x$, which agrees with eq.
(3.14) of ref. \cite {Casahorran91}.
The energy eigenvalues for $V_{1(0)}$ are $E_{n
(0)}^{(1)} = A^2 - (A - n)^2$ and one has unbroken SUSY.  The potentials
generated by $W_1(x)$ are $V_{1(1)} = (A-1)^2 - A(A+1)\sech^2 x, ~V_{2(1)}=
(A-1)^2 - A(A-1)\sech^2x + 2\cosech^2 x$.  These potentials are shape
invariant \cite {Dutt88,Gendenshtein83}
and their eigenvalues are $E_{n(1)}^{(1)} = (A-1)^2 -
(A-n)^2,~E_{n(1)}^{(2)} = (A-1)^2 - (A-2n-3)^2$, indicating partial
degeneracy, as expected for symmetric potentials.  Similarly, our previous
experience indicates that the partner potentials constructed from $W_2(x)$
will have no degeneracy.  Thus, we have not only reproduced some families
of singular superpotentials previously considered in the literature, but
given the general method to construct new ones.

In conclusion, we have shown how excited state wave functions can be used
to construct singular superpotentials in SUSY QM.  Our method provides a
complete and unified picture of the origin of negative energy states and
the presence or absence of degeneracy.  Although our technique is perfectly
general, our examples were taken from the class of shape invariant
potentials, since these are analytically solvable.

\subsection{Bound States in the Continuum}

In 1929, Von Neumann and Wigner \cite{von Neumann29} realized that it was
possible to
construct potentials which have quantum mechanical bound states embedded in
the classical energy continuum (BICs). Further developments, by many authors
\cite {Stillinger75,Meyer-Vernet82,Friedrich85,Robnik79}
have produced more examples and a better understanding of the kind of potential
that can have such bound states, although there is not as yet a fully
systematic approach. These authors have also suggested possible applications to
atoms and molecules.
Capasso et al. \cite{Capasso92} have recently reported direct
evidence for BICs by constructing suitable potentials using semiconductor
heterostructures grown by molecular beam epitaxy. Finally, it is interesting
to note that BICs  have found their way into a recently
written text \cite{Ballentine90}.

In this subsection, we show how one can start from a potential
with a continuum of energy eigenstates, and use
the methods of SUSY QM to generate
families of potentials with bound states in the continuum [BICs]
\cite{Pappademos93}. Basically, one is using the technique of generating
isospectral potentials (discussed in Sec. 7) but this time
starting from states in the continuum. The method preserves the spectrum of the
original potential except it adds these discrete BICs at selected energies.
As illustrative examples,
we compute and graph potentials which have bound states in the
continuum starting from a null potential representing a free particle and the
Coulomb potential.

{\bf (a) One Parameter Family of BICs}

Consider any spherically symmetric potential $V(r)$ which vanishes
as $ r \rightarrow \infty $.
The radial $s$-wave Schr\"odinger equation for the reduced wave function $u(r)$
(in units where $ \hbar=2m=1$) is
\begin{equation} \label{111}
-u''+V(r)~ u(r) = E~ u(r),
\end{equation}
where we have scaled the energy and radial variables such that
all quantities are dimensionless. Eq. (\ref{111}) has a classical
continuum of positive energy solutions which are clearly not normalizable.

As we have seen in Sec. 7, the Darboux \cite{Darboux82} procedure for
deleting and then reinstating the ground state $u_0(r)$ of a potential $V(r)$,
generates a family of potentials $\hat{V}(r;\lambda )$ which have the
same eigenvalues as $V(r)$. These isospectral potentials are labeled by a real
parameter $\lambda$ in the ranges $\lambda > 0$ or $\lambda < -1$.
The isospectral potential $\hat{V}(r;\lambda )$  is given  in terms of the
original potential $V(r)$ and the original ground state wave function $u_0(r)$
by \cite{Nieto84,Khare89a,Keung89}
\begin{equation} \label{vhat}
\hat{V}(r;\lambda)=
V(r)-2[\ln(I_0+\lambda)]''=
V(r)-\frac{4u_0u_0'}{I_0+\lambda}+\frac{2u_0^4}{(I_0+\lambda)^2},
\end{equation}
where
\begin{equation} \label{i0}
I_0(r) \equiv \int_0^r u_0^2(r')dr'.
\end{equation}

Except in this section, $u_0$ was taken to be the nodeless, normalizable ground
state wave function of the starting potential $V(r)$. However, it is easy to
generalize the above equations to the case where $u_0(r)$
is any solution of eq. (\ref{111}) with arbitrary energy $E_0.$ If $u_0(r)$ has
nodes, this leads to singular superpotentials and to singularities in the
partner potential $V_2(r).$ However, when the original state at $E_0$ is
re-inserted, the resulting family of potentials  $\hat{V}(r;\lambda)$ is free
of singularities \cite{Panigrahi93}. Our results are best
summarized in the following statement:

{\bf Theorem:} Let $u_0(r)$ and $u_1(r)$ be any two nonsingular solutions of
the Schr\"odinger equation for the potential $V(r)$ corresponding to
arbitrarily selected energies $E_0$ and $E_1$ respectively. Construct a new
potential $\hat{V}(r;\lambda)$ as prescribed by eq. (\ref{vhat}). Then, the two
functions
\begin{equation}
\hat{u}_0(r;\lambda)=\frac{u_0(r)}{I_0+\lambda},
\label{u0hat}
\end{equation}
and
\begin{equation}  \label{u1hat}
\hat{u}_1(r;\lambda)= (E_1-E_0)u_1 + \hat{u}_0 ~W(u_0,u_1),
\end{equation}
[where $W$ denotes the Wronskian, $W(u_0,u_1) \equiv u_0u_1'-u_1u_0'$]
are solutions of the Schr\"odinger equation for the new potential
$\hat{V}(r;\lambda)$ corresponding to the same energies $E_0$ and $E_1$.

While the new potential in eq. (\ref{vhat}) and the new wave functions in eq.
(\ref{u0hat}) were  originally inspired by SUSY QM, the easiest proof of the
above theorem  is by direct substitution. One simply computes
$-\hat{u}_i''+\hat{V}(r;\lambda)\hat{u}_i$ (i=0,1), with the wave functions
$\hat{u}_i$ given in the theorem. After straightforward but tedious algebraic
manipulations, one gets $E_i\hat{u}_i$, thus establishing the theorem. The
algebra is considerably simplified by using the following identity for the
Wronskian of two solutions of the Schr\"odinger equation:
\begin{equation}
\frac{d}{dr} W(u_0,u_1) = (E_0-E_1) u_0u_1.\label{wder}
\end{equation}

Let us now take $u_0$ to be a scattering solution at a positive
energy $E_0=k^2$ of a potential $V(r)$ which vanishes at $r$=$\infty$. Taking
$u_0(r=0)=0$ satisfies one of the required boundary conditions, but clearly
$u_0$ oscillates as $r \rightarrow \infty$ and has an amplitude which does not
decrease. Consequently, the integral $I_0(r)$ in eq. (\ref{i0}) now
grows like $r$ at
large $r$ and $\hat{u}_0$ is now square integrable for $\lambda >0$, while the
original wave function $u_0$ was not. Negative values of $\lambda $ are no
longer allowed.  Therefore, we see that all the potentials
$\hat{V}(r;\lambda)$ have a BIC with energy $E_0$. Note from eq. (\ref{u0hat})
that $\hat{u}_0$ has the same zeros as the original $u_0$. At zeros of $u_0$,
$\hat{V}(r;\lambda)$ and $V(r)$ are equal. All the other
oscillatory solutions of the Schr\"odinger equation with $V(r)$ get
transformed into oscillatory solutions to the new Schr\"odinger equation with
$\hat{V}(r;\lambda)$ with the same energy. In particular, note that
$\hat{u}_1(r;\lambda)$ remains a non-normalizable scattering solution of the
corresponding Schr\"odinger equation.

We note that the new potential $\hat{V}(r;\lambda)$ in eq. (\ref{vhat}) and the
BIC at energy $E_0$ are formed using the corresponding wave function $u_0(r).$
Any other state, say $u_1(r)$, is transformed into a solution of the new
Schr\"odinger equation by the operation given in eq. (\ref{u1hat}) which
involves both $u_0$ and $u_1$. The central column of Table 12.2  gives a
convenient overview of the relationship of the potentials $V$ and $\hat{V}$
and the solutions of the corresponding Schr\"odinger equations.

We now give two examples to explicitly illustrate how one applies the above
procedure to obtain potentials possessing one BIC.

{\bf Example 1: Free Particle on the Half Line.} Consider a free particle on
the half line ($V\equiv 0$ for  $0\le r < \infty$). We
choose $u_0=\sin kr$, the spherical wave solution, corresponding to energy
$E_0=k^2>0,$ which vanishes at $r=0.$ The integral $ I_0$ given in
eq.(\ref{i0}) becomes \begin{equation} \label{i0sp}
I_0=[2kr-\sin(2kr)]/(4k).
\end{equation}
We observe that $ I_0 \rightarrow r/2 $ as $ r \rightarrow \infty $.

The potential family $\hat{V}$, defined in eq.(\ref{vhat}) becomes
\begin{equation} \label{v0hsp}
\hat{V}(r;\lambda)
=\frac{32~k^2~\sin^4kr}{D_0^2}-\frac{8~k^2\sin(2kr)}{D_0}
\end{equation}
with
\begin{equation} \label{d0sp}~~~
D_0(r;\lambda)=2kr-\sin(2kr)+ 4k\lambda.
\end{equation}
$\hat{V}$ has a BIC at energy $E_0=k^2$ with wave function
\begin{equation}  \label{u0hsp} ~~~
\hat{u}_{0}(\lambda)=4k~\sin kr/D_0.
\end{equation}

For special values of the parameters $k$ and $\lambda$, the potential $
\hat{V}$
and its BIC wave functions are shown in Figs. 12.3a and 12.3b. The original
null
potential has now become an oscillatory potential which asymptotically has a
1/$r$ envelope. The new wave function at $ E_0=k^2$ also has an additional
damping factor of 1/$r$ which makes it square integrable. As $ u_{0}$ appears
in
the numerator of $ \hat{V}$, eq. (\ref{vhat}), every node of $\hat{u}_0$ is
associated with a node of $\hat{V}$ but not every node of $\hat{V}$ produces a
node of $\hat{u}_0$. The value of the eigenenergy $ E_0$ is clearly above the
asymptotic value, zero, of the potential. Evidently, the many oscillations of
this potential, none of them able to hold a bound state, conspire in such a way
as to keep the particle trapped.

The parameter $\lambda$ which appears in the denominator function
$D_0(r;\lambda)$ plays the role of a damping distance; its magnitude indicates
the value of r at which the monotonically growing integral $I_0$ becomes a
significant damping factor, both for the new potential and for the new wave
function. This is illustrated graphically in
Figs. 12.3a and 12.3b which are drawn
for very different values of $\lambda$. [Note that the wave functions
shown in the figures are not normalized].
The parameter $\lambda$ must be
restricted to values greater than zero in order to avoid infinities in
$\hat{V}$ and in the wave functions. In the limit $\lambda \rightarrow \infty$,
$\hat{V}$ becomes identical to $V$.

{\bf Example 2: Coulomb Potential.} Here $ V =
Z/r,$
and the unbound, reduced $l$ = 0 wave function satisfies the Schr\"odinger
equation eq.(\ref{111}), which can be written in standard form
\begin{equation} \label{couleq}
u_{0}''+(1-2\frac{\tilde{\eta}}{\rho})u_{0}=0
\end{equation}
with $ \rho =\sqrt{E} r$ and $ \tilde {\eta} =Z/2 \sqrt{E}.$

The solutions involve
confluent hypergeometric functions which in the asymptotic limit approach sine
waves phase-shifted by a logarithmic term. Useful expressions for these
solutions in the regions near and far from the origin are available in the
literature \cite{Abramowitz64,NBS52}. Stillinger
and Herrick \cite{Stillinger75}, following the
method of Von Neumann and Wigner \cite{von Neumann29}, have
constructed BIC potentials
and wave functions for the case of the repulsive Coulomb potential. Here we use
our theorem to construct a one-parameter family of isospectral potentials
containing a BIC. The procedure is the same for both positive and negative $Z$;
the only difference being in the sign of $ \tilde {\eta}$. The formal
expressions for the BIC potentials and wave functions have been given above,
eqs.(\ref{vhat}) and (\ref{u0hat}), in terms of $u_{0}$.

The positive
energy solution of eq.(\ref{couleq}) can be written in the usual form
\cite{Abramowitz64,NBS52,Stillinger75} as the real function
\begin{equation}
u_{0}(\rho)=C_0(\tilde {\eta})~e^{-i\rho}M(1-i\tilde {\eta},2,2i\rho),
\end{equation}
where
\begin{equation}
C_0(\tilde {\eta})=(e^{-\pi\tilde{\eta}/2})\mid\Gamma(1+i\tilde {\eta})\mid
\end{equation}
and $M(a,b;z)$ is
Kummer's function. Using tabulated expressions for the Coulomb wave functions
\cite{NBS52} and doing the integral for $ I_0$ numerically,
we have obtained the
BIC wave functions for representative values of $\lambda.$ The corresponding
one-parameter family of potentials obtained by the SUSY procedure is given in
eq. (\ref{vhat}) with $ V_0~=~Z/r$.

The results are displayed in Fig. 12.4.  Fig. 12.4a shows the BIC partner to
the attractive Coulomb potential for $\lambda$ = 1, $k$ = 1, and $Z$ = -2.
Fig. 12.4b shows the (unnormalized)
wave function of the bound state in the continuum for this
potential at $E_0$ = $k^2$. For comparison the original Coulomb
potential and wave function are also shown dotted. It is seen that the
potential which holds a bound state of positive energy shows an oscillatory
behavior about the Coulomb potential, $ V_C$, as is also evident from the form
of eq. (\ref{vhat}) for $\hat{V}$. Since the oscillating component vanishes
whenever $u_0$ vanishes, we have $\hat{V}~=~V$ at each node of $u_0$. Compared
to the original, unnormalizable wave function, the BIC wave function in both
cases shows a damped behavior due to the denominator function. This is also
seen in the figures.

A similar behavior is also expected for other spherically symmetric potentials
with a continuous spectrum of positive eigenvalues. For arbitrary
one-dimensional potentials, where the range extends from $-\infty$ to
$+\infty$,
the situation is not so clear cut. Our method works for the Morse
potential which is steeply rising on the negative $x$-axis with correspondingly
damped wave functions. It also works for the case of a particle in a constant
electric field for similar reasons. For potentials, such as $V(x)= - V_0
{}~\sech^2 x$, the integral $I_0$ in eq. (\ref{i0}) is not
convergent if the starting
point is chosen at $ -\infty $, and it gets negative contributions if the
starting point is selected at finite $x$-values. This leads to a vanishing
denominator function in the expressions for some wave functions which makes
them unacceptable.

{\bf (b) Two Parameter Family of Potentials}

In the previous section
(a), we have seen how to generate a one-parameter family of
potentials with one BIC.
We now show how this procedure can be
extended to construct two-parameter families which contain two BICs.

In constructing the new wave functions for the one-parameter family, eq.
(\ref{vhat}), we observe that the denominator function given in eq.
(\ref{u0hat})  was all that was needed to create the BIC, while the operation
in eq. (\ref{u1hat}) ensured that the wave functions for all the other states,
there represented by $\hat{u}_1$, are a solution to the new potential. Note
again, there is nothing special about the ordering of the two energy values nor
the relative magnitude of $E_0$ and $E_1$, therefore we can repeat this
procedure by applying the theorem to the wave functions and the potential of
the one-parameter family, but this time we transform the state at $E_1$ into a
BIC. The state at $E_0$, which already is a BIC, is transformed in the step of
eq. (\ref{u1hat}), suitably modified, to become a solution to the new
potential. In this way we obtain the two parameter family of potentials

\begin{equation} \label{v2hat}
\hat{\hat{V}}(r;\lambda ,\lambda_1)
=\hat{V}-2[\ln(\hat{I}_1+\lambda_1)]'' =
\hat{V}-\frac{4\hat{u}_1\hat{u}_1'}{\hat{I}_1+\lambda_1}+
\frac{2\hat{u}_1^4}{(\hat{I}_1+ \lambda_1)^2}
\end{equation}
with the solutions of the corresponding Schr\"odinger equation
\begin{equation} \label{u02hat}
\hat{\hat{u}}_0= (E_0-E_1)\hat{u}_0 + \hat{\hat{u}}_1
{}~W(\hat{u}_1,\hat{u}_0),
\end{equation}
\begin{equation} \label{u12hat}
\hat{\hat{u}}_1= \frac{\hat{u}_1}{\hat{I}_1+\lambda_1},
\end{equation}
and
\begin{equation}
\hat{I}_1 \equiv \int_0^r \hat{u}_1^2(r')dr'.
\end{equation}

\noindent The precise relationship of the new potential and its wave functions,
which are now both BICs, is illustrated in the last column of Table 12.2.

While the compact form of Eqs. (\ref{v2hat} - \ref{u12hat}) explicitly shows
the method of construction, it is useful to observe that the integral
$\hat{I}_1$ can be conveniently re-cast into a simpler form which contains
integrals of the form
\begin{equation} \label{ii}
I_i = \int_0^r u_i^2(r')dr',
\end{equation}
involving the original wave functions only. Making
use of eq. (\ref{u1hat}) for $\hat{u}_1$, we get
\begin{equation}
\hat{I}_1 =
\int_0^r \left[(E_1-E_0)^2 u_1^2 +\frac{u_0^2 W^2}{(I_0+\lambda )^2}
+2(E_1-E_0)\frac{u_0u_1}{(I_0+\lambda )}W  \right]dr'. \label{iu1hat}
\end{equation}
The second term is integrated by parts as
\begin{equation}
\int_0^r\frac{u_0^2}{(I_0+\lambda )^2} W^2(r') dr' =
\left. \frac{-W^2}{I_0+ \lambda }
\right|_0^r +\int_0^r \frac{2WW'}{(I_0+\lambda )}dr' .
\end{equation}
We
now use eq.~(\ref{wder}) for the derivative of a Wronskian of two solutions of
the Schr\"odinger equation to rewrite the second term and observe, that it
exactly cancels the last term in eq. (\ref{iu1hat}). We therefore have
\begin{equation} \hat{I}_1 (r) = \frac{-W^2(r)}{I_0+\lambda } +(E_1-E_0)^2
{}~I_1(r).
\end{equation}
Here we have made use of the fact that our boundary
conditions imply that $W(0)=0.$

As an example, we evaluate the two-parameter potential
\begin{equation}
\hat{\hat{V}}=V-2 \left[ \ln \left\{ (I_0+\lambda )
{[(E_1-E_0)^2I_1-\frac{W^2(r)}{I_0+\lambda } +\lambda_1]}  \right\} \right]
''.
\end{equation}
The argument of the logarithm can be rewritten as
\begin{equation}
(E_1-E_0)^2
I_0I_1 - W^2(r) +\lambda \lambda_1 +\lambda (E_1-E_0)^2I_1 +\lambda_1 I_0.
\label{arg}
\end{equation}

We happen to have transformed first the state at energy $E_0$ into a BIC and
then, in the second step, the state at $E_1,$ which introduced the parameters
$\lambda$ and $\lambda_1 .$ Let us now consider applying our procedure in the
reverse order, that is let us first transform the state at energy $E_1$ into a
BIC and then the state at energy $E_0$, producing the parameters $\mu$ and
$\mu_1$. For this situation, the argument corresponding to eq.~(\ref{arg}) is
\begin{equation} \label{arg2}
(E_1-E_0)^2 I_0I_1 - W^2(r) +\mu \mu_1 +\mu_1 (E_1-E_0)^2I_0
+\mu I_1.
\end{equation}
Clearly, one expects symmetry. This is
guaranteed if the parameters are related by
\be
 \mu =\lambda (E_1-E_0)^2 ~,~\mu_1 =\lambda_1 / (E_1-E_0)^2.
\ee
This also leads to the same two-parameter
wave functions.  We also note that transforming any state twice by eq.
(\ref{u0hat}) does not create a second denominator or anything else new, but
simply changes the value of the parameter $\lambda$ as shown in
ref. \cite{Khare89a}.
Finally, relation (\ref{u02hat}) ensures that all other eigenstates will be
solutions to the new potentials.

Shown in Fig. 12.5 is a potential with two BICs at energies $E_0=1$, $E_1=4$.
Clearly, the above procedure can be readily extended to obtain multi-parameter
families with multiple BICs at arbitrarily selected energies.

Our discussion of BICs has been restricted to
effectively one dimensional problems, and as
stated before there is now some experimental evidence for the existence of
BICs under appropriately chosen conditions \cite{Capasso92}. Recently, there
has also been a computation which claims the existence of BICs in QED in
three dimensions \cite{Spence91}. These authors have also speculated that
the BIC energies they compute in $e^{+}e^-$ scattering in QED are in fact
in reasonable agreement with unusual peaks observed in recent heavy ion
experiments \cite{Cowan86}.

\section{Parasupersymmetric Quantum Mechanics and Beyond}
\label{sec13}

In the last few years, exotic quantum statistics have been widely discussed
in the literature. For example, in two space dimensions,
one can have a one-parameter family of statistics interpolating between Bose
and Fermi statistics \cite{Leinaas77}. On the other hand, in three and
higher space dimensions parafermi and parabose statistics
\cite{Green53,Volkov60,Greenberg65,Ohnuki82}
are the natural extensions of the usual Fermi and Bose statistics. In
particular, whereas Fermi and Bose statistics describe the two
one-dimensional representations of the permutation group,
parafermi and parabose statistics describe the higher dimensional
representations of the same group. In view of the fact that the SUSY
has provided us with an elegant symmetry between fermions and bosons, it is
natural to enquire if there exists a generalization which includes the above
exotic statistics. Such a question was raised many years ago in the context of
parastring models \cite{Ardalan74}, but the specific symmetry algebra was still
the usual SUSY one.

In this section, we study the possibility of having a
symmetry between bosons and parafermions. We shall
construct parasupersymmetric quantum mechanics (PSQM) of a boson and a
parafermion of order $p (= 1,2,3...)$. It turns out that whereas in the
usual SUSY QM, the symmetry generators obey structure relations that involve
bilinear products, in PSQM of order $p$, the structure relations involve
products of $(p+1)$ parasupersymmetry
(PARASUSY) charges. Various consequences of this algebra are
also discussed.

It is worth adding here that historically the PSQM of order 2 was
introduced first \cite{Rubakov88} and its various consequences were
discussed \cite{Beckers89,Durand89,Spiridonov91,Andrianov91a,Merkel90}.
Initially it was felt that the
generalization to order $p$ was not possible in the sense that the PSQM
of order $p$ cannot be characterized with one universal algebraic
relation \cite{Durand91}. However, later on, one was indeed able to
construct such a PSQM of order $p$ \cite{Khare92}.

Very recently, new forms of quantum statistics called orthofermi and
orthobose statistics, have been constructed \cite{Mishra91}. It is
then natural to construct orthosupersymmetric quantum mechanics (OSQM) where
there is symmetry between a boson and an orthofermion of order $p$
\cite{Khare93c}. Unlike the PSQM of order $p$, one finds that the
structure relations now involve only bilinear products of the
symmetry generators. Various consequences of SUSY QM, PSQM and OSQM are also
discussed in this section. In particular, it is worth pointing out that
whereas
in SUSY QM and OSQM of order $p$, the energy eigenvalues are  necessarily
nonnegative, in PSQM of order $p$ they need not be so.

\subsection{Parasupersymmetric Quantum Mechanics}
\label{sec13.1}

In order to motivate the algebra of PSQM, let us recall that in SUSY QM
the symmetry between a boson and a fermion is characterized
by the algebra
\be \label{eq13.1}
Q^2 = 0 =Q^{\dag 2} ,~ [H,Q] = 0 ,~ QQ^{\dag}+Q^{\dag}Q=2H .
\ee
Note that there is an extra factor of 2 on the right hand side compared to
the algebra given in Sec. 2. This results from assuming $m = 1$ in this
section, in conformity with the notation followed by the various
authors in this field. The SUSY QM algebra is easily motivated by recalling
that
the fermionic operators $a,a^{\dag}$ satisfy the algebra \be \label{eq13.2}
a^2 = 0 = a^{\dag 2} , \{a,a^{\dag}\} = 1 .
\ee
A useful representation of $a$ and $a^{\dag}$ is in terms of the $2 \times
2$ matrices
\be \label{eq13.3}
a =\pmatrix{0 & 1\cr 0 & 0\cr} , a^{\dag} =\pmatrix{0 & 0\cr 1 & 0\cr} .
\ee

Let us now consider the parafermi operators $b,b^{\dag}$ of  order $p$
(=1,2,...
 )
which are known to satisfy the algebra \cite{Ohnuki82}
\be \label{eq13.4}
  (b)^{p+1}  = 0 = (b^{\dag})^{p+1} ,
\ee
\be \label{eq13.5}
[[b^{\dag},b],b]  = -2b , [[b^{\dag},b],b^{\dag}] = 2b^{\dag} .
\ee
On letting
\be
J_+ = b^{\dag} , J_- = b , J_3 = {1\over 2}[b^{\dag},b] ,
\ee
it immediately follows from eqs. (\ref{eq13.4}) and (\ref{eq13.5}) that
the operators
$J_{\pm}$ and $J_3$ satisfy the SU(2) algebra
\be
[J_+,J_-]= 2J_3 , [J_3, J_{\pm}]= \pm J_{\pm} .
\ee

Let us now choose $J_3$ to represent the third component of the spin
${p\over 2}$ representation of the SU(2) group as given by
\be \label{eq13.6}
J_3 = diag ({p\over 2},{p\over 2}-1,...-{p\over 2}+1,-{p\over 2}).
\ee

It is now easily seen \cite{Khare92} that the
operators $b$ and $b^{\dag}$ can be represented by the following
$(p+1)\times(p+1)$ matrices $[\alpha,\beta = 1,2..., (p+1)]$
\be
(b)_{\alpha \beta}=C_{\beta} \delta_{\alpha,\beta+1} ;
(b^{\dag})_{\alpha \beta} = C_{\beta} \delta_{\alpha+1,\beta}
\ee
where
\be
C_{\beta} = \sqrt{\beta(p-\beta+1)} = C_{p-\beta+1} .
\ee

It is easily checked that the operators $b$ and $b^{\dag}$ indeed satisfy
the algebra as given by eqs. (\ref{eq13.4}) and (\ref{eq13.5}).
One can now ask as to what multilinear relation
is satisfied by $b$ and $b^{\dag}$ apart from the one given by
eq. (\ref{eq13.4})? It turns out that the nontrivial relation is
\be
b^pb^{\dag}+b^{p-1}b^{\dag}b +...+bb^{\dag}b^{p-1} + b^{\dag}b^p = {1\over
6}p(p
 +1)(p+2)b^{p-1},
\ee
where one has $(p+1)$ terms on the left hand side. As expected, for $p = 1$
this
reduces to the bilinear relation for the fermionic operators
given in eq. (\ref{eq13.2}).

This relation between $b$ and $b^{\dag}$ strongly suggests that one may have
an analogous multilinear relation in the algebra of PSQM of order $p$.
To that purpose, let us choose the PARASUSY charges $Q_1$ and $Q^{\dag}_1$
as $(p+1)\times (p+1)$ matrices as given by
\be
(Q_1)_{\alpha\beta}=(P-iW_{\beta}) \delta_{\alpha,\beta+1} ;
(Q_1^{\dag})_{\alpha\beta}=(P+iW_{\alpha}) \delta_{\alpha+1,\beta}
\ee
where $\alpha,\beta = 1,2,..., (p+1)$, so that $Q_1$ and $Q^{\dag}_1$
automatically satisfy
\be \label{eq13.12}
Q^{p+1}_1 = 0 = (Q^{\dag}_1)^{p+1} .
\ee
Further, it is easily shown that the Hamiltonian $(\hbar = m = 1)$
\be \label{eq13.13}
(H)_{\alpha\beta} = H_{\alpha} \delta_{\alpha\beta},
\ee
where $(r = 1,2,..., p)$
\bea \label{eq13.14}
H_r = {P^2 \over 2}+{1 \over 2}(W^2_r-W'_r)+{1 \over 2} C_r \nonumber \\
H_{p+1} = {P^2 \over 2}+{1 \over 2}(W^2_p+W'_p)+{1 \over 2} C_p
\eea
commutes with the PARASUSY charges $Q_1$ and $Q^{\dag}_1$
provided $(s = 2,3..., p)$
\be \label{eq13.15}
W^2_{s-1} + W'_{s-1} + C_{s-1} = W^2_s - W'_s + C_s .
\ee
Here $C_1,C_2,...,C_p$ are arbitrary constants with the dimension of
energy. It turns out that the nontrivial relation between
$Q_1,Q^{\dag}_1$ and $H$ is given by \cite{Khare92}
\be \label{eq13.16}
Q_1^pQ^{\dag}_1+Q^{p-1}_1
Q^{\dag}_1Q_1+...+Q_1Q^{\dag}_1Q_1^{p-1}+Q^{\dag}_1Q^p_1 = 2pQ^{p-1}_1 H ,
\ee
and Hermitian-conjugated relations (which we shall not write explicitly)
provided \be
C_1 + C_2+ ...+ C_p = 0 .
\ee

An example is in order at this stage to illustrate the structure of PSQM of
orde
 r
$p$. If one chooses
\be
W_1 = W_2 = ... = W_p = \omega x,
\ee
then it follows from eq. (\ref{eq13.15}) that in this case $(r =1,2,..., p)$
\be
C_{r+1} - C_r = 2 \omega ,
\ee
and the Hamiltonian (\ref{eq13.13}) takes a very simple form given by
\be
H = {P^2\over 2} + {1\over 2} \omega^2 x^2 - J_3 \omega ,
\ee
where $J_3$ is as given by eq. (\ref{eq13.6}). This $H$ describes the
motion of a particle with spin $p/2$ in an oscillator potential and a
uniform magnetic field. The spectrum of this Hamiltonian is
\be
E_{n,m}  = (n+{1\over 2}-m)\omega,
\ee
where $n = 0,1,2,...$ and $m = p/2, p/2-1,..., -p/2$ so that the ground
state energy of the system is negative unlike in the usual SUSY QM. It is also
clear from here that whereas the ground state is nondegenerate, the
first excited state is two-fold degenerate, etc. and finally the $p$'th and
higher excited states are $(p+1)$-fold degenerate. Of course for $p = 1$
we recover the well known results of SUSY QM.

Several comments are in order at this stage:

\begin{enumerate}
\item For arbitrary $W_r$
too one can show that the spectrum of the PSQM Hamiltonian is
$(p+1)$-fold degenerate at least starting from the $p$'th and higher excited
states. The nature of the ground and the first $(p-1)$ excited states would
however depend on the specific form of $W_r$.

\item With the PSQM Hamiltonian as given by eqs. (\ref{eq13.13})
and (\ref{eq13.14}), one can associate $p$
supersymmetries characterized by the corresponding $p$ superpotentials
$W_r(r=1,2,..., p)$ and the corresponding $p$ SUSY QM Hamiltonians are given
by
\be \label{eq13.22}
H^{(r)}_{SUSY} = \pmatrix{H_r-{1\over 2}C_r & 0 \cr 0 &
H_{r+1}-{1\over 2} C_r\cr} ,
\ee
where $H_r$ is given by eq. (\ref{eq13.14}).

\item Apart from $Q_1$, there are $(p-1)$ other conserved,
independent PARASUSY charges $Q_2,Q_3,..., Q_p$ defined by $(s = 2,3,... p)$
\bea
Q_s &=& (P-iW_{\beta})\delta_{\alpha,\beta+1} \quad {\rm if} \quad \beta
\not = s, \nonumber \\
    &=& - (P-iW_{\beta})\delta_{\alpha,\beta+1} \quad {\rm if} \quad \beta = s,
\eea
all of which commute with the Hamiltonian (\ref{eq13.14}) and also
satisfy the PSQM algebra as given by eqs. (\ref{eq13.12})
and (\ref{eq13.16}) \cite{Khare92}.

\item There is an interesting application of PSQM in nonrelativistic quantum
mechanics. As we have seen in Sec. 3, given any Hamiltonian $H_1$ with $p$
bound states with energies $E_1,E_2,..., E_p$ and the corresponding
eigenfunctions $\psi_1,\psi_2,..., \psi_p$, one can always generate $p$ other
Hamiltonians $H_2,H_3,..., H_{(p+1)}$ with the same spectrum as $H_1$ except
that $1,2,..., p$ levels respectively are missing from them. Further
in that case there are $p$ supersymmetries with the corresponding
Hamiltonians being precisely given by eq. (\ref{eq13.22}). Besides,
the $p$ constants $C_1, C_2,..., C_p$ are related to the energy
eigenvalues by $(r = 1,2,..., p)$
\be
C_r = {2\over p} [E_p+E_{p+1}+...+E_{r+1}-(p-1)E_r+E_{r-1}+...+E_1] .
\ee
Thus instead of associating the symmetry algebra $sl(1/1)\otimes
SU(2)$ or $U(1)\otimes SU(2)$ as we did in Sec. 3, one can also
associate parasupersymmetry of order $p$ to the hierarchy of
Hamiltonians $H_1,H_2,..., H_p$ \cite{Khare92}.

\item What is the most general solution of the relation
(\ref{eq13.15}) which must be satisfied in order to have the PSQM of order
$p$? Treating $W_r\pm W_{r-1}$ as the two variables, eq. (\ref{eq13.15})
can be reduced to a simple nonlinear equation which can be
immediately solved \cite{Rubakov88}. This is however not very useful
as it gives us $W_r+W_{r-1}$ for a given $W_r-W_{r-1}$. Ideally, we
would like to know the most general solution for $W_2$ for a given
$W_1$ and then using this $W_2$, one would recursively obtain
$W_3,W_4,..., W_p$. Unfortunately, this problem is still unsolved.
Of course shape invariant potentials satisfy
eq. (\ref{eq13.15}) but shape invariance is clearly not necessary. As
discussed above, given any $H$ with $p$ bound states, one can always
construct $W_1,W_2,....W_p$ which will satisfy eq. (\ref{eq13.15}).
\end{enumerate}

One unsatisfactory feature of the PSQM of order $p$ is that except
for $p = 1$, $H$ cannot be directly expressed in terms of the PARASUSY
charges $Q_1$ and $Q^{\dag}_1$. This is because, in eq. (\ref{eq13.16}), $H$
is multiplied by $Q^{p-1}_1$ whose inverse does not exist. Another
unsatisfactory feature is that unlike in SUSY QM, the energy eigenvalues
are not necessarily nonnegative and there is no connection between
the nonzero (zero) ground state energy and the broken (unbroken)
PARASUSY. It turns out that in case all the constants
$C_1,C_2....C_p$ are chosen to the zero, then one can take care of
both of these unsatisfactory features \cite{Khare93b}. In
particular, in that case $H$ as given by eq.(\ref{eq13.14}) can be
expressed in terms of any one of the $p$ PARASUSY charges $Q_s$ by
$(s=1,2,..., p)$
\be \label{eq13.25}
H = {1\over 2}[(Q_s^{\dag}
Q_s-Q_sQ^{\dag}_s)^2+Q^{\dag}_sQ_sQ_sQ^{\dag}_s]^{1/2
 }
\ee

It is immediately clear from here that all the energy eigenvalues are
necessarily nonnegative and that the ground state energy being 0
($ >$ 0) corresponds to unbroken (broken) PARASUSY. Further, one
can also show that in this case all the excited states are always
$(p+1)$-fold degenerate while the nature of the ground state will
depend on the specific form of $W_s$ \cite{Khare93b}. In the limit
of $p\rightarrow \infty$ it has been shown that the relation
(\ref{eq13.25}) reduces to
\be
H = {1\over 2} Q_s Q^{\dag}_s,
\ee
which can be termed as the PSQM of infinite order whose algebra
corresponds to that of Greenberg's infinite statistics
\cite{Greenberg90}.

There is an interesting application of this version of PSQM to the
strictly isospectral Hamiltonians discussed in Sec. 7. In particular,
one can show that any potential $V_1\equiv W^2_1 - W'_1$ with at least
one bound state, forms PSQM of order $p$ along with its SUSY partner
potential $V_2 = W^2_1+W'_1$ and with the strictly isospectral
potential families $V_2 (x,\lambda_2),
V_2(x,\lambda_3),...,V_2(x,\lambda_p)$ where
$\lambda_2,\lambda_3,..., \lambda_p$ are arbitrary parameters which
are either $> 0$ or $< - 1$ \cite{Khare93e}.

Eq. (\ref{eq13.25})  shows an explicit expression for the Hamiltonian $H$ which
involves just one charge $Q_s$ but an overall square root. There is an
alternative expression for $H$ where it is not necessary to take the square
root, but which involves all the $p$ PARASUSY charges \cite{Khare93b}:
\be \label{eq13.27}
2H = Q_rQ^{\dag}_r+Q^{\dag}_rQ_r+{1\over
4}\sum^p_{s=1}(Q^{\dag}_rQ_s+Q_s^{\dag}
 Q_r-2Q^{\dag}_rQ_r)
\ee
where, $r,s=1,2,...p$ and $ s\not = r$ . Further, $H$ and $Q_r$ also
satisfy the simpler  relation
\be
Q_r Q^{\dag}_r Q_r= 2Q_r H.
\ee
In case one chooses $W_{\alpha}$ to be of form $(\alpha = 1,2..., p)$
\be
W_{\alpha} = - {\lambda+\alpha-1\over x},
\ee
then one has a model for conformally invariant PSQM of order $p$. In
this case one can show that the dilatation operator $D$ and the conformal
operator $K$ defined by
\be
D = -{1\over 4} (xP + Px), K = {1\over 2} x^2
\ee
satisfy relations analogous to those given in eqs. (\ref{eq13.25}),
(\ref{eq13.27}) and (\ref{eq13.16}) in terms of the PARASUSY charges
$Q_{\alpha}$ and parasuperconformal charges $S_{\alpha}$ defined by
$[\alpha = 1,2,..., p; i,j = 1,2,..., (p+1); r = 2,3,..., p]$
\be
(S_1)_{ij} = - x\delta_{i+1,j}
\ee
\bea
(S_r)_{ij} & = & - x\delta_{i+1,j} \quad {\rm if} \quad i \not = r
+1,\nonumber \\
           & = & x\delta_{i+1,j} \quad {\rm if} \quad i  = r +1.
\eea

\subsection{Orthosupersymmetric Quantum Mechanics}

Recently, Mishra and Rajasekaran \cite{Mishra91} have introduced new
forms of quantum statistics called orthofermi and orthobose
statistics. Orthofermi statistics contain a new exclusion
principle which is more stringent than the Pauli exclusion principle:
an orbital state shall not contain more than one particle, whatever
be the spin direction. The wave function is thus antisymmetric in
spatial indices alone with the order of the spin indices frozen. In
an analogous way, one can also define orthobose statistics. All these
properties follow provided the corresponding creation and
annihilation operators $C^{\dag}$ and $C$ satisfy
\be \label{eq13.32}
C_{k\alpha}C^{\dag}_{m\beta}\pm\delta_{\alpha\beta}
\sum^p_{\gamma=1}C^{\dag}_{m\
 gamma
}C_{k\gamma}=\delta_{km}\delta_{\alpha\beta},
\ee
\be \label{eq13.33}
C_{k\alpha}C_{m\beta} \pm C_{m\alpha}C_{k\beta}=0,
\ee
where the upper and the lower signs lead to the orthofermi and the orthobose
cases respectively and the Latin indices $k,m,...$ and the Greek
indices $\alpha,\beta,\gamma,...$ correspond to space and spin
indices respectively.

For constructing the quantum mechanics of a boson and an orthofermion, we
ignore the spatial indices in eqs. (\ref{eq13.32}) and (\ref{eq13.33})
and obtain
\be \label{eq13.34}
C_{\alpha} C^{\dag}_{\beta}+\delta_{\alpha\beta}
\sum^p_{\gamma=1}C^{\dag}_{\gamma}
C_{\gamma} = \delta_{\alpha\beta} ,
\ee
\be \label{eq13.35}
C_{\alpha}C_{\beta} = 0.
\ee
Eq. (\ref{eq13.34}) implies that
\be
C_1C^{\dag}_1 = C_2C^{\dag}_2 = ... = C_p C^{\dag}_p.
\ee

Following the discussion of the last subsection it is easy to see
that a useful representation of these operators is in terms of the
$(p+1)\times (p+1)$ matrices defined by $(\alpha,\beta=1,2,..., p; r,s =
1,2,..., (p+1))$
\be
(C_{\alpha})_{rs}=\delta_{r,1}\delta_{s,\alpha+1};
(C^{\dag}_{\alpha})_{rs}=\delta_{s,1}\delta_{r,\alpha+1}
\ee

Let us now try to write down the algebra for the OSQM of order $p$ where
there is a symmetry between a boson and an orthofermion of order $p$
\cite{Khare93c}. On comparing the algebra for the fermionic and the
orthofermionic operators as given by eqs. (\ref{eq13.2}), (\ref{eq13.34}) and
(\ref{eq13.35}) and remembering the SUSY QM algebra as given by
eq.(\ref{eq13.1})  it is easy to convince oneself that the $p$ ORTHOSUSY
charges
$Q_{\alpha},Q^{\dag}_{\alpha}$ and the Hamiltonian must satisfy the
algebra $(\alpha,\beta = 1,2,..., p)$
\be \label{eq13.38}
Q_{\alpha}Q^{\dag}_{\beta}+\delta_{\alpha\beta}
\sum^p_{\gamma=1}Q^{\dag}_{\gamma
 }
Q_{\gamma} = 2 \delta_{\alpha\beta} H,
\ee
\be \label{eq13.39}
Q_{\alpha}Q_{\beta} = 0 , [ H,Q_{\alpha} ] = 0.
\ee
Note that for $p = 1$ we recover the usual SUSY QM algebra. It is easily
checked that if we choose the $p$ ORTHOSUSY charges $Q_{\alpha}$ as
$(p+1)\times (p+1)$ matrices as given by $(r=1,2,..., (p+1))$
\be
(Q_{\alpha})_{rs}=(P-iW_{\alpha})\delta_{r,1}\delta_{s,\alpha+1};
(Q_{\alpha}^{\dag})_{rs}=(P+iW_{\alpha})\delta_{s,1}\delta_{r,\alpha+1}
\ee
and the Hamiltonian $H$ as
\be
(H)_{rs} = H_r \delta_{rs},
\ee
where $(r = 1,2,..., p)$
\bea
H_1      =  {P^2\over 2}+{1\over 2}(W^2_1+W'_1)\nonumber \\
H_{r+1}  =  {P^2\over 2}+{1\over 2}(W^2_1-W'_1)
\eea
then the OSQM algebra as given by eqs. (\ref{eq13.38}) and
 (\ref{eq13.39}) is indeed satisfied provided
\be
W^2_{\alpha} + W'_{\alpha} = W^2_{\beta} + W'_{\beta} .
\ee
Note that this condition directly follows from the OSQM relations
\be
Q_1Q^{\dag}_1 = Q_2Q^{\dag}_2 ... = Q_pQ^{\dag}_p ,
\ee
which follow from the OSQM relation (\ref{eq13.38}). Thus, unlike PSQM, the
constants $C_1,C_2,...$ are not allowed in OSQM. The various consequences of
the OSQM have been discussed in detail in \cite{Khare93c}. At this point, it
may
be worthwhile to make a relative comparison of SUSY QM with PSQM and OSQM.

\begin{enumerate}
\item First of all, the close similarity in structure between
OSQM and PSQM must be noted. For order $p$, both are based on
$(p+1)\times (p+1)$ matrices and the structure of $H$ in the two cases
is very similar. The chief difference between the two is the absence
of the constants $C_1,C_2...,C_p$ in the former while in the latter
they may or may not be zero.

\item Whereas the ground state energy is zero $(> 0)$ in
case  the SUSY or the ORTHOSUSY is unbroken (spontaneously broken), in general,
there is no such restriction in the PSQM case and the
energy eigenvalues can even be negative. However, in the special case
when $C_1,C_2,..., C_p$ are all zero then the PSQM has similar prediction
as the other two.

\item Whereas in the SUSY QM, all the excited states are necessarily
two-fold degenerate, in OSQM of order $p$ they are necessarily
$(p+1)$-fold degenerate. On the other hand, in the PSQM of order $p$, all
the levels starting from the $p$'th excited state (and above) are necessarily
$(p+1)$-fold degenerate except when all the constants are zero in which
case all the excited states are necessarily $(p+1)$-fold degenerate.
\end{enumerate}

\section{Omitted Topics}
\label{sec14}

So much work has been done in the area of SUSY QM in the last 12 years that
it is almost impossible to cover all the topics in such a review. We have
therefore decided to give a brief description of some of the omitted
topics. For each topic, a few
references are provided, so that interested readers can go back and trace other
references and get a good idea of the developments.

(i) Supermathematics
\cite{Berezin65}.

(ii)SUSY in Atomic Physics
\cite{Kostelecky84,Rau86}

In a series of papers, Kostelecky et al. \cite{Kostelecky84}
have discussed the relationship
between the physical spectra of different atoms and ions using
SUSY QM. In particular, they have suggested that the helium and hydrogen
spectra come from SUSY partner potentials. This connection has been
commented upon by Rau \cite{Rau86}.

(iii)SUSY in Condensed Matter and Statistical Physics
\cite{Sourlas85,Cardy85,Shapir85,Nambu85,Efetov82,Brezin84}

Ideas from SUSY have been used to give insight into random magnetic
fields in Ising-like models, polymers, electron localization
in disordered media and ferromagnets. These topics are covered in a set
of review talks found in the Proceedings of the Conference on Supersymmetry
held at CNLS in 1983 \cite{Kostelecky85}.

(iv)Index Theorem and SUSY
\cite{Atiyah63,Alvarez83,Friedan85,Alvarez84,Alvarez85}

The Atiyah-Singer index theorem can be related to understanding
the index of the Dirac operator on suitably defined spaces. From our
previous discussion of SUSY and the Dirac equation, it is not
surprising that the index of the Dirac operator is related to the
Witten index of the related SUSY QM. Using
techniques similar to calculating the fermion propagator in an external
field, it has been possible to give a proof of the Atiyah-Singer
index theorem using the supersymmetric representation of the index theorem.
This work is best described in lectures of Alvarez-Gaume
given at the Bonn Summer School \cite{Alvarez85}. The question
of axial anamolies are best phrased using these methods.

(v)Factorization Method and Solvable Potentials
\cite{Schrodinger40,Infeld51,Stahlhofen89}

The method of factorization which can be traced as far back as
Bernoulli(1702)  and Cauchy(1827) can be shown to be
exactly equivalent to solving potentials by SUSY and shape invariance with
translation. The reader interested in the history as well as
an  excellent presentation of the factorization method and its connection
with SUSY is referred to the work of Stahlhofen \cite{Stahlhofen89}.

(vi)Group Theory Method and Solvable Potentials
\cite{Alhassid83}

In the doctoral thesis
of Jainshi Wu, it is proven that there is a one to one correspondence between
using the differential realizations of the $S0(2,1)$ potential group and the
factorization method of Infeld and Hull \cite{Infeld51}. Thus all the
solvable potentials found
by Infeld and Hull can be also obtained using group theory methods. By
extending
the symmetry group to $SO(3,1)$ and $SU(3,1)$ the Yale group was also able to
study several three dimensional scattering problems.

(vii)Quasi-Solvable Potentials
\cite{Shifman89,Turbiner88}

These potentials, for which a finite number of
states can be determined analytically, are intermediate between
non-solvable and analytically solvable potentials discussed in this review.
Quasi-solvable potentials have been discussed by Shifman and Turbiner
\cite{Shifman89,Turbiner88} and using the techniques of SUSY QM
a large number of new quasi-solvable problems have been
discovered \cite{Jatkar89}.

(viii)SUSY Breaking and Instantons
\cite{Witten82,Salomonson82,Cooper83,Abbott83,Khare84b}

One of the least understood problems in supersymmetric field theories is
that of the origin of SUSY breaking. One suggestion for SUSY breaking is that
it is of
dynamical origin and that instantons are
responsible for it.  As a testing ground of these ideas, the role of instantons
 in the dynamical breaking of SUSY has been studied extensively in
various quantum mechanical models.

(ix)Propagators for SUSY Partner Potentials
\cite{Das90,Jayannavar93}

Since the propagator for a system with a given potential is determined
from the Hamiltonian, and the various Hamiltonians in a heirarchy of
SIP are related by
\be
Q(a_s)H^{(s)} = H^{(s+1)} Q(a_s),
\ee
one can derive recursion relations for the propagators of the heirarchy.
If one member the heirarchy has a known propagator (such as the harmonic
oscillator or the free particle) then it is easy to use these recursion
relations to derive the propagators for the heirarchy.  This also allows one
to calculate path integrals for SIP.

(x) SUSY and N-body Problem
\cite{Shastry92,Freedman90}

A novel algebraic structure is found for $1/r^2$ family of many body problems
which provide a novel link between SUSY and quantum
integrability \cite{Shastry92}. Also of considerable interest is a
supersymmetric generalization of a N-particle quantum mechanical
model with combined harmonic and repulsive forces \cite{Freedman90}.

(xi) Coherent states for SIP
\cite{Fukui93}

Recently a Lie algebraic treatment of the SIP has been given and using it
coherent states have been constructed for these potentials.

(xii) SUSY and New Soliton Solutions
\cite{Kumar87}

It is well known that the stability equation for the kink solutions in scalar
field theories in 1+1 dimensions is a Schr\"{o}dinger-like equation. It has
been
shown that new scalar field theories with kink solutions can be obtained by
considering the isospectral deformation of this problem.

(xiii) SWKB and Tunneling
\cite{Sil93,Dutt91b}

Expression for transmission coefficient T through potential barriers has been
obtained within the SWKB approximation and it has been shown that
the analytic continuation of T for the
inverted SIP (with translation) leads to the exact bound state spectrum.

(xiv) Time-Dependent Pauli Equation
\cite{Kost93}

Kostelecky et al. \cite{Kost93} have
succeeded in  factorizing the time dependent Pauli Equation and utilizing the
SUSY to solve the Pauli Equation for a time dependent spatially uniform
magnetic
induction.

(xv) SUSY and Nuclear Physics
\cite{Iachello85,Baye87,Amado91}.

Relations between the spectra of even-even and neighbouring even-odd nuclei
has been obtained by using the concepts of SUSY \cite{Iachello85}. Also, Baye
\cite{Baye87} has shown that the deep and shallow nucleus-nucleus
potentials which have been successfully used in the past are in fact SUSY
partner potentials. It has also been suggested that the relationships
seen between the energy levels of adjacent superdeformed nuclei can be
understood in terms of SUSY \cite{Amado91}

{\bf Acknowledgments} It is a pleasure to thank all of our many collaborators
without whose efforts this work would not have been accomplished. We would
especially
like to thank Joe Ginocchio for carefully reading the manuscript. We would
like to thank Los Alamos National Laboratory for its hospitality during the
writ
ing
of the manuscript. We gratefully acknowledge the U.S. Department of
Energy for partial support.

\pagebreak

\newpage

\noindent{\bf FIGURE CAPTIONS}\\
\vspace{0.2in}

\noindent Figure 2.1: The energy levels of two supersymmetric partner
potentials  $V_1$ and $V_2$. The figure corresponds to unbroken SUSY.
The energy levels are degenerate except that $V_1$ has an extra state at zero
energy $E_0^{(1)} = 0$. The action of the operators $A$ and $A^{\dag}$ in
connecting eigenfunctions is shown.
\vspace{0.2in}

\noindent Figure 2.2:  The infinite square well potential $V=0$ of width
$L=\pi$
and its supersymmetric partner potential $2{\rm cosec}^2 x$ in units
$\hbar=2m=1$. The ground state of the infinite square well has energy $1$. Note
the degenerate higher energy levels at energies $2^2,3^2,4^2,...$
\vspace{0.2in}

\noindent Figure 4.1: Self-similar superpotentials $W(x)$ for various values of
the deformation parameter $q$. The curve labeled H.O. (harmonic oscillator)
corresponds to the limiting case of $q=1$. Note that only the range $x \ge 0$
is plotted since the superpotentials are antisymmetric $W(x)=-W(-x)$.
\vspace{0.2in}

\noindent Figure 4.2: Self-similar potentials $V_1(x)$ (symmetric about $x=0$)
corresponding to the superpotentials shown in Fig. 4.1 . The curves are taken
from ref. \cite{Barclay93}.
\vspace{0.2in}

\noindent Figure 4.3: A double well potential $V_1(x)$ (solid line) and its
single well supersymmetric partner $V_2(x)$ (dotted line). Note that these two
potentials are shape invariant \cite{Barclay93} with a scaling change of
parameters. The energy levels of $V_1(x)$ are clearly marked.
\vspace{0.2in}

\noindent Figure 5.1: Diagram showing how all the shape invariant potentials of
Table 4.1 are inter-related by point canonical coordinate transformations.
Potentials on the outer hexagon have eigenfunctions which are hypergeometric
functions whereas those on the inner triangle have eigenfunctions which are
confluent hypergeometric functions. These two types are related by suitable
limiting procedures. The diagram is from ref. \cite{Gangopadhyaya94}.
\vspace{0.2in}

\noindent Figure 7.1: Selected members of the family of potentials with energy
spectra identical to the one dimensional harmonic oscillator with $\omega=2$.
The choice of units is $\hbar=2m=1$. The curves are labeled by the value
$\lambda_1$, and cover the range $0<\lambda_1\le\infty$. The curve
$\lambda_1=\infty$ is the one dimensional harmonic oscillator. The curve marked
$\lambda_1=0$ is known as the Pursey potential \cite{Pursey86} and has one
bound
state less than the oscillator.
\vspace{0.2in}

\noindent Figure 7.2: Ground state wave functions for all the potentials shown
in Fig. 7.1, except the Pursey potential.
\vspace{0.2in}

\noindent Figure 7.3: A schematic diagram showing how SUSY
transformations are used for deleting the two lowest states of a potential
$V_1(x)$ and then re-inserting them, thus producing a two-parameter
$(\lambda_1,\lambda_2)$ family of potentials isospectral to $V_1(x)$.
\vspace{0.2in}

\noindent Figure 7.4: The pure three-soliton solution of the KdV equation as a
function of position $(x)$ and time $(t)$. This solution results from
constructing the isospectral potential family starting from a reflectionless,
symmetric potential with bound states at energies $E_1=-25/16,
E_2=-1,E_3=-16/25$. Further details are given in the text and ref.
\cite{Wang90}.
\vspace{0.2in}

\noindent Figure 9.1: A ``deep" symmetric double well potential $V_1(x)$ with
minima at $x=\pm x_0$ and its supersymmetric partner potential $V_2(x)$.
\vspace{0.2in}

\noindent Figure 9.2: Supersymmetric partner potentials $V_1(x)$ and $V_2(x)$
corresponding to two choices of the parameter $x_0$ for the potentials given in
eq. (\ref{u10}). For a detailed discussion, see the text and ref.
\cite{Keung88}.
\vspace{0.2in}

\noindent Figure 9.3: The energy splitting $t=E_1-E_0$ as a function of the
separation $2x_0$ of the superposed Gaussians in the ground state wave
function $\psi_0(x)$. This figure, taken from ref. \cite{Keung88}, shows the
remarkable accuracy of the SUSY-based energy splitting computations for
a double well potential.
\vspace{0.2in}

\noindent Figure 12.1: The harmonic oscillator potential $V_{1(1)}$ and its
partner potential $V_{2(1)}$ as given by eq. (\ref{eq14}). These potentials
arise from a singular superpotential $W_1=x-1/x$ corresponding to the choice
$\phi=\psi_1(x),~ \epsilon=E_1$ in Table 12.1 . The energy levels for both
potentials are shown. Notice the negative energy state of $V_{1(1)}$ at energy
$-2$, and the partial degeneracy of the eigenvalue spectra coming from the even
parity of the potentials \cite{Panigrahi93}.
\vspace{0.2in}

\noindent Figure 12.2: The Morse and its supersymmetric partner potential [eq.
(\ref{eq20})] coming from the singular superpotential $W_1$ given in eq.
(\ref{eq19}). The singularity at $x=-{\rm ln}~3.5$ breaks $V_{2(1)}$ into two
disjoint pieces, and the eigenvalue spectra have no degeneracy
\cite{Panigrahi93}.
\vspace{0.2in}

\noindent Figure 12.3: Potentials $\hat{V}(r)$ (solid lines) possessing one
bound state in the continuum (BIC) obtained by starting from a free particle
and constructing a one-parameter ($\lambda$) family. The BIC wave function
$\hat{u}_0(r)$ is at energy $E=1$, and is shown by the dashed line. Fig. (a)
corresponds to $\lambda=0.5$ and fig. (b) corresponds to a much larger value
$\lambda=5$ .
\vspace{0.2in}

\noindent Figure 12.4: (a) A potential (solid line) with one BIC at
energy $E=0.25$ obtained by starting from an attractive Coulomb potential
(dotte
 d
line). (b) A plot of the wave functions corresponding to the potentials in part
(a).
\vspace{0.2in}

\noindent Figure 12.5:(a) An example of a potential with two BICs at energies
$E_0=1$ and $E_1=4$. (b) The wave functions at energies $E_0=1$ (dashed line)
and $E_1=4$ (dotted line) \cite{Pappademos93}.
\vspace{1in}

\noindent{\bf TABLE CAPTIONS}\\
\vspace{0.2in}

\noindent Table 4.1: All known shape invariant potentials in which
the parameters
$a_2$ and $a_1$ are related by a translation $(a_2=a_1+\alpha)$. The energy
eigenvalues and eigenfunctions are given in units $\hbar=2m=1$. The constants
$A,B,\alpha,\omega,l$ are all taken $\ge 0$. Unless otherwise stated, the range
 of
potentials is $-\infty\le x \le \infty,~ 0 \le r \le \infty$. For spherically
symmetric potentials, the full wave function is
$\psi_{nlm}(r,\theta,\phi)=\psi_{nl}(r)Y_{lm}(\theta,\phi)$. Note that the wave
functions for the first four potentials (Hermite and Laguerre polynomials) are
special cases of the confluent hypergeometric function while the rest (Jacobi
polynomials) are special cases of the hypergeometric function. Fig. 5.1,
taken from ref. \cite{Gangopadhyaya94},
shows the inter-relations between all the SIPs
in the table via point canonical coordinate transformations.

\vspace{0.2in}

\noindent Table 6.1: Comparison of the lowest order WKB and SWKB predictions
for
the bound state spectrum of the Ginocchio potential for different values of the
parameters $\lambda,\nu$ and several values of the quantum number $n$. The
exact
answer is also given. Units corresponding to $\hbar=2m=1$ are used throughout.
\vspace{0.2in}

\noindent Table 9.1: Comparison of the three lowest energy eigenvalues obtained
 by
a variational method with the exact results.
 \vspace{0.2in}

\noindent Table 12.1: Choice of the constant $\epsilon$ and the corresponding
solution $\phi(x)$ in eq. (\ref{eq6}) determines the choice of the
superpotential $W_{\phi}$ which produces any given non-singular potential
$\tilde{V}(x)$ [eq. (\ref{eq7})]. The table shows how certain choices of
$\epsilon$ give rise to singular superpotentials, negative energy eigenstates
and a breakdown of the degeneracy theorem. We have taken $\phi(x=-\infty)=0$
for convenience \cite{Panigrahi93}.
\vspace{0.2in}

\noindent Table 12.2: One and two parameter families of potentials with bound
states in the continuum. These families are generated by applying the theorem
described in the text to scattering states $u_0$ and $u_1$ at energies $E_0$
and $E_1$ in succession \cite{Pappademos93}.

\end{document}